\documentclass{article}
\pdfoutput=1

\usepackage{PRIMEarxiv}

\usepackage[utf8]{inputenc} 
\usepackage[T1]{fontenc}    
\usepackage{hyperref}       
\usepackage{url}            
\usepackage{booktabs}       
\usepackage{amsfonts}       
\usepackage{nicefrac}       
\usepackage{microtype}      
\usepackage{lipsum}
\usepackage{fancyhdr}       
\usepackage{graphicx}       
\graphicspath{{media/}}     
\usepackage{amsmath}
\usepackage[numbers]{natbib}
\usepackage{bm} 
\usepackage{rotating}
\usepackage{float} 

\newcommand{\indep}{\perp \!\!\! \perp} 
\newcommand{\beginsupplement}{%
        \setcounter{table}{0}
        \renewcommand{\thetable}{S\arabic{table}}%
        \setcounter{figure}{0}
        \renewcommand{\thefigure}{S\arabic{figure}}%
        \setcounter{equation}{0}
        \renewcommand{\theequation}{S\arabic{equation}}%
     }

\title{{Hypothesis-driven mediation analysis for compositional data: an application to gut microbiome}
}

\author{ \href{https://orcid.org/0000-0001-7859-0483}{\includegraphics[scale=0.06]{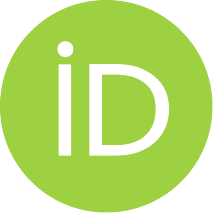}\hspace{1mm}Noora Kartiosuo}$^{1,2,3}$, 
\href{https://orcid.org/0000-0001-6295-0245}{\includegraphics[scale=0.06]{orcid.pdf}\hspace{1mm}Jaakko Nevalainen}$^{4}$,
\href{https://orcid.org/0000-0001-9365-3702}{\includegraphics[scale=0.06]{orcid.pdf}\hspace{1mm}
Olli Raitakari}$^{2,3,5}$, 
\href{https://orcid.org/0000-0001-9338-4397}{\includegraphics[scale=0.06]{orcid.pdf}\hspace{1mm}Katja Pahkala}$^{2,3,6}$, 
\href{https://orcid.org/0000-0002-8162-2260}{\includegraphics[scale=0.06]{orcid.pdf}\hspace{1mm}Kari Auranen}$^{1,7}$\\
}

\begin{document}
\maketitle

\noindent {
For a revised version and its published version, please refer to: \\
Kartiosuo, N., Nevalainen, J., Raitakari, O., Pahkala, K., \& Auranen, K. (2024). Hypothesis-driven mediation analysis for compositional data: an application to gut microbiome. Biostatistics \& Epidemiology, 8(1). \url{https://doi.org/10.1080/24709360.2024.2360375} \\ 
 \\
 
1 Department of Mathematics and Statistics, University of Turku, Turku, Finland \\
2 Research Centre of Applied and Preventive Cardiovascular Medicine, University of Turku, Turku, Finland \\
3 Centre for Population Health Research, University of Turku, Turku, Finland  \\
4 Health Sciences Unit, Faculty of Social Sciences, Tampere University, Tampere, Finland\\
5 Department of Clinical Physiology and Nuclear Medicine, University of Turku and Turku University Hospital, Turku, Finland\\
6 Paavo Nurmi Centre \& Unit for Health and Physical Activity, University of Turku, Turku, Finland\\
7 Department of Clinical Medicine, University of Turku, Turku, Finland\\
 } \\

\begin{abstract}
Biological sequencing data consist of read counts, e.g. of specified taxa and often exhibit sparsity (zero-count inflation) and overdispersion (extra-Poisson variability). As most sequencing techniques provide an arbitrary total count, taxon-specific counts should ideally be treated as proportions under the compositional data-analytic framework. There is increasing interest in the role of the gut microbiome composition in mediating the effects of different exposures on health outcomes. Most previous approaches to compositional mediation have addressed the problem of identifying potentially mediating taxa among a large number of candidates. 
We here consider causal inference in compositional mediation when \textit{a priori} knowledge is available about the  hierarchy for a restricted number of taxa, building on a single hypothesis structured in terms of contrasts between appropriate sub-compositions. Based on the theory on multiple contemporaneous mediators and the assumed causal graph, 
 we define non-parametric estimands for overall and coordinate-wise mediation effects, 
and show how these indirect effects can be estimated from empirical data based on simple parametric linear models.
The mediators have straightforward and coherent interpretations, related to specific causal questions about the interrelationships between the sub-compositions.
We perform a simulation study focusing on the impact of sparsity and overdispersion on estimation of mediation. While unbiased, the precision of the estimators depends, for any given magnitude of indirect effect,  on sparsity and the relative magnitudes of exposure-to-mediator and mediator-to-outcome effects in a complex manner. 
  We demonstrate the approach on empirical data, finding an inverse association of fibre intake on insulin level, mainly attributable to direct rather than indirect effects. 
\end{abstract}


\section{Introduction}
With improved sequencing methods, biological measurements consisting
of read counts, such as those recording the bacterial makeup of the gut flora, have become increasingly common. Such data consist  of counts of multiple and sometimes thousands of different taxa and often exhibit sparsity (excess of zero counts) as well as overdispersion (extra-Poisson variability).  Most of the current sequencing techniques provide data with an arbitrary total count. As all information carried by the taxon-specific counts relates to their proportion of the total, the observed data should be ideally interpreted as a composition where a change in the abundance of one part inevitably changes the abundances of the other parts of the composition \Citep{Gloor2017}.

Approaches based on compositional data analysis  have been endorsed for analysing  microbiome data  \Citep{Gloor2017, Silverman2017}.
In this framework, the compositionality of  observations is accounted for by scaling taxon-specific read counts by their total into proportions of the whole \Citep{Pawlowsky-Glahn2015, Pawlowsky2011}. 
Due to the resulting unit-sum constraint, data based on proportions cannot be validly analysed with standard statistical methods. Their use is enabled, however, by transforming count-based data into logratios, which contrast distinct parts of the composition against each other \Citep{Pawlowsky2011}. A common choice is the isometric logratio transformation (ilr), where the ratios are based on  pre-specified contrasts \Citep{Pawlowsky2011}. Unlike proportions, the resulting ilr coordinates can be presented in the Euclidean space and have been shown to be asymptotically multivariate normally distributed in the case of multinomial counts \Citep{Graffelman2011}. 
It has been suggested that approppriate contrasts in the analysis of microbiome data can be based on prior knowledge of the bacterial taxonomy or phylogenetic systematics \Citep{Silverman2017}, results from data-driven approaches \Citep{Gordon-Rodriguez2021}, or a specific interest in a particular taxon  \Citep{Zhang2019a}.  Such \textit{a priori} knowledge  may also be employed to restrict the analysis to concern only a smaller-dimensional subset of the taxa.

There is an increasing interest in  the role of the gut microbiome composition  in mediating the effects of different exposures on health outcomes. Mediation analysis in general aims at discovering and quantifying  causal pathways between an exposure ($X$) and a response ($Y$), and the extent to which their association is mediated by an intermediate variable (mediator $M$) \Citep{Pearl2014, VanderWeele2016}. 
Figure \ref{fig1} (a) presents a directed acyclic graph (DAG)  depicting a simple model with one mediator.  The exposure  affects the response both directly via pathway (i) and indirectly through the mediator via pathways (ii) and (iii). Figure \ref{fig1} (b) presents a DAG with a compositional mediator, such as the microbiome, where two different types of indirect effects may be of interest: the mediating effect through the entire composition (light arrows), and the mediating effects through the individual components  (e.g. arrows (ii.1) and (iii.1)). 

Different data-driven methods have been proposed to assess the mediating role of the gut microbiome and to discover specific mediating taxa \Citep{Zhang2019a, Sohn2019, Wang2020, Fu2023}. Sohn and Li \Citep{Sohn2019} derived expressions for taxon-specific as well as microbiome-wise indirect effects. Their approach was based on contrasting individual taxa against a chosen reference taxon and compositional algebra constructed in the simplex space, thus precluding the use of orthonormal (Euclidean) coordinates. Wang et al.  \Citep{Wang2020} developed a similar yet regularised model in the simplex space, further accounting for the sparsity of the read counts using Dirichlet regression.
Zhang et al.  used the ilr transformation with contrasts of a single component against the other parts of the composition  \Citep{Zhang2019a}. While the ilr transformation allows interpreting the coordinates as log-ratios of  the distinct parts of the composition, the authors only considered the mediating effect  of each component at a time, while other components were considered as nuisance and their effects were regularised using a penalisation criterion. In a somewhat similar approach, Fu et al. \Citep{Fu2023} utilised the ilr transformation  within a joint model for the response and the microbial composition to find  mediating taxa and to estimate indirect effects mediated through their relative abundances.

Although  data-driven methods are useful when the aim is to search candidates for particular mediating taxa, incorporation of available  biological knowledge in the assessment of mediators has been advocated, especially if the aim is to interpret  mediation through a causal lens \Citep{Blum2020}.  Instead of  applying a sequence of large number of different and often regularised models, the analysis should then be based on a pre-specified causal structure. 
The presence of multiple,  possibly correlated mediators then poses the key analytical problem. In our application to the microbiome, these mediators will be  ilr coordinates.

In the presence of dependencies between multiple mediators, special attention is required  when defining  and estimating mediator-specific effects. Methods to deal with multiple, causally ordered mediators  \Citep{VanderWeele2014c, Daniel2015, Imai2013} or correlated yet not causally dependent mediators  \Citep{Wang2013, Kim2019, Taguri2018} have been presented before. 
VanderWeele and Vansteelandt demonstrated for correlated mediators that the indirect effects are different when estimated based on one multivariable mediation analysis compared to applying mediation analysis to single mediator at a time \Citep{VanderWeele2014c}. 
With a focus on causally dependent mediators, these authors suggested an incremental approach to define and estimate the additional contribution of each added mediator \Citep{VanderWeele2014c}. 
However, this approach does not enable obtaining  mediator-specific indirect effects based on a single model of mediation.

Focusing on a binary response variable, Wang et al. derived a non-parametric decomposition of the overall mediated effect into coordinate-specific effects mediated via individual coordinates  of a vector-valued mediator  \Citep{Wang2013}. Their decomposition, based on a single causal model, allows interpreting the mediating effect of each coordinate in a coherent manner. In particular, the coordinate-wise indirect effects sum up to the overall indirect effect.  Kim et al. presented an alternative approach,  considering contributions attributable to each of the mediators as well as their combinations  \Citep{Kim2019}. However, under their non-parametric formulation, the coherence between the overall and mediator-specific  indirect effects does not hold.

In this article, we consider causal inference about the mediating role of the microbiome composition
when sufficient knowledge is available to form appropriate constrasts between a small number of subcompositions. We consider the taxonomic structure of the microbiome as such {\sl a priori} information although any other knowledge could be used to built the structure of causal influences. Similarly to Zhang et al. \Citep{Zhang2019a} and Fu et al. \Citep{Fu2023}, we base our approach on Euclidean (ilr) coordinates but formulate the mediated effects in the counterfactual framework. 
Following the non-parametric definitions of indirect effects for multiple contemporaneous mediators as presented by Wang et al \Citep{Wang2013}, we  consider the overall as well as mediator-specific indirect effects for each of the ilr coordinates. 
Based on a single causal {\sl a priori}  hypothesis coded as a DAG, where each coordinate pertains to comparing the relative proportions of two subcompositions of interest, these effects are straightforward to interpret. 
This hypothesis-driven approach enables estimating all mediated effects in a single mediation analysis scheme, in contrast to investigating multiple mediation analyses with altering contrast matrices \Citep{Zhang2019a}.  We then build a parametric model of mediation and specify conditions that allow estimating the mediated effects from empirical observations in terms of standard linear regression models.
As opposed to  previous  research on microbial mediation, in our simulation study we demonstrate the estimation of the indirect effects in presence of sparsity and overdispersion, often present in data on microbiome, investigating their effect on the precision and power of estimation.

The structure of the paper is as follows. In Section 2,  we review the construction of ilr coordinates based on a hypothesis-driven approach. Section 3 defines the non-parametric estimands of overall and coordinate-wise indirect effects, specifies the necessary conditions for their identifiability and presents the parametric forms of the estimands. In Section 4, we build a simulation study to illustrate the proposed method of mediation analysis. In Section 5, we analyse the mediating role of the gut  microbiome between fibre intake and insulin level.  The discussion in Section 6 includes a comparison of the proposed approach with other analysis methods in compositional mediation regarding the microbiome.

\section{Compositional counts and ilr coordinates}\label{Coda}
Consider an observation of counts  $K_{j}$, $j = 1, \ldots, J+1$, with $J+1$ distinct classes. Denote the sum of these counts, i.e., the total count, as $K$. As we are interested of the composition of the counts rather than the counts themselves, we scale the counts by their total into proportions  $p_{j} = K_{j}/K, j=1, \ldots, J+1$.  The class-specific proportions are considered as parts (i.e. components) of a composition  with the unit-sum constraint  $\sum_{j=1}^{J+1} p_{j} = 1$. To be able to use standard statistical models, the proportions need to be transformed from the $(J+1)$-dimensional simplex to a $J$-dimensional Euclidean space. We accomplish this using the isometric logratio (ilr) transformation.

\subsection{Contrasts and the ilr coordinates}
The first step in transforming proportions into ilr coordinates is to define an appropriate set of contrasts between the $J+1$ components. We here build on the idea of deriving the  contrasts from prior  biological knowledge and review how the ilr coordinates are calculated in practice \Citep{Pawlowsky-Glahn2015, Pawlowsky2011}. 

The three-level dendrogram  of Figure \ref{fig2} (a) presents an example of  an \textit{a priori} known  hierarchy, such as taxonomic knowledge, that can be used  as the basis for defining  contrasts between the five taxa using sequential binary partition (SBP) \Citep{Pawlowsky-Glahn2015}. For example, we may build an $(J+1) \times J$ SBP matrix for the 5 classes using the following \textit{balances} \Citep{Filzmoser2018}
 \begin{align}\label{taxmatrix}
\boldsymbol{\Psi} = (\psi_{jk})= \begin{bmatrix}
+1 &  0  & 0 &   0 \\
-1 & +1 & +1 & 0 \\
-1 & +1 & -1 &  0 \\
-1 & -1 & 0 & +1 \\
-1 & -1 & 0 & -1 \\
\end{bmatrix},
 \end{align}
where each column defines how the components indicated by +1 are contrasted against those indicated by -1. The first contrast compares taxon $A$ against all other taxa. Taxonomic groups $\lbrace B, C \rbrace$ and  $\lbrace D, E \rbrace$ are contrasted against each other (second column), whereas the taxa within both groups are contrasted against each other (B against C, third column; D against E, fourth column). Figure \ref{fig2} (b) describes the hierarchy that is encoded by matrix \ref{taxmatrix}. Whereas Figure \ref{fig2} (a) describes the taxonomic hierarchy, Figure  \ref{fig2} (b)  shows the hierarchy that is hypothesised to carry causally mediated effects. For example, even though the  groups ${A}$, $\lbrace B, C \rbrace$ and  $\lbrace D, E \rbrace$  are on the same taxonomic level,  the first contrast in the matrix entails the assumption that the relative proportion of taxon ${A}$ against the other groups carries the mediating effect. 
Depending on the empirical research question and other \textit{a priori} knowledge at hand, the matrix could be built in a different manner as well. 

Let $n_k^+$ and $n_k^-$ be the numbers of cells with +1 and -1, respectively, in the $k$th column of the SBP matrix $\boldsymbol{\Psi}$, and let gm() 
denote the geometric mean. The ilr transformation of proportions  $\boldsymbol{p}$ is defined as follows \Citep{Pawlowsky-Glahn2015}:
\begin{align}\label{ilrvek}
\boldsymbol{M}= \hbox{ilr}(\boldsymbol{p}) = \textbf{V}^T \hbox{log}(\boldsymbol{p}),
\end{align}
where the elements of a $(J+1) \times J$ contrast matrix \textbf{V} are:
\begin{align*}
v_{jk} = \sqrt{\frac{n^+_k n^-_k}{n^+_k + n^-_k}} 
\left[\frac{1}{n_k^+}\right]^{\hbox{I}[\hbox{sign}(\boldsymbol{\Psi}_{jk})=+]}
\left[\frac{1}{n_k^-}\right]_.^{\hbox{I}[\hbox{sign}(\boldsymbol{\Psi}_{jk})=-]} 
\end{align*}
Thus, the ilr coordinates are 
\begin{align}\label{ilrkaava}
M_{k} = \sqrt{\frac{n^+_k n^-_k}{n^+_k + n^-_k}}  \hbox{log} \frac{\hbox{gm}(p_{k}^+)}{\hbox{gm}(p_{k}^-)}, k = 1, \ldots, J.
\end{align}
In summary, the ilr coordinates presented here correspond to a scaled log-ratio of geometric means of proportions indicated by the chosen contrast matrix. 
 Whereas the original proportions $p_{j}$ lie in the simplex, the ilr-transformed coordinates reside in  the Euclidean space. Each ilr coordinate pertains to a log-ratio between two sets of components, entailing a specific hypothesis and carrying thus a straightforward interpretation.

A special way of building balances is contrasting each of the components against the remaining components. Here, we call this approach \textit{pivotal} 
\Citep{Filzmoser2018}. The pivotal SBP matrix is
\begin{equation}\label{pivotmat}
\boldsymbol{\Psi} = 
\begin{bmatrix}
+1 &  0  & 0 & \ldots & 0 \\
-1 & +1 & 0 & \ldots & 0 \\
-1 & -1 & +1 & \ldots & 0 \\
\vdots & \vdots & \vdots  & \ddots & \vdots  \\
-1 & -1 & -1 & \vdots & +1 \\
-1 & -1 & -1 & \vdots & -1 \\
\end{bmatrix}.
\end{equation}
It follows from Equation (\ref{ilrkaava}) that this choice of contrasts   leads to the following ilr coordinates \Citep{Barabesi2020}:
\begin{equation*}\label{ilrkaavapivot}
M_{k} = \sqrt{\frac{(J+1)-k}{(J+1)-k+1}} \hbox{log} \frac{p_{k}}{\sqrt[(J+1)-k]{\prod_{h=k+1}^{(J+1)} p_{h}}}, k=1, \ldots, J.
\end{equation*}
Having fixed a specific order of the $J+1$ classes, the pivotal SBP matrix $\boldsymbol{\Psi}$ leads to contrasting the first component against the remainder, then the second component against the remaining components excluding the first one, and so forth. Each class is thus used as a pivot against the remaining classes of the composition. 

Some previous analyses of compositional mediation have applied
pivot coordinates without a  strict causal \textit{a priori} structure. In practice, each of the potentially mediating taxa is then used as a first pivot in turn \Citep{Zhang2019a}. This means that a set of ilr coordinates is built altogether $J+1$ times, alternating the order of the components at each time. In any of the $J+1$ sets of coordinates, the mediating effect of only the first coordinate is of interest, with an interpretation  as the relative proportion of a single taxon contrasted to the other taxa. Unlike the balances based on a clear SBP matrix derived based on \textit{a priori} knowledge, pivot coordinates do not necessarily convey other meaningful hypotheses than contrasting the pivot element to the rest. 
To distinguish this  approach from ours, we will  refer to it as \textit{alternating pivot} coordinates. If a single balance matrix can be built as as sequence of pivots, following the structure of matrix (\ref{pivotmat}), we refer the analysis as based on pivot SBP coordinates.

\section{Compositional mediation analysis}
In this section, we first introduce the non-parametric causal quantities of interest. We then define the parametric models for the exposure, compositional mediator (i.e., the ilr coordinates) and the response and present the identifiability conditions. This leads to parametric expressions for the direct as well as overall and coordinate-wise indirect effects.

\subsection{Causal effects with a compositional mediator}\label{sec:codamed}
We  treat the total, direct and indirect effects of exposure within the potential outcomes framework \Citep{Pearl2014}. 
 Denote the levels of the binary exposure $X$ as $x$ and $x^\ast$ and let $Y$ be a continuous response variable.
 Let $Y(x^\prime,{\bm m})$ denote the response if exposure was set to level $x^\prime$ and a vector-valued mediator ${\bm M}$ to value ${\bm m}$.  In our application, this will be the $J$-vector of the ilr coordinates. 
 Then, $Y(x^\prime,{\bm M}(x^{\prime\prime}))$ is the response if $X$ was set to level $x^\prime$ and the mediator ${\bm M}$ would obtain the value it would naturally have when exposure is set at  $x^{\prime\prime}$. Here $x^\prime,x^{\prime\prime} \in \lbrace x,x^\ast\rbrace$. 

To define coordinate-wise effects, we also need more detailed manipulations of the vector-valued mediator $\bm M$. Specifically, for any $k=1,\ldots,J$, let $Y(x_0,{\bm M}_{k-}(x_1),M_{k}(x_2),{\bm M}_{k+}(x_3))$ denote the response if exposure is set at $x_0$, and $(M_1,\ldots,M_{k-1})$, $M_k$ and $(M_{k+1},\ldots,M_{J})$ would obtain values they would have when exposure is set at $x_1$, $x_2$, and $x_3$, respectively. Here $x_0,x_1,x_2,x_3\in\lbrace x,x^\ast\rbrace$. 
We assume that for each coordinate such an assignment under treatment $X$ can be done independently of the other coordinates. In our application, the plausibility of this assumption is based on the nature of a sequential binary partition, in which one balance can be set without affecting the rest of the balances. Note that with $x_0=x^\prime$ and $x_1=x_2=x_3=x^{\prime\prime}$ (and for any $k$) we obtain
as a special case that $Y(x^\prime,{\bm M}_{k-}(x^{\prime\prime}),M_k(x^{\prime\prime}),
M_{k+}(x^{\prime\prime}))$ $= Y(x^\prime,{\bm M}(x^{\prime\prime}))$.

Following the standard approach to causal mediation analysis   \Citep{Pearl2014,  VanderWeele2016}, we  define the natural direct effect (NDE),  
the natural overall indirect effect (OIE) and the total effect (TE) as follows:
\begin{align}\label{ndeniete}
\begin{array}{ll}
\hbox{NDE} &= \hbox{E}[Y(x,{\bm M}(x^\ast))] - \hbox{E}[Y(x^\ast, {\bm M}(x^\ast))],\\[3mm]
\hbox{OIE} &= \hbox{E}[Y(x,{\bm M}(x))] - \hbox{E}[Y(x,{\bm M}(x^\ast))],\\[3mm]
\hbox{TE} &=  \hbox{E}[Y(x,{\bm M}(x))] - \hbox{E}[Y(x^\ast, {\bm M}(x^\ast))].
\end{array}
\end{align}

As usual, the above effects are average direct and indirect natural effects where the average is taken with respect to the population of units (individuals) \Citep{Pearl2001}. The total effect equals the sum of the natural direct and natural indirect effects, i.e., $\hbox{TE} = \hbox{NDE} + \hbox{OIE}$. The indirect effect above refers to the effect on exposure mediated through the entire vector ${\bm M}$. 
 NDE is the expected change in the response $Y$ when the value of exposure changes from $x^\ast$ to $x$ while the mediator is set to the value it would have under $X=x^\ast$. OIE is the expected change in the response $Y$ when the mediator is changed from the level it would have at $X=x^\ast$ to the level it would have at $X=x$ while holding the  level of exposure at $X=x$.

Based on Wang et al. \Citep{Wang2013}, we define the coordinate-wise indirect effects sequentially, concerning one coordinate at a time:
\begin{align}\label{Wang_CIE}
  \hbox{CIE}_k = 
&\hbox{E}[Y(x,{\bm M}_{k-}(x),M_k(x),{\bm M}_{k+}(x^\ast)]-\\[4mm]
&\hbox{E}[Y(x,{\bm M}_{k-}(x),M_k(x^\ast),{\bm M}_{k+}(x^\ast)],~k=1,...,J. \nonumber
\end{align}

An alternative definition considers each coordinate with all remaining coordinates set at their reference levels \Citep{Kim2019}:
\begin{align}\label{Kim_CIE}
\overline{\hbox{CIE}}_k =  
&\hbox{E}[Y(x,{\bm M}_{k-}(x),M_k(x),{\bm M}_{k+}(x))]-\\[4mm]
&\hbox{E}[Y(x,{\bm M}_{k-}(x),M_k(x^\ast),{\bm M}_{k+}(x)],~k=1,...,J. \nonumber
\end{align}

In definition (\ref{Wang_CIE}), the order of  coordinates is important when interpreting the coordinate-wise indirect effects. Each $\hbox{CIE}_k$ corresponds to the expected change in  response $Y$  when an individual coordinate $M_k$  is changed from the level it would have at $X=x^\ast$ to the level it would have at $X=x$, keeping the perceding coordinates at the exposed level ($x$) and the rest at the base level ($x^\ast$).  Of note, this sequential approach complies  with the sequential binary partition used in building the ilr coordinates. Based on definition (\ref{Wang_CIE}),  the coordinate-wise indirect effects sum up to the overall indirect effect, i.e., $\hbox{OIE} = \sum_{k=1}^J \hbox{CIE}_k$ (see Supplementary material \ref{IE_selitys}).

In definition (\ref{Kim_CIE}), $\overline{\hbox{CIE}}_k$ is the expected change in the response $Y$  when an individual component $M_k$ of the mediator is changed from the level it would have at $X=x^\ast$ to the level it would have at $X=x$, while all other mediators are set to the same level $X=x^\ast$. Thus, the order of the mediators is not of importance. Under this definition, the coordinate-wise effects do not sum up to the overall indirect effect. Nevertheless, under the linear dependence of the response on the mediators parameterised in Section \ref{dgm} both of these non-parametric definitions lead to the same parametric expression.

It is important to note that  changing the level of exposure $x$ inevitably changes the entire composition due to the unit-sum constraint. This means that indirect effects on the \textit{component-wise} level  cannot be defined in a counterfactual manner. However, it is possible, and even meaningful, to consider the coordinate-wise indirect effects with this counterfactual interpretation. In particular, each of the CIE coordinates, as coded on the basis of a hierarchical hypothesis-driven contrast matrix, pertains to a specific causal question on the mediating role of the partition into two subcompositions, with the relative proportions of all other parts remaining intact.

\subsection{Parametric models}\label{dgm}
We here  define parametric models for the dependencies between the exposure, mediator and response (Figure \ref{fig1} (b)). In what follows, we denote the individual explicitly. For example, for individual $i$, the count observation is $\boldsymbol{K}^{(i)} = (K^{(i)}_1,...,K^{(i)}_{J+1})$ and the mediator $\boldsymbol{M}^{(i)} = (M^{(i)}_1,...,M^{(i)}_{J})$. 
There are three sets of   variables that confound causal influences between 
$X^{(i)}$, $Y^{(i)}$ and $\boldsymbol{M}^{(i)} $, namely  exposure-outcome confounders $C^{XY(i)}$,  exposure-mediator confounders $C^{XM(i)}$ and mediator-outcome confounders $C^{MY(i)}$. 
We here consider these confounders as categorical so that the regression models can be stratified by their levels. Let $H$ denote the set of strata pertaining to the variables $C = \{ C^{XM}, C^{XY}, C^{MY} \} $ and $P(h)$ be the probability of a unit belonging to stratum $h \in H$. The number of exposed and unexposed participants in each stratum $h$ may vary but are assumed to be at least one. 
The set of an individual's levels of confounders is denoted as $C^{(i)}$ and the stratum based on the confounders as $h^{(i)}$. 

\textbf{Parametric models for the compositional mediator and response}
As the read counts often exhibit overdispersion in the total count as well as in the class-specific counts, we define the distribution of the counts for individual $i$ $(K_{i1},  \ldots, K_{i,J+1})$ hierarchically as follows:
\begin{align} \label{mixtuuri}
&K^{(i)} \sim \hbox{NegativeBinomial}(\mu, \theta) ,\\
&(\pi_{1}^{(i)}, \ldots, \pi_{J+1}^{(i)}) \sim \hbox{Dirichlet}(\alpha_1, \ldots, \alpha_{J+1}), \nonumber \\
&(K_{1}^{(i)}, \ldots, K_{J+1}^{(i)} \vert K^{(i)}; \pi_{1}^{(i)}, \ldots, \pi_{J+1}^{(i)})  \sim \hbox{Multinomial}(K^{(i)}, \pi_{1}^{(i)}, \ldots, \pi_{J+1}^{(i)}). \nonumber
\end{align}
The sum $\alpha_S=\sum_{j=1}^{J+1}\alpha_j$ controls the sparsity of counts. In particular, with small values of $\alpha_S$ the individual's class-specific counts are sparse in the sense that they tend to concentrate in a single class while the other classes may have very small or even zero counts. The dominating class then also varies across individuals. When $\alpha_S \rightarrow \infty$, the taxon-specific counts approach the multinomial model.  Parameter $\theta$ controls the variation of the total count across individuals. When $\theta=0$, the distribution of the total count is Poisson whereas large values of $\theta$ correspond to overdispersion with respect to the Poisson model. Conditionally on the total count $K^{(i)} = \sum_{j=1}^{J+1} K_{j}^{(i)}$, the distribution of $(K_{1}^{(i)}, \ldots, K_{J+1}^{(i)})$ is Dirichlet-multinomial. 

The distribution of the counts depends on the individuals' exposure and confounders. Specifically, 
we let the parameters of the Dirichlet distribution in (\ref{mixtuuri}) depend on exposure $X^{(i)}$ as follows:
\begin{align}\label{alfaefektit}
\alpha_{j}^{(i)} \vert X^{(i)},C^{(i)}   = a_{h_0j}^{(i)} + a_{1j}X^{(i)},  j=1, \ldots, J+1.
\end{align}

The ilr coordinates, i.e., the mediators, are taken to depend on the exposure as follows: 
\begin{align}\label{mediator}
M_{k}^{(i)} \vert X^{(i)}, C^{(i)}   = \beta_{h^{(i)}0k} + \beta_{h^{(i)}1k} X^{(i)} +  \varepsilon_{k}^{(i)}, k=1 , \ldots, J, 
\end{align}
where the stratum-specific regression parameters $\beta_{h^{(i)}0k}$ and $\beta_{h^{(i)}1k}$, pertaining to the expected values of the ilr coordinates,  follow from models (\ref{mixtuuri}) and (\ref{alfaefektit})  through transformation  (\ref{ilrkaava}). However, they lack analytical expressions under sparsity and overdispersion. 
Importantly, the variation in the class probabilities and the counts across individuals is  reflected in the variation of the mediator. In addition, for any one individual, the error terms  $\varepsilon_{k}^{(i)}$ for $ k=1 , \ldots, J$, are correlated, reflecting the dependence between the ilr coordinates. This dependence is based on the choice of the contrast matrix and the correlation between the proportions.   
 We note that the linear dependence of the Dirichlet parameters on the confounders in equation (\ref{alfaefektit}) does not induce a linear model for the ilr coordinates in equation (\ref{mediator}). Therefore, for convenience, we have assumed  that all confounders are categorical or can be categorized so that the linear dependence on them remains trivially in equation (6). We return to discuss the implications of this assumption in the Discussion. 
 In our model, the exposure and  confounders are not assumed to affect the total count $K^{(i)}$. Instead, they only affect the individual-specific proportions and, subsequently, the ilr coordinates.

Lastly, the exposure and mediators are assumed to affect the response in a linear way: 
\begin{align}\label{response}
Y^{(i)} \vert X^{(i)}, \boldsymbol{M}^{(i)}, C^{(i)}  = \gamma_{h^{(i)}0} + \sum_{k=1}^{J} \gamma_{1k} M_{k}^{(i)}+ \gamma_2 X^{(i)} + e^{(i)},
\end{align} 
where the error term $ e^{(i)} $ is assumed to be normally distributed with variance $\sigma_2^2$. Of note, in this model we assume no exposure-mediator interaction.

\textbf{Identifiability} 
To enable the estimation of the natural direct and indirect effect from empirical observations, a set of assumptions is needed \Citep{Pearl2014}.  Here, we require the existence of set of variables $C$ such that the following sequentical ignorability assumptions hold \Citep{Wang2013}: 

$$\begin{array}{ll}
\hbox{(i)}~\lbrace{Y^{(i)}(x,{\bf m}),{\bm M^{(i)}}}\rbrace \indep X^{(i)}\vert C^{(i)},\\[2mm]
\hbox{(ii)}~Y^{(i)}(x^\prime,{\bm m})\indep M^{(i)}_k(x^\prime)\vert X^{(i)}=x^\prime,C^{(i)};~~ k=1,\ldots,J; x^\prime\in\lbrace x,x^\ast\rbrace .\\[2mm]
\end{array}
$$
The identifiability assumptions are discussed in more detail in Supplementary material \ref{IE_selitys}.

\textbf{Parametric expressions for the mediated effects} 
Based on the linear models (\ref{mediator}) and (\ref{response}), the non-parametric definitions (\ref{ndeniete}) and (\ref{Wang_CIE}), and the identifiability conditions,  the parametric expressions of the direct and indirect effects are  found to be: 
\begin{align}\label{par_exp}
&\hbox{NDE} = \gamma_2(x-x^\ast), \nonumber\\
&\hbox{OIE}  = \sum_{k=1}^{J} \gamma_{1k}\sum_{h\in H} \beta_{h1k}\hbox{P}(h), \\
&\hbox{CIE}_k =  \gamma_{1k}  \sum_{h \in H} \beta_{h1k}\hbox{P}(h).\nonumber 
\end{align}
The equations leading to these parametric forms are derived in Supplementary material \ref{IE_selitys}. With the above definitions, the coordinate-wise indirect effects sum up to the overall indirect effect, i.e., $\hbox{OIE} =\sum_{k=1}^{J}\hbox{CIE}_k$. 
While the OIE  describes the  overall influence transmitted through the composition (arrows (ii) and (iii) in Figure \ref{fig1} (b)), each  CIE$_k$ describes the effect of a single coordinate   (e.g. arrows ii.1 and iii.1 in Figure \ref{fig1} (b)). As the coordinate-wise effects may counteract, it is important to consider both the OIE and the CIE$_k$'s. This also means that individual CIE$_k$'s may be larger than the OIE. 
 Note that with the chosen linear parametric model for the dependence of the response on the mediators, both definitions (\ref{Wang_CIE}) and (\ref{Kim_CIE}) lead to the same parametric expressions and thus, in this parametric model, the additivity holds regardless of which of the two non-parametric definitions of the coordinate-wise indirect effects is used. 

In practice, the total effect is estimated through linear regression of the exposure on the response. The direct effect is estimated as the effect of exposure on the response controlling for the mediators based on model (\ref{response}). The indirect effects are estimated based on the multivariate linear model (\ref{mediator}) of the dependence of the ilr coordinates on the exposure and model (\ref{response}) of the dependence of the response on ilr coordinates. The standard errors of the total, direct, and indirect effects are based on the delta method (Supplementary material \ref{s_e_t}).

\textbf{A remark on a single vs. several contrast matrices.} 
In this article, we address overall and coordinate-wise indirect effects as based on a single causal DAG or, equivalently, a single SBP matrix. It is important to note,  that the relationship between the coordinate-wise and overall effects only holds in this case under the parametrisations as defined above. If alternating pivot coordinates are built, so that each of the coordinates is used as the pivot in turn, relying each time on a different SBP matrix, the sum of individual indirect pivot effects would not sum up to the overall indirect effect. Nevertheless, under linear models, regardless of how the contrast matrix is built the OIE  remains the same, as choosing a different contrast matrix is equivalent to choosing another orthonormal basis for the simplex, and the information contained in all of the coordinates still captures the information contained in the composition. Hence, the coordinates $\boldsymbol{M}$ corresponding to a contrast matrix are rotations of those with another contrast matrix and can be obtained by multiplication by an orthogonal matrix $\textbf{P}$. When regression equations (\ref{mediator}) and (\ref{response}) are formulated in terms
of the new coordinates $\tilde{\boldsymbol{M}} = \textbf{P}
\boldsymbol{M}$, the regression parameters are transformed to $\tilde{\boldsymbol{\gamma}} = \textbf{P} \boldsymbol{\gamma}$ and $\tilde{\boldsymbol{\beta}} =  \textbf{P} \boldsymbol{\beta}$, where $\boldsymbol{\gamma}$ and $\boldsymbol{\beta}$ are the vectors of parameters $\gamma_{11}, \ldots, \gamma_{1,J}$ and $\beta_{h11}, \ldots, \beta_{h1,J}$, respectively.
 It follows that $\tilde{\boldsymbol{\gamma}}^T \tilde{\boldsymbol{\beta}} = \boldsymbol{\gamma}^T  \textbf{P}
^T \textbf{P}
 \boldsymbol{\beta} = \boldsymbol{\gamma}^T \boldsymbol{\beta} $, i.e., the inner product of the two sets of coefficients is invariant under the orthogonal transformation of the basis. We demonstrate empirically that the individual coordinate-wise effects differ when using different matrices while the OIE remains the same in Section  \ref{performance} (part “Misspecification of the contrast matrix”).

\section{Estimation of mediated effects in compositional data under sparse and overdispersed counts }
To demonstrate the estimation of effects  mediated through a compositional entity we set up a simulation study. Specifically, we investigate the effect of overdispersion and sparsity of the underlying counts on the estimation. 

\subsection{Simulation set-up}
We investigated the estimation of the OIE and the CIE's in different settings with varying the extents of sparsity  $\alpha_S$ and overdispersion 
$\theta$.  Table \ref{parametritaulu} presents the parameters in the simulation study.  The sample size of each simulated dataset was $n=1000$. The expected value $\mu$ of each individual's total count was fixed at 10000, a realistic value for total count in read-count data.

The amount of overdispersion, controlled by  $\theta$, ranged from no overdispersion to approximately 1000-fold or 5000-fold excess variance of the total count relative to the standard Poisson distribution.  The amount of sparsity was controlled by parameter $\alpha_S$. 
We considered no sparsity leading to fixed class-specific multinomial probabilities, intermediate sparsity ($\sim$200-fold excess variance in the class-specific counts, $\alpha_S = 50$) and extreme sparsity ($\sim$5000-fold excess variance, $\alpha_S = 1$).

Data were simulated following the DAG of Figure \ref{fig1} (b) 
 assuming  5 classes of the underlying counts. We assumed  two binary confounders, $C_1$ and $C_2$, that  affect the exposure, mediator and  response, and divide the data accordingly into four strata.  We considered three   scenarios for how the total, direct and indirect effects come about.
 In scenario 1, we assumed  the underlying taxonomic hierarchy of Figure \ref{fig2} (b). We chose the parameters of model (\ref{alfaefektit}) so that the difference in the expected class-specific proportions ($\pi$ in equation (\ref{mixtuuri})) between groups $X=0$ and $X=1$ was clear-cut.       Scenario 2 was otherwise similar,  but the difference in the expected proportions between groups  $X=0$ and $X=1$ was  moderate.   Scenario 3 demonstrates 
the use of coordinates based on a single pivotal SBP matrix in which  exposure $X$ affects the proportion of the first component, while the remaining components compensate so that their relative probabilities remain the same.

Given  $\mu$, $\theta$ and $\alpha_S$, read counts were simulated from model (\ref{mixtuuri}) and further scaled  into ilr coordinates through Equation (\ref{ilrvek}). For scenarios 1 and 2, the ilr coordinates were built following the contrasts of matrix (\ref{taxmatrix})
 whereas in scenario 3 a single  pivotal SBP matrix with the mediating part of the composition as the pivot  was used.  Given models (\ref{mixtuuri}) and (\ref{alfaefektit}), the compositions and ilr coordinates and, subsequently, the true values of  the stratum-specific parameters $\beta_{h1k}, k=1, \ldots, J,$ of Equation (\ref{mediator}) are defined. However, lacking an analytical expression,
we approximated the true values of these parameters as differences in the mean values of the ilr coordinates under $X=1$ and $X=0$, as realised by Monte Carlo simulation, separately  for each scenario and each combination of $\theta$ and $\alpha_S$.

For each scenario, we fixed OIE at 0.10 and the CIE's as presented in Table \ref{parametritaulu}. The OIE of 10 \% corresponds to 20 \% of the total effect. In scenario 1, we fixed each of the coordinates to carry a positive indirect effect. In scenario 2, the magnitudes of the CIE's were the same as in scenario 1, but the $\beta_{ik}$ parameters were smaller due to the smaller difference between the $X=0$ and $X=1$ groups.  In scenario 3, OIE = $\hbox{CIE}_1$ = 0.10.
 Under each scenario and parameter setting, 
  the $\gamma_{1k}$ parameters were chosen so that the products $\gamma_{1k}\beta_{ik}$ were equal to the corresponding coordinate-wise CIE's. A weighted average value $\beta_{1k}$ over all four strata of the $\beta_{1kh}$ values was used (see Equation (\ref{par_exp}) and  Table \ref{trueparameters_all} in Supplementary material).   Finally, continuous responses were  simulated  according to model (\ref{response}). 
  
\subsection{Estimation of mediated effects}
In scenarios 1 and 2, the indirect effects (CIE and OIE) were estimated using  ilr coordinates that were based on the same contrast matrix  used to simulate the data. In scenario 3, we applied two approaches for building the ilr coordinates when assessing  mediation.
First, we employed the correct contrast matrix with the first part of the composition as the pivot (model 3A). Second, we used a misspecified contrast matrix by setting the pivot  as the last part of the composition (model 3B). Figures \ref{fig3} (a) and \ref{fig3} (b) show the DAGs corresponding to these two assumed causal structures. 
When deriving the ilr coordinates, zero-counts were replaced by 0.5, which is a preprocessing step commonly used in the analysis of microbial count data  \Citep{Sohn2019, Zhang2019a}.

We estimated the CIE's and OIE's in each stratum following models (\ref{mediator}) and (\ref{response})  and  derived the effects adjusted for confounding by calculating weighted averages over the strata. The dependencies between the ilr coordinates were taken into account by assuming a multivariate linear model for the coordinates. The parameter estimation was conducted using \texttt{R} (version 4.1.1).
 For each parameter setting, we generated 1000 data sets. The performance of estimation was assessed   by bias,  average standard error (mean of the standard error estimates) and empirical standard error (standard deviation of the estimates).  The ability to identify the mediated effects was assessed by statistical power and coverage probability of the 90 \% confidence interval (CI).

\subsection{Performance of the estimation}\label{performance}
 Tables \ref{tulos1_10000} and \ref{tulos2_10000} present the results of the simulation study for scenarios 1 and 2. The results of analyses 3A and 3B under Scenario 3 are presented in  Supplementary material (Tables \ref{tulos3a_10000} and  \ref{tulos3b_10000}, respectively).

\textbf{Bias.} The estimation of the coordinate-wise and overall indirect effects was generally unbiased. Under the multinomial counts, i.e., when $\alpha_S \rightarrow \infty$, the estimation appears occasionally slightly biased. This stems from the large variances of the estimators of the $\gamma_{1k}$ parameters under multinomial counts and the relatively small sample size ($n=1000$). The overdispersion term $\theta$ had practically no effect on the bias.

\textbf{Standard errors.} The average standard error and the standard deviation of the 1000 replicate estimates were essentially the same, even under  extreme sparsity (Tables \ref{tulos1_10000}, \ref{tulos2_10000}, \ref{tulos3a_10000}, \ref{tulos3b_10000}).  To understand how the standard error depends on the parameters,  consider the asymptotic variance of the estimator of a single coordinate-wise indirect effect,
\begin{equation}\label{se_erilainen}
\sigma^2_{\gamma_{k}} \sigma^2_{\beta_{k}} \left[ 
\left(  \frac{\beta_{1k}}{\sigma_{\beta_{k}}}\right)^2 + 
\left( 
\frac{\gamma_{1k}}{\sigma_{\gamma_{k}}}
\right)^2 
\right],
\end{equation}
where $\sigma^2_{\gamma_{k}}$ and $\sigma^2_{\beta_{k}}$
are the variances of the estimators of $\beta_{1k}$ and $\gamma_{1k}$, $k=1,\ldots,J$ (see Supplementary material A2 and Equation  \ref{cie_se} therein). Based on the simulations, we found that the product $\sigma^2_{\gamma_{k}} \sigma^2_{\beta_{k}}$ remains essentially constant. 
This is obviously due to the fact that when the variance of the ilr coordinates (mediators) is large, the $\beta_k$'s are estimated less precisely but at the same time the large variability of the ilr coordinates leads to more precise estimation of the $\gamma_k$'s.
 The variance (\ref{se_erilainen}) thus mainly depends on the sum of the squared  ratios $(\beta_{1k}/\sigma_{\beta_{k}})^2$ and $(\gamma_{1k}/\sigma_{\gamma_{k}})^2$. 
Table \ref{ievartaulu} presents these ratios and their product for the first CIE under a number of different parameter settings.

Higher sparsity means larger between-individual heterogeneity of the ilr coordinates and, subsequently, larger $\beta_{1k}$ parameters due to the larger difference between the mean ilr values in the unexposed and exposed groups (cf. Equation (7)).  However, with increasing sparsity, the variance $\sigma^2_{\beta_{k}}$ increases faster than  $\beta_{1k}^2$, and thus the ratio  $(\beta_{1k}/\sigma_{\beta_{k}})^2$ decreases. The effect of sparsity on $\gamma_{1k}^2$ and $\sigma^2_{\gamma_{k}}$ is reciprocal and thus the opposite occurs with the ratio $(\gamma_{1k}/\sigma_{\gamma_{k}})^2$ (Table \ref{ievartaulu}).

   The reciprocal behaviour of the two ratios in (\ref{se_erilainen}) appears to imply that the asymptotic variances of the estimators of the indirect effects are smallest at intermediate levels of sparsity. However, where the optimum lies also depends on the relative magnitudes of the two ratios and how fast they decrease/increase as sparsity increases. For example, the larger the contrast between the two exposure group ($X=0$ vs. $X=1$) is, the larger the value $\beta_{1k}$ and the corresponding ratio and, due to the fixed value of CIE$_k$, the opposite is true for $\gamma_{1k}$ and the other ratio.
 We found that under scenarios 1 and 3, where the contrasts between the exposure groups were large, the variances of the indirect effects were smallest under the highly sparse settings and largest in the multinomial settings (Tables \ref{tulos1_10000} and \ref{tulos3a_10000})). By contrast, under scenario 2 with smaller contrasts between the exposure groups, the variances were smallest at intermediate levels of sparsity (Table \ref{tulos2_10000}) .

The impact of overdispersion on the standard errors of the indirect effects remained moderate. Higher  overdispersion appears only slightly to increase  the standard errors  compared to the case with a Poisson variance in scenario 2. With intermediate sparsity, overdispersion did not seem to have a strong effect on the standard errors in either of the scenarios, whereas under multinomial counts overdispersion slightly decreased the standard errors.

\textbf{Power.} In scenarios 1 and 3, the best statistical power was obtained under extreme sparsity (Tables \ref{tulos1_10000} and \ref{tulos3a_10000}) whereas in scenario 2 intermediate sparsity yielded the best power (Table \ref{tulos2_10000}). These findings result from the differences in the standard errors as described above. Despite the slight impact of the overdispersion parameter $\theta$ on the standard errors of the indirect effects, it only affected the statistical power in an inconsistent and marginal manner.

\textbf{Misspecification of the contrast matrix.} 
In scenario 3, the first ilr coordinate carries the entire mediated effect. We analysed the simulated data using the correct contrast matrix  (model 3A, Table \ref{tulos3a_10000}) or a misspecified matrix (model 3B, Table \ref{tulos3b_10000}). While the OIE remains the same in both models, in agreement with the argument given in Section 3,  the CIE's are different.
Under model 3A, the  indirect effects of the first part of the composition were identified correctly. 
 However, under model 3B the true contrasts and the coordinates that were chosen to represent the composition differed, so that  each of the four ilr coordinates  will capture some of the influence of the  true mediating coordinate $M_{1A}$. 
With misspecified contrasts, each coordinate appears to carry a mediating effect (Table \ref{scen3_betagamma}). One can thus be led to infer that the contrasts of each element of the composition against $p_1$ and other remaining parts of the composition play a role in carrying the mediating effects.
Importantly, if the pivotal contrast matrix was used so that each of the components were used as the first pivot in turn, the first coordinate-wise indirect effects for these alternations would be 0.10, 0.01, 0.01, 0.01 and 0 summing up to 0.13 (in an example scenario with $\theta=0$ and $\alpha_S=0$; data not shown). This demonstrates that the sum of individual first pivot effects does not equal the true overall indirect effect, and thus the alternating pivot approach cannot be used to estimate the OIE. Of note, these alternating approaches aiming at searching of signals of causal mediation are deliberately based on trying multiple (often a very large number of) missspecified  pivots.
It is evident that the estimated effects from a misspecified model would not correspond to the true effects based on the correctly specified model. Furthermore, defining the “true” values for the effects under the misspecified model would not be reasonable. For this reason, statistics relying on the true values  (i.e., bias, power and coverage probability)  have been omitted from the Table \ref{tulos3b_10000}.

\section{Estimation of the effect of fibre intake on insulin level via the gut microbiome}

\subsection{Background}
 
We illustrate the methods of Sections 2 and 3 by exploring the mediating role  of the gut microbiome in the association between fibre intake and serum insulin level. Sufficient fibre intake is known to bring numerous health benefits,  such as healthy and diverse gut microflora and metabolic health including insulin sensitivity \Citep{Barber2020}. Furthermore, based on a recent literature review, the gut microbiome plays a role in the pathophysiology of type 2 diabetes \Citep{Gurung2020}. 
Thus, it is of interest to investigate whether fibre intake affects the  insulin level through the gut microbiome.  The potential mediating pathway of fibre intake on the insulin level via the microbiome stems from the production of short-chain fatty acids (SCFA)  by microbial fermentation of dietary fibre. 
SCFAs in turn may play a role in insulin sensitivity via the gut barrier function and reduced inflammation \Citep{Barber2020, Chambers2018}.

 In addition to numerous specific genera, the \textit{Actinobacteria} phylum has been suggested as an example of a taxonomically high level that might be associated with type 2 diabetes (e.g. \Citep{Gurung2020}). Here, we investigate whether the relative proportion of  the \textit{Actinobacteria} compared to the other phyla mediates the effect of fibre intake on the insulin level. Furthermore,  we investigate whether the  genera within the \textit{Actinobacteria}  phylum mediate such effects.
 
\subsection{Data description}
 
We retrieved data from the most recent (26-year) follow-up visit conducted in  the Special Turku Coronary Risk  Factor Intervention Project (STRIP) (Supplementary material \ref{STRIP} and  \Citep{Pahkala2020, Ronnemaa2008a}). Of the 546 participants who participated in the follow-up study, 325 provided data on diet, insulin levels and had  successfully-sequenced faecal samples. 
We excluded participants with type 1 diabetes and those who had not fasted before blood sampling or  had had antibiotic treatment during the past three months. In addition, we excluded obese participants (BMI $\geq 30 \hbox{kg/m}^2$) to avoid a potential collider bias (see Supplementary material \ref{strip_restrict} for further details).  The resulting sample size was $n = 264$.

  The participants were initially recruited at age 6 months in well-baby clinics  and randomised into intervention and control groups. 
The intervention group received dietary counselling on   biannual visits throughout  childhood and adolescence until 20 years of age, whereas the control group visited the research clinic twice a year until the age of seven and after that once a year but did not receive counselling. The intervention included promoting the use  of unsaturated fats, whole-grain products, fruits and vegetables and lowering the intake of saturated fat and salt. The 26-year follow-up visit was carried out six years after the intervention period had ended.

 The raw gut microbiome sequence data were processed into an amplicon sequence variant (ASV) table using the \textit{DADA2} pipeline \Citep{Callahan2016}. Prior to transforming the counts into ilr coordinates, the taxa that appeared in less than 10 $\%$ of the samples were filtered out. As the ilr transformation does not allow zero-counts, we replaced all remaining zero counts  by the maximum rounding error of 0.5.
 The observed insulin level was transformed using the natural logarithm.
 The fibre intake was assessed based on food diary and quantified using the Micro Nutrica programme \Citep{Hakala1996}. The optimal fibre intake was defined as $\geq 25 $ grams per day, as per Finnish dietary recommendations in 2014  \Citep{ruokavirasto2014}. 
Of the 264 participants, 63 ($ 23.9 \%$) achieved this target. 
Belonging to the intervention group and sex were considered as potential confounders, and the sample was stratified accordingly. The stratum-specific sample sizes were 68, 82, 59, and 55  for intervention females, control females, intervention males and control males, respectively.

\subsection{Conduct of the mediation  analysis and results}\label{empwithin}
We investigated two mediation questions on different phylogenetic levels of the microbiome. The first question concerned the mediating role of the genera within the \textit{Actinobacteria} phylum in the association between sufficient fibre intake and $\log$(insulin) level. The second question  utilised a pivotal SBP matrix,  focusing on the mediating effect of the relative proportion  of the \textit{Actinobacteria} in the association between sufficient fibre intake and $\log$(insulin) level. 

Models (\ref{mediator}) and (\ref{response}) were fitted separately for each stratum defined by the confounders. We  report  estimates averaged over the strata (see Equation \ref{par_exp}). The stratum-specific results are available in Supplementary material D.3. 

The average total counts were approximately 4200 and 180 000 for the hypotheses within the \textit{Actinobacteria} phylum and over the different phyla, respectively, whereas the amounts of excess variance in the total count were 12000-fold and 54000-fold, respectively. In terms of the simulation setting and model (\ref{mixtuuri}), these correspond to values 3 and 0.3 of parameter $\theta$. For each count, we also calculated a moment-based estimate of $\alpha_s$ based on the taxon-specific variances and proportions and the total count. While the $\alpha_S$ parameter obtained via this method varied between the taxa, the median $\alpha_S$ was 36 within the phylum-specific counts and 33 within the genus-specific counts.

\subsubsection*{Mediation through  the genera within the \textit{Actinobacteria}}
After the preprocessing, the data on the \textit{Actinobacteria} phylum included one class, which was divided into three orders and within these, four families. In the four families, there were altogether 11 genera  (Figure \ref{fig4}). We first investigated the mediating effect of this subcomposition of the microbiome.  Table \ref{phylocount} presents the corresponding counts and  proportions  in the two fibre groups by genera.  Based on the  taxonomy (Figure \ref{fig4}) we built the SBP matrix as follows: 
\begin{equation*}\label{matriisiemphierar}
\begin{matrix}
Bifidobacterium \\
Actinomyces     \\
Rothia          \\
Enterorhabdus \\
Eggerthella \\
Collinsella \\
Adlercreutzia \\
Slackia \\
Senegalimassilia  \\
Olsenella \\
Gordonibacter \\ 
\end{matrix}  
\begin{bmatrix}
-1 & 0 & +1 & 0 & 0 & 0 & 0 & 0 & 0 & 0 \\
+1 & +1 & 0 &  0 & 0 & 0 & 0 & 0 & 0 & 0 \\
+1 & -1 & 0 &  0 & 0 & 0 & 0 & 0 & 0 & 0 \\
-1 & 0 & -1 & +1  & 0 & 0 & 0 & 0 & 0 & 0 \\
-1 & 0 & -1 & -1  & +1 & 0 & 0 & 0 & 0 & 0 \\
-1 & 0 & -1 & -1  & -1 & +1 & 0 & 0 & 0 & 0 \\
-1 & 0 & -1 & -1  & -1 & -1 & +1 & 0 & 0 & 0 \\
-1 & 0 & -1 & -1  & -1 & -1 & -1 & +1 & 0 & 0 \\
-1 & 0 & -1 & -1  & -1 & -1 & -1 & -1 & +1 & 0 \\
-1 & 0 & -1 & -1  & -1 & -1 & -1 & -1 & -1 & +1 \\
-1 & 0 & -1 & -1  & -1 & -1 & -1 & -1 & -1 & -1 \\
\end{bmatrix}.
\end{equation*}
Each of the 10 columns thus corresponds to one ilr coordinate $M_k$, $k=1, \ldots, 10$, describing a single contrast. For example, the fourth column contrasts the \textit{Enterorhabdus} genus against the other genera within the \textit{Coriobacteriales} order.

The total effect of fibre on log(insulin) was estimated at $-0.106$ (90$\%$ CI $[-0.202; -0.011]$) whereas the direct effect was  $\hat{\gamma}_2 = -0.095$ (90$\%$ CI $[-0.202; 0.012]$). 
 The sufficient fibre intake thus seems to reduce insulin levels mostly via the direct route. 
 The parameters  $\hat{\beta}_{1k}$ describing the effects of sufficient fibre intake on the ilr coordinates are presented in Table \ref{phylotulos}. For example, sufficient fibre intake reduced the second coordinate, which corresponds to the relationship between relative proportions of \textit{Actinomyces} and \textit{Rothia} ($\hat{\beta}_{12} = -0.255$, $[-0.501; -0.009]$).  By contrast, sufficient fibre intake increased $M_4$, which corresponds to  the relative proportion of \textit{Enterorhabdus} compared to the other genera within the \textit{Coriobacteriales} ($\hat{\beta}_{14} = 0.409$,  $[0.060; 0.759]$). 

The $\hat{\gamma}_{1k}$ parameters in Table \ref{phylotulos} describe the effects of each ilr coordinate on the log(insulin) level in a multivariate linear regression. For example, a higher level of $M_3$ decreased the insulin level, i.e., the higher the relative proportion of \textit{Bifidobacterium} contrasted to \textit{Coriobacteriales}, the higher the insulin levels ($\hat{\gamma}_{13} = 0.025$,  $[0.002; 0.048]$).
While $M_4$ (relative proportion of \textit{Enterorhabdus} compared to other genera within the \textit{Coriobacteriales}) decreased the log(insulin) level ($\hat{\gamma}_{14} = -0.041$, $[-0.074; -0.008]$), $M_5$ (relative proportion of \textit{Eggerthella} contrasted to remaining genera within the \textit{Coriobacteriales})  increased the log(insulin) level ($\hat{\gamma}_{15} = 0.038$,  $[0.015; 0.060]$).

The CIE's  $\hat{\beta}_{1k} \hat{\gamma}_{1k}$  with their 90 \% confidence intervals are presented in Table \ref{phylotulos}. 
 The largest mediated effect, yet inconclusive, acted through  coordinate $M_4$, as sufficient fibre intake increased the  relative proportion of \textit{Enterorhabdus} compared to other genera within the \textit{Coriobacteriales}, and the relative proportion of   \textit{Enterorhabdus} had a decreasing effect on the log(insulin) level ($\hat{\beta}_{14} \hat{\gamma}_{14} = -0.017$, $[-0.036; 0.003]$).  
  The OIE was  $\sum_{k=1}^{10}\hat{\beta}_{1k} \hat{\gamma}_{1k} = -0.011$ (90 $\%$ CI $[-0.073; 0.051]$), corresponding to  approximately  10\% of the total effect. These results demonstrate our previous note  that the coordinate-wise indirect effects may counteract, making some of the CIE's  larger than  OIE in absolute value.

\subsubsection*{Mediation through the \textit{Actinobacteria} phylum}
We investigated whether the entire 
  \textit{Actinobacteria} phylum in comparison to the other phyla has a mediating role using a pivotal SBP matrix (\ref{pivotmat}). After the preprocessing stage altogether 11 phyla remained. The contrasts of interest were encoded through a pivotal SBP contrast matrix where the pivot coordinate describes the  proportion of the \textit{Actinobacteria} against the proportions of the other phyla. Those who had sufficient fibre intake had a smaller proportion of \textit{Actinobacteria} (1.6\%) as compared to those who had insufficient fibre intake (2.7\%).

Based on model (\ref{mediator}), the average effect of sufficient fibre intake on  the first ilr coordinate contrasting the \textit{Actinobacteria} phylum  to the other phyla was  $\hat{\beta}_{11}$ =$ -0.451$ (90 \% CI $[-0.762; -0.140]$).  Furthermore, the higher  proportion of the \textit{Actinobacteria} phylum against the other phyla may have a positive association with higher insulin level, as the effect of the pivot coordinate on log(insulin)  was $\hat{\gamma}_{11} = 0.042$ (90 \% CI $[-0.004; 0.088]$).  The CIE of the coordinate of interest was  thus $\hat{\beta}_{11} \hat{\gamma}_{11}$ = $-0.019$ (90 \% CI $[-0.043; 0.006]$). The total effect of sufficient fibre intake on log(insulin) level  remained as   $-0.106$ (90 $\%$ CI $[-0.202; -0.011]$) whereas the direct effect was  $ -0.10 $ (90 $\%$ CI $[-0.206; 0.005]$). The OIE was $-0.006$, i.e., only  5\% of the total effect.

\section{Discussion}
In this study, we proposed and investigated a hypothesis-driven approach to mediation analysis with a compositional mediator  based on sparse and overdispersed microbial count data. 
 We specified non-parametric causal estimands for both coordinate-wise and overall indirect effects relying on an \textit{a priori} defined structure and presented  conditions under which they can be identified from empirical data based on simple parametric models.  The method can be applied when some \textit{a priori} knowledge about the structure between the parts of the composition is available and  allows simultaneous estimation of coordinate-wise indirect effects (CIE's) and the overall indirect effect (OIE).  The estimation of mediated effects was found to be generally unbiased, even under extreme amounts of sparsity and overdispersion in the underlying count data. However, sparsity and the magnitudes of the path coefficients were found to affect the precision of the estimation in a complex way.

The use of ilr coordinates transforms proportions in the simplex  into Euclidean coordinates. A standard statistical approach for mediation analysis can thus be applied to estimate  indirect effects \Citep{VanderWeele2016}.  Importantly to the causal interpretation, the ilr coordinates have a one-to-one correspondence with a sequence of partitions of the entity of interest  into subcompositions. Each single ilr coordinate corresponds  to a  partition of a specific subcomposition into two parts, while the relative proportions of all other parts remain intact, and the coordinate can thus be interpreted as representing a causal relationship encoded by that partition. Furthermore, with the assumed purely
linear dependence of the response on the mediators and no exposure-mediator interaction,  the overall effect is the sum of all coordinate-wise effects and quantifies mediation via the entire composition of interest.

The theory on multiple mediators mainly focuses on two scenarios, causally dependent mediators \Citep{VanderWeele2014c, Daniel2015, Imai2013} and mediators that are correlated but do not affect each other causally \Citep{Wang2013, Kim2019, Taguri2018}. 
Our treatment of mediation through  ilr coordinates is an example of the latter. 
We considered two alternative ways to define causal effects non-parametrically. 
The definition  of Wang et al. \Citep{Wang2013} leads to coherent estimands for the overall indirect effects and coordinate-wise indirect effects.
For the counterfactual interpretation of the indirect effects, the order of the mediators is important, as the exposure levels of the other mediators are assigned sequentially.  Of note, these non-parametric definitions correspond to the sequential build-up of the ilr coordinates. 
In contrast, in the alternative  non-parametric definition  of Kim et al. \Citep{Kim2019}, the order of the mediators plays no role.  While this formulation does not lead to a coherent decomposition of the overall indirect effect into the coordinate-wise indirect effects, under linear models and no interaction between the mediators the parametric formulations are the same under both alternative non-parametric definitions, and the coordinate-wise indirect effects sum up to the overall indirect effect.

In Wang et al. \Citep{Wang2013}, the parametric dependence of the response on the mediators was based on logistic regression. 
The parametric expressions of the indirect effects then depend on the joint  distribution of the mediators. This in turn requires an \textit{a priori} specification of their correlation structure.
As our application here focuses on a linear dependence of the response on the mediators, the parametric expressions only depend on the estimable marginal distributions of the mediators.

Given that the aims and scope of our proposed method fundamentally differs from other studies that have addressed compositional mediation via the microbiome \Citep{Zhang2019a, Sohn2019, Wang2020, Fu2023}, a formal comparison of the methods' performance was not carried out.
Instead, we here provide a comparison of the main characteristics of these approaches. 
Table \ref{paperivertailut} summarises some key features of these  studies.
The previous studies have focused on identifying signals of mediation by  different taxa  and rely on testing a large number of tentative (causal) associations between the components of the microbiome.  As argued in the present study, many of such associations are potentially misleading if the aim is to answer to a structured causal question.
  In particular, when causal effects based on relevant biological mechanisms are of special interest,
a  more targeted analysis based on \textit{a priori} knowledge may be called for  in contrast to data-driven approaches that rather emphasise the predictive performance of the model. 
  An analysis that is conditioned on a given set of contrasts can be thought as the next step on the pathway of building evidence about mediation.

 All previously presented methods to analyse compositional mediation  through the microbiome have been based on some log-ratio transformation.  Sohn and Li \Citep{Sohn2019} and Wang \Citep{Wang2020} used the additive logratio which differs from the isometric logratio in interpretation, whereas Zhang et al. \Citep{Zhang2019a} and Fu et al. \Citep{Fu2023} relied on the isometric logratio using alternating pivot coordinates.
  While  we applied the isometric logratio, we deviate from these methods in three aspects. Firstly, instead of sieving for signals of mediation (i.e. testing for multiple null hypotheses), we aim at estimating mediated effects within a predefined  \textit{a priori}  partition, which leads to investigating multiple contrasts of interest simultaneously without additional re-fitting of ilr transformations using alternating pivots. Secondly, we investigated how the sampling variation among the units, caused by sparsity and overdispersion, affects the estimation of mediated effects. 
Thirdly,  our approach is confined to low-dimensional problems. Thus, regularisation or Bayesian model selection approaches are not needed. In addition, due to the lower dimension, our approach is free of multiple comparison problems of sieving approaches especially present in Zhang et al. \Citep{Zhang2019a}, and as such allows  unbiased estimation of the overall indirect effect. Lastly, the coordinate-wise effects defined in Fu et al. \Citep{Fu2023} were basically regression-based, i.e., they relied on a parametric formulation and thus lacked proper definition in terms of causal contrasts. Consequently, the parametric forms of the indirect effects are not similar to those of Wang et al. \Citep{Wang2013} or those considered in our paper.

 Our approach is based on a  pre-specified contrast matrix that contains the \textit{a priori} causal assumptions encoded as a DAG. Thus, any conclusions that are drawn will depend on the specification of the matrix, i.e.,  the analysis of mediation is conditional on the assumed contrasts. 
 The idea of utilising knowledge about taxonomic distances  in building  ilr coordinates
has been presented earlier, although in a data-driven manner  \Citep{Silverman2017}. 
    Other possible sources of taxonomic knowledge include     biological functionalities of different taxa, or hypotheses based on previous (data-driven) studies. 
      The  pivotal approach of Zhang et al. \Citep{Zhang2019a}, where the coordinates are alternated $J+1$ times to identify the mediating role of each part of the composition, corresponds to testing $J+1$ different  hypotheses, each concerning a  role of a specific part of the composition.

Sparsity is an important feature of sequencing data, following from true biological variation and reflecting actual heterogeneity in the microbiome across individuals. Larger sparsity means larger variability of the ilr coordinates. To investigate the effect of sparsity, we simulated data at the count level, as opposed to some previous research on compositional mediation where  mediators were simulated directly as logratio coordinates based on a multivariate normal distribution \Citep{Zhang2019a, Sohn2019}.  Sparsity was induced into the counts by simulating the class-specific probabilities from a Dirichlet distribution.

With any given level of  CIE, we found that with larger sparsity the variation in the coordinates increased, and thus the variability of the $\beta$ parameters increased, and correspondingly the variability of the $\gamma$ parameters decreased (see the discussion following equation (\ref{se_erilainen})). 
 When the effects of exposure on the ilr coordinates ($\beta_{1k}$) were large, the  variance of the estimators of the indirect effects  was  smallest under extreme sparsity (scenarios 1 and 3).  In contrast, when the effect of exposure was smaller, the smallest variance of the estimators of the indirect effects  was found under intermediate sparsity (scenario 2).

In terms of the precision of estimated mediation effects, it would be ideal to conduct the  analysis under an intermediate level of sparsity. When designing a study, however, the components affecting the variance of the  CIE's --- that is, the magnitudes of the pathway coefficients ($\beta_{1k}$ and $\gamma_{1k}$) and their variability  --- cannot be controlled in advance.
The amount of sparsity varies in  microbiome data and also depends on the taxonomic level of interest. At  deeper taxonomic levels, such as genus level, the  number of taxa is larger and data are more likely sparse. However, if the mediated effects of interest are considered at a higher level, each  taxon can be considered as an amalgamation of lower-level taxa within that taxon, which  potentially reduces sparsity.  Thus, while the sparsity present in the data cannot be controlled within any chosen level of taxonomy, restricting the analysis to a higher taxonomic level may result in less sparse counts. Additionally, the most extremely sparse counts are often removed while preprocessing  the data. In general, one should pay specific attention to the sparsity of the counts and be cautious of its implications to the estimation of indirect effects.

Overdispersion of the total read count is another feature of microbiome data and reflects technical rather than biological variation across individuals. When count data are treated as compositions and scaled into proportions, the impact of overdispersion is mostly removed.  Nevertheless, in some scenarios of our simulation study, overdispersion increased the variability of the ilr coordinates. In turn, under multinomial counts, this  affected slightly the variances of the pathway coefficients thus decreasing variances of the estimators of the indirect effects. 
 
We demonstrated the estimation of compositional mediation with an empirical data set, investigating the potential indirect effect of the gut microbiome in the association between dietary fibre intake and insulin level.  The total effect was found to be dominated by the direct effect, but we cannot rule out the possibility that certain parts of the gut microbiome, such as the \textit{Enterorhabdus}, carry a small mediating role. 
 Nonetheless,  our empirical data analysis should be considered as a demonstration rather than a comprehensive analysis. For example, we did not consider other confounders apart from sex and intervention status.   In addition, even though reverse causality is implausible in the light of current biological knowledge, the temporal ordering of the exposure (fibre intake), mediator (gut microbiome) and response (insulin level) was obscure as each  was measured at the same  visit.  
 Although the direct and overall effects were of similar magnitudes 
  whether analysing mediation of \textit{Actinobacteria} at a phylum-level contrasted to other phyla or of the genera  within the \textit{Actinobacteria}, 
 the models reflect different levels of  taxonomic hierarchy and thus potentially different biological mechanisms. The levels of sparsity ($\alpha_S$) and overdispersion ($\theta$) in the empirical data indicate that our choices of these parameters in the simulation study, as well as the chosen magnitudes of the indirect effects, were  reasonable.

There are some limitations to our approach.  Firstly, as in any mediation analysis based on DAGs, it is possible to misspecify the causal structure  e.g. by misspecifying the contrast matrix or not accounting for relevant covariates.  
As shown by the simulation example, a misspecified contrast matrix may result in estimates that do not reflect the true causal effects and lead to erroneous inferences on the mediating roles of the parts of the composition. 
This emphasises the importance of utilising expert knowledge or \textit{a priori} information when coding the hypotheses into the contrast matrix. The \textit{a priori} knowledge can be based on e.g. sieving studies, such as those suggested by e.g. Zhang et al., Sohn and Li or Wang et al. \Citep{Zhang2019a, Sohn2019, Wang2020}, or on biological knowledge of the roles of different parts of the composition, such as the relationship between \textit{Bacteroidetes} and \textit{Firmicutes} in human gut microbiome.  It is of note that the contrast matrix depends on the question in hand, and thus no general guidelines on building the contrasts in a hypothesis-driven manner can be given.  In addition, if there are unobserved or unadjusted confounders, the causal effects may not be estimated correctly. 

Secondly, while we investigated the impact of sampling variation of the class probabilities and counts on the estimation of mediated effects in the simulation study, in the empirical analyses we simply assumed that the ilr coordinates are normally distributed. However, the normality may not always hold, as the variation in the class probabilities and counts is reflected in the variation of the ilr coordinates. Regardless, we showed that the estimated standard errors corresponded on average to the true ones as found by simulation.

Thirdly, we formulated the approach on the presupposition that confounders are categorical or can be categorised. Without this assumption, the linear dependence of the Dirichlet parameters on the confounders  would not induce a linear dependence of the ilr coordinates on the confounders (see Equations (\ref{alfaefektit}) and (\ref{mediator})).  Nevertheless, the general idea of basing causal mediation analysis on  \textit{a priori} defined hierarchy coded through a single SBP matrix remains valid, and there is nothing in principle that would preclude using  continuous covariates in model (\ref{mediator}).
   In our analysis, the assumption of categorical confounders was used in the simulation study when assessing the performance of estimating mediated effects. 
  Previous studies on compositional mediation analysis in the context of microbiome data have simulated data directly at the log-ratio level assuming a multivariate normal distribution and linear dependence between coordinates and covariates \Citep{Sohn2019,Zhang2019a}. Such approaches fail to account for the variation across individuals in the actual counts. In this respect, our analysis can be viewed as more comprehensive despite the assumption of categorical covariates. Although not studied here, it is also conceivable that our results on the estimability of the indirect effects remain essentially the same even if one would apply continuous covariates.

In summary, we here have combined the framework of compositional data analysis with causal mediation analysis with specific focus on theory on multiple mediators and the causal interpretation of the coordinates based on the microbiome composition. 
The contribution of our work is threefold. First, we  suggested incorporating  \textit{a priori} or expert knowledge in microbial mediation and presented microbial mediation analysis with specific focus on the contrasts between sub-compositions of interest rather than individual taxa. 
Second, we have combined the sequential approach of defining mediator-specific indirect effects \Citep{Wang2013}  with the hypothesis-driven sequential binary partition in the context of microbial phylogenetics, and thus defined coordinate-wise and overall indirect effects that are coherent in their non-parametric definitions. Furthermore, these effects are based on a single causal mediation model and their empirical derivation does not require multiple testing, regularisation or model selection.
Third, unlike in other simulation studies on microbial mediation, we investigated the effect of sparsity and overdispersion on the estimation of the mediated effects.

In addition to the analysis of the  microbiome, the approach adopted in this study  is applicable for a wide range of research questions where the mediator can be considered compositional.
 For example, the method could be utilised in other types of biological read-count data, such as epigenetic RNA counts. Although  RNA molecules do not follow similar taxonomic phylogenies as the microbiome, they can be classified e.g. based on their type \Citep{Hombach2016}. Such classifications can be utilised as \textit{a priori} knowledge to formulate meaningful contrasts. 
Compositional data analysis, and the proposed approach, also provide opportunities within the context of e.g. dietary composition, blood lipid composition or composition of the daily activity levels, 
and using \textit{a priori} knowledge of the hierarchies between the parts of composition may also provide interesting insights into the mediating role of these compositions. 
However, the applicability of compositional transformations under different sources of variation and different types of data still needs further research.

\section*{Tables}

\begin{center}
\begin{table*}[t]%
\caption{Parameters of the simulation study. \label{parametritaulu}}
\centering
\begin{tabular}{l l l }
\toprule
\textbf{Parameter} & \textbf{Value} & \textbf{Note} \\
\midrule
\multicolumn{3}{l}{\underline{\textbf{Distribution of the counts}}}\\
E($K_i$)= $\mu$ &  100000. \\
Dispersion $\theta$ & $\rightarrow 0$; 0.1; 0.5 \\
Sparsity $\alpha_S$ & 1; 50; $\alpha_S \rightarrow \infty$ \\
\underline{\textbf{Binary confounders}} & & \\
$p(C_1 = 1)$ & 0.50 & \\
$p(C_2 = 1)$ & 0.50 & \\
\underline{\textbf{Binary exposure}} & & \\
$p(X=1 |C_1, C_2)$ & $0.25+0.05C_1+0.05C_2$ & \\
\underline{\textbf{Mediator}} & & \\
& (0.02, 0.01, -0.01, -0.01, -0.01) & Effect of $C_1$ on $\boldsymbol{\alpha}$ \\
& (-0.01, -0.01, 0.02, 0.00, 0.00) & Effect of $C_2$ on $\boldsymbol{\alpha}$ \\
\multicolumn{3}{l}{$(\alpha_1/\alpha_S, \ldots, \alpha_5/\alpha_S) | X=0$} \\
& (0.60, [0.15, 0.10], [0.10, 0.05]) & scenario 1 \\ 
& (0.60, [0.15, 0.10], [0.08, 0.07]) & scenario 2 \\ 
& (0.60, 0.10, 0.10, 0.10, 0.10)  & scenario 3 \\ 
\multicolumn{3}{l}{$(\alpha_1/\alpha_S, \ldots, \alpha_5/\alpha_S) | X=1$} \\
& (0.50, [0.20, 0.05], [0.10, 0.15]) & scenario 1 \\ 
& (0.58, [0.13, 0.13], [0.10, 0.06]) & scenario 2 \\ 
& (0.50, 0.125, 0.125, 0.125, 0.125) & scenario 3 \\ 
\underline{\textbf{Response}} & & \\
$(\gamma_0,\gamma_2,\gamma_{C_1},\gamma_{C_1})$ & (2.0, 0.40, 0.05, -0.05) \\
\underline{\textbf{Indirect effects}} & & \\
OIE; CIE & 0.10;  (0.04, 0.01, 0.03, 0.02) & scenario 1 \\
&0.10;  (0.04, 0.01, 0.03, 0.02) &  scenario 2  \\
&0.10;  (0.10, 0.00, 0.00, 0.00) & scenario 3 \\
\bottomrule
\end{tabular}
\end{table*}
\end{center}

\begin{center}
\begin{table*}[t]%
\caption{Simulation results for scenario 1. The true and estimated indirect effects, bias, average standard errors ($\widehat{\hbox{SE}}$), standard deviation of the estimates ($\hbox{SE}_{\hbox{est.}}$), and power and coverage probability of the 90 \% confidence intervals
of the coordinate-wise (CIE) and overall (OIE) indirect effects with varying sparsity $\alpha_S$ and overdispersion $\theta$.  \label{tulos1_10000} }
\centering
\begin{tabular}{llllllllllll}
\toprule
 $\alpha_S$ & $\theta$  & \textbf{Eff.} & \textbf{True} & \textbf{Est.} & \textbf{Bias} & $\widehat{\hbox{\textbf{SE}}}$ & $\hbox{\textbf{SE}}_{\hbox{est.}}$ & \textbf{Power} & \textbf{Coverage} \\ 
\midrule
 	1	&	$\rightarrow 0$	&	CIE$_1$	&	0.04	&	0.04	&	0.00	&	0.015	&	0.015	&	0.91	&	0.91	\\
 	1	&	$\rightarrow 0$	&	CIE$_2$	&	0.01	&	0.01	&	0.00	&	0.017	&	0.016	&	0.12	&	0.93	\\
 	1	&	$\rightarrow 0$	&	CIE$_3$	&	0.03	&	0.03	&	0.00	&	0.017	&	0.016	&	0.56	&	0.93	\\
 	1	&	$\rightarrow 0$	&	CIE$_4$	&	0.02	&	0.02	&	0.00	&	0.020	&	0.020	&	0.27	&	0.91	\\
 	1	&	$\rightarrow 0$	&	OIE	&	0.10	&	0.10	&	0.00	&	0.036	&	0.035	&	0.90	&	0.92	\\
 	50	&	$\rightarrow 0$	&	CIE$_1$	&	0.04	&	0.04	&	0.00	&	0.044	&	0.044	&	0.24	&	0.91	\\
 	50	&	$\rightarrow 0$	&	CIE$_2$	&	0.01	&	0.01	&	0.00	&	0.061	&	0.060	&	0.09	&	0.91	\\
 	50	&	$\rightarrow 0$	&	CIE$_3$	&	0.03	&	0.03	&	0.00	&	0.057	&	0.058	&	0.16	&	0.89	\\
 	50	&	$\rightarrow 0$	&	CIE$_4$	&	0.02	&	0.02	&	0.00	&	0.055	&	0.054	&	0.12	&	0.91	\\
 	50	&	$\rightarrow 0$	&	OIE	&	0.10	&	0.10	&	0.00	&	0.089	&	0.091	&	0.30	&	0.88	\\
 	$\rightarrow \infty$	&	$\rightarrow 0$	&	CIE$_1$	&	0.04	&	0.05	&	-0.01	&	0.592	&	0.595	&	0.11	&	0.89	\\
 	$\rightarrow \infty$	&	$\rightarrow 0$	&	CIE$_2$	&	0.01	&	0.00	&	0.01	&	0.774	&	0.801	&	0.11	&	0.89	\\
 	$\rightarrow \infty$	&	$\rightarrow 0$	&	CIE$_3$	&	0.03	&	0.03	&	0.00	&	0.742	&	0.737	&	0.11	&	0.90	\\
 	$\rightarrow \infty$	&	$\rightarrow 0$	&	CIE$_4$	&	0.02	&	0.03	&	-0.01	&	0.728	&	0.700	&	0.09	&	0.91	\\
 	$\rightarrow \infty$	&	$\rightarrow 0$	&	OIE	&	0.10	&	0.11	&	-0.01	&	1.212	&	1.202	&	0.10	&	0.91	\\
 	1	&	0.1	&	CIE$_1$	&	0.04	&	0.04	&	0.00	&	0.015	&	0.015	&	0.89	&	0.90	\\
 	1	&	0.1	&	CIE$_2$	&	0.01	&	0.01	&	0.00	&	0.017	&	0.016	&	0.13	&	0.92	\\
 	1	&	0.1	&	CIE$_3$	&	0.03	&	0.03	&	0.00	&	0.017	&	0.017	&	0.55	&	0.91	\\
 	1	&	0.1	&	CIE$_4$	&	0.02	&	0.02	&	0.00	&	0.020	&	0.019	&	0.27	&	0.91	\\
 	1	&	0.1	&	OIE	&	0.10	&	0.10	&	0.00	&	0.036	&	0.035	&	0.88	&	0.91	\\
 	50	&	0.1	&	CIE$_1$	&	0.04	&	0.04	&	0.00	&	0.043	&	0.043	&	0.22	&	0.92	\\
 	50	&	0.1	&	CIE$_2$	&	0.01	&	0.01	&	0.00	&	0.061	&	0.059	&	0.09	&	0.91	\\
 	50	&	0.1	&	CIE$_3$	&	0.03	&	0.03	&	0.00	&	0.057	&	0.056	&	0.13	&	0.90	\\
 	50	&	0.1	&	CIE$_4$	&	0.02	&	0.02	&	0.00	&	0.055	&	0.053	&	0.12	&	0.92	\\
 	50	&	0.1	&	OIE	&	0.10	&	0.10	&	0.00	&	0.090	&	0.090	&	0.29	&	0.89	\\
 	$\rightarrow \infty$	&	0.1	&	CIE$_1$	&	0.04	&	0.05	&	-0.01	&	0.563	&	0.566	&	0.10	&	0.90	\\
 	$\rightarrow \infty$	&	0.1	&	CIE$_2$	&	0.01	&	0.00	&	0.01	&	0.736	&	0.746	&	0.11	&	0.89	\\
 	$\rightarrow \infty$	&	0.1	&	CIE$_3$	&	0.03	&	0.04	&	-0.01	&	0.707	&	0.726	&	0.11	&	0.89	\\
 	$\rightarrow \infty$	&	0.1	&	CIE$_4$	&	0.02	&	0.00	&	0.02	&	0.692	&	0.683	&	0.09	&	0.91	\\
 	$\rightarrow \infty$	&	0.1	&	OIE	&	0.10	&	0.09	&	0.01	&	1.152	&	1.189	&	0.11	&	0.89	\\
 	1	&	0.5	&	CIE$_1$	&	0.04	&	0.04	&	0.00	&	0.015	&	0.015	&	0.90	&	0.90	\\
 	1	&	0.5	&	CIE$_2$	&	0.01	&	0.01	&	0.00	&	0.017	&	0.017	&	0.14	&	0.91	\\
 	1	&	0.5	&	CIE$_3$	&	0.03	&	0.03	&	0.00	&	0.017	&	0.017	&	0.55	&	0.91	\\
 	1	&	0.5	&	CIE$_4$	&	0.02	&	0.02	&	0.00	&	0.020	&	0.020	&	0.26	&	0.89	\\
 	1	&	0.5	&	OIE	&	0.10	&	0.10	&	0.00	&	0.036	&	0.036	&	0.89	&	0.88	\\
 	50	&	0.5	&	CIE$_1$	&	0.04	&	0.04	&	0.00	&	0.043	&	0.043	&	0.22	&	0.91	\\
 	50	&	0.5	&	CIE$_2$	&	0.01	&	0.01	&	0.00	&	0.061	&	0.060	&	0.09	&	0.91	\\
 	50	&	0.5	&	CIE$_3$	&	0.03	&	0.03	&	0.00	&	0.057	&	0.057	&	0.13	&	0.90	\\
 	50	&	0.5	&	CIE$_4$	&	0.02	&	0.02	&	0.00	&	0.055	&	0.054	&	0.11	&	0.91	\\
 	50	&	0.5	&	OIE	&	0.10	&	0.09	&	0.01	&	0.089	&	0.087	&	0.27	&	0.91	\\
 	$\rightarrow \infty$	&	0.5	&	CIE$_1$	&	0.04	&	0.05	&	-0.01	&	0.431	&	0.427	&	0.09	&	0.91	\\
 	$\rightarrow \infty$	&	0.5	&	CIE$_2$	&	0.01	&	0.01	&	0.00	&	0.562	&	0.578	&	0.12	&	0.89	\\
 	$\rightarrow \infty$	&	0.5	&	CIE$_3$	&	0.03	&	0.03	&	0.00	&	0.541	&	0.565	&	0.11	&	0.89	\\
 	$\rightarrow \infty$	&	0.5	&	CIE$_4$	&	0.02	&	0.03	&	-0.01	&	0.528	&	0.510	&	0.08	&	0.92	\\
 	$\rightarrow \infty$	&	0.5	&	OIE	&	0.10	&	0.11	&	-0.01	&	0.879	&	0.868	&	0.10	&	0.90	\\
\bottomrule
\end{tabular}
\end{table*}
\end{center}

\begin{center}
\begin{table*}[t]%
\caption{Simulation results for scenario 2. The true and estimated indirect effects, bias, average standard errors ($\widehat{\hbox{SE}}$), standard deviation of the estimates ($\hbox{SE}_{\hbox{est.}}$), and power and coverage probability of the 90 \%  confidence intervals
of the coordinate-wise (CIE) and overall (OIE) indirect effects with varying sparsity $\alpha_S$ and overdispersion $\theta$. 
\label{tulos2_10000}}
\centering
\begin{tabular}{llllllllllll}
\toprule
$\alpha_S$ & $\theta$  & \textbf{Eff.} & \textbf{True} & \textbf{Est.} & \textbf{Bias} & $\widehat{\hbox{\textbf{SE}}}$ & $\hbox{\textbf{SE}}_{\hbox{est.}}$ & \textbf{Power} & \textbf{Coverage} \\ 
\midrule
 	1	&	$\rightarrow 0$	&	CIE$_1$	&	0.04	&	0.04	&	0.00	&	0.034	&	0.034	&	0.33	&	0.91	\\
 	1	&	$\rightarrow 0$	&	CIE$_2$	&	0.01	&	0.00	&	0.01	&	0.102	&	0.102	&	0.10	&	0.90	\\
 	1	&	$\rightarrow 0$	&	CIE$_3$	&	0.03	&	0.03	&	0.00	&	0.013	&	0.012	&	0.81	&	0.90	\\
 	1	&	$\rightarrow 0$	&	CIE$_4$	&	0.02	&	0.02	&	0.00	&	0.011	&	0.010	&	0.57	&	0.93	\\
 	1	&	$\rightarrow 0$	&	OIE	&	0.10	&	0.09	&	0.01	&	0.110	&	0.109	&	0.21	&	0.91	\\
 	50	&	$\rightarrow 0$	&	CIE$_1$	&	0.04	&	0.04	&	0.00	&	0.015	&	0.015	&	0.93	&	0.91	\\
 	50	&	$\rightarrow 0$	&	CIE$_2$	&	0.01	&	0.01	&	0.00	&	0.015	&	0.014	&	0.14	&	0.92	\\
 	50	&	$\rightarrow 0$	&	CIE$_3$	&	0.03	&	0.03	&	0.00	&	0.024	&	0.024	&	0.33	&	0.91	\\
 	50	&	$\rightarrow 0$	&	CIE$_4$	&	0.02	&	0.02	&	0.00	&	0.019	&	0.018	&	0.26	&	0.92	\\
 	50	&	$\rightarrow 0$	&	OIE	&	0.10	&	0.10	&	0.00	&	0.037	&	0.034	&	0.89	&	0.92	\\
 	$\rightarrow \infty$	&	$\rightarrow 0$	&	CIE$_1$	&	0.04	&	0.05	&	-0.01	&	0.123	&	0.122	&	0.12	&	0.91	\\
 	$\rightarrow \infty$	&	$\rightarrow 0$	&	CIE$_2$	&	0.01	&	0.01	&	0.00	&	0.026	&	0.026	&	0.14	&	0.90	\\
 	$\rightarrow \infty$	&	$\rightarrow 0$	&	CIE$_3$	&	0.03	&	0.03	&	0.00	&	0.320	&	0.328	&	0.11	&	0.89	\\
 	$\rightarrow \infty$	&	$\rightarrow 0$	&	CIE$_4$	&	0.02	&	0.02	&	0.00	&	0.244	&	0.238	&	0.09	&	0.91	\\
 	$\rightarrow \infty$	&	$\rightarrow 0$	&	OIE	&	0.10	&	0.11	&	-0.01	&	0.417	&	0.419	&	0.12	&	0.90	\\
 	1	&	0.1	&	CIE$_1$	&	0.04	&	0.04	&	0.00	&	0.035	&	0.037	&	0.34	&	0.88	\\
 	1	&	0.1	&	CIE$_2$	&	0.01	&	0.02	&	-0.01	&	0.106	&	0.105	&	0.10	&	0.90	\\
 	1	&	0.1	&	CIE$_3$	&	0.03	&	0.03	&	0.00	&	0.013	&	0.012	&	0.83	&	0.92	\\
 	1	&	0.1	&	CIE$_4$	&	0.02	&	0.02	&	0.00	&	0.011	&	0.010	&	0.57	&	0.92	\\
 	1	&	0.1	&	OIE	&	0.10	&	0.11	&	-0.01	&	0.114	&	0.113	&	0.23	&	0.90	\\
 	50	&	0.1	&	CIE$_1$	&	0.04	&	0.04	&	0.00	&	0.015	&	0.015	&	0.91	&	0.89	\\
 	50	&	0.1	&	CIE$_2$	&	0.01	&	0.01	&	0.00	&	0.014	&	0.014	&	0.15	&	0.91	\\
 	50	&	0.1	&	CIE$_3$	&	0.03	&	0.03	&	0.00	&	0.025	&	0.024	&	0.36	&	0.90	\\
 	50	&	0.1	&	CIE$_4$	&	0.02	&	0.02	&	0.00	&	0.020	&	0.019	&	0.25	&	0.92	\\
 	50	&	0.1	&	OIE	&	0.10	&	0.10	&	0.00	&	0.037	&	0.036	&	0.85	&	0.90	\\
 	$\rightarrow \infty$	&	0.1	&	CIE$_1$	&	0.04	&	0.04	&	0.00	&	0.117	&	0.113	&	0.11	&	0.91	\\
 	$\rightarrow \infty$	&	0.1	&	CIE$_2$	&	0.01	&	0.01	&	0.00	&	0.025	&	0.025	&	0.12	&	0.90	\\
 	$\rightarrow \infty$	&	0.1	&	CIE$_3$	&	0.03	&	0.02	&	0.01	&	0.305	&	0.307	&	0.10	&	0.90	\\
 	$\rightarrow \infty$	&	0.1	&	CIE$_4$	&	0.02	&	0.02	&	0.00	&	0.232	&	0.230	&	0.10	&	0.90	\\
 	$\rightarrow \infty$	&	0.1	&	OIE	&	0.10	&	0.09	&	0.01	&	0.397	&	0.391	&	0.10	&	0.90	\\
 	1	&	0.5	&	CIE$_1$	&	0.04	&	0.04	&	0.00	&	0.035	&	0.034	&	0.33	&	0.92	\\
 	1	&	0.5	&	CIE$_2$	&	0.01	&	0.01	&	0.00	&	0.128	&	0.128	&	0.10	&	0.91	\\
 	1	&	0.5	&	CIE$_3$	&	0.03	&	0.03	&	0.00	&	0.013	&	0.012	&	0.84	&	0.93	\\
 	1	&	0.5	&	CIE$_4$	&	0.02	&	0.02	&	0.00	&	0.011	&	0.011	&	0.56	&	0.91	\\
 	1	&	0.5	&	OIE	&	0.10	&	0.10	&	0.00	&	0.136	&	0.134	&	0.20	&	0.90	\\
 	50	&	0.5	&	CIE$_1$	&	0.04	&	0.04	&	0.00	&	0.015	&	0.015	&	0.91	&	0.90	\\
 	50	&	0.5	&	CIE$_2$	&	0.01	&	0.01	&	0.00	&	0.015	&	0.014	&	0.14	&	0.93	\\
 	50	&	0.5	&	CIE$_3$	&	0.03	&	0.03	&	0.00	&	0.025	&	0.024	&	0.34	&	0.92	\\
 	50	&	0.5	&	CIE$_4$	&	0.02	&	0.02	&	0.00	&	0.019	&	0.020	&	0.28	&	0.91	\\
 	50	&	0.5	&	OIE	&	0.10	&	0.10	&	0.00	&	0.037	&	0.036	&	0.88	&	0.91	\\
 	$\rightarrow \infty$	&	0.5	&	CIE$_1$	&	0.04	&	0.04	&	0.00	&	0.090	&	0.088	&	0.12	&	0.92	\\
 	$\rightarrow \infty$	&	0.5	&	CIE$_2$	&	0.01	&	0.01	&	0.00	&	0.020	&	0.018	&	0.11	&	0.92	\\
 	$\rightarrow \infty$	&	0.5	&	CIE$_3$	&	0.03	&	0.04	&	-0.01	&	0.234	&	0.229	&	0.10	&	0.91	\\
 	$\rightarrow \infty$	&	0.5	&	CIE$_4$	&	0.02	&	0.02	&	0.00	&	0.177	&	0.178	&	0.12	&	0.89	\\
 	$\rightarrow \infty$	&	0.5	&	OIE	&	0.10	&	0.11	&	-0.01	&	0.304	&	0.299	&	0.11	&	0.90	\\
\bottomrule
\end{tabular}
\end{table*}
\end{center}

\begin{center}
\begin{table*}[t]%
\caption{Empirical variances of the estimators of $\beta_{11}$ and $\gamma_{11}$ and their product under different scenarios and varying degrees of overdispersion ($\theta$) and sparsity ($\alpha_S$).
 The last two columns present the ratios between the squared estimators and the variances for for the first coordinate wise indirect effect (CIE$_1$).  For each setting, the values are based on 1000 replications of simulated data. \label{ievartaulu}}
\centering
\begin{tabular}{llllllllllll}
\toprule
\textbf{Scenario} & $\theta$ & $\alpha_S$ & $\beta_{11}$	&	$\gamma_{11}$ &		$\hat{\sigma}^2_{\beta_1}$	&	$\hat{\sigma}^2_{\gamma_1}$		&$\hat{\sigma}^2_{\beta_1}\hat{\sigma}^2_{\gamma_1}$	& 	$\beta_{11}^2/\hat{\sigma}^2_{\beta_1}$ &	$\gamma_{11}^2/\hat{\sigma}^2_{\gamma_1}$ & var(IE)
\\
\midrule
1	&	$\rightarrow 0$	&	1	&	-0.87	&	-0.05	&	0.0263	&	0.0002	&	0.000005	&	28.72	&	11.46	&	0.0002	\\
1	&	$\rightarrow 0$	&	50	&	-0.36	&	-0.11	&	0.0004	&	0.0139	&	0.000006	&	323.99	&	0.86	&	0.0018	\\
1	&	$\rightarrow 0$	&	$\rightarrow \infty$	&	-0.34	&	-0.15	&	0.0000	&	3.0048	&	0.000005	&	64007.84	&	0.01	&	0.3494	\\
1	&	0.1	&	1	&	-0.86	&	-0.05	&	0.0262	&	0.0002	&	0.000005	&	28.24	&	11.31	&	0.0002	\\
1	&	0.1	&	50	&	-0.36	&	-0.10	&	0.0004	&	0.0139	&	0.000006	&	323.92	&	0.75	&	0.0018	\\
1	&	0.1	&	$\rightarrow \infty$	&	-0.34	&	-0.15	&	0.0000	&	2.7107	&	0.000005	&	57500.11	&	0.01	&	0.3152	\\
1	&	0.5	&	1	&	-0.85	&	-0.05	&	0.0258	&	0.0002	&	0.000005	&	28.21	&	11.33	&	0.0002	\\
1	&	0.5	&	50	&	-0.36	&	-0.10	&	0.0004	&	0.0138	&	0.000006	&	319.80	&	0.79	&	0.0018	\\
1	&	0.5	&	$\rightarrow \infty$	&	-0.34	&	-0.14	&	0.0000	&	1.5932	&	0.000006	&	31618.72	&	0.01	&	0.1853	\\
2	&	$\rightarrow 0$	&	1	&	-0.19	&	-0.21	&	0.0252	&	0.0002	&	0.000005	&	1.46	&	226.34	&	0.0011	\\
2	&	$\rightarrow 0$	&	50	&	-0.07	&	-0.55	&	0.0004	&	0.0142	&	0.000005	&	14.31	&	21.43	&	0.0002	\\
2	&	$\rightarrow 0$	&	$\rightarrow \infty$	&	-0.07	&	-0.64	&	0.0000	&	3.0710	&	0.000006	&	2603.26	&	0.13	&	0.0151	\\
2	&	0.1	&	1	&	-0.19	&	-0.22	&	0.0290	&	0.0002	&	0.000006	&	1.27	&	234.20	&	0.0014	\\
2	&	0.1	&	50	&	-0.07	&	-0.54	&	0.0004	&	0.0143	&	0.000005	&	13.90	&	20.49	&	0.0002	\\
2	&	0.1	&	$\rightarrow \infty$	&	-0.07	&	-0.53	&	0.0000	&	2.7673	&	0.000006	&	2459.66	&	0.10	&	0.0136	\\
2	&	0.5	&	1	&	-0.19	&	-0.22	&	0.0245	&	0.0002	&	0.000005	&	1.49	&	233.19	&	0.0011	\\
2	&	0.5	&	50	&	-0.07	&	-0.54	&	0.0004	&	0.0142	&	0.000005	&	14.39	&	20.89	&	0.0002	\\
2	&	0.5	&	$\rightarrow \infty$	&	-0.07	&	-0.51	&	0.0000	&	1.6191	&	0.000006	&	1418.24	&	0.16	&	0.0080	\\
3	&	$\rightarrow 0$	&	1	&	-0.95	&	-0.11	&	0.0274	&	0.0002	&	0.000005	&	32.96	&	61.93	&	0.0005	\\
3	&	$\rightarrow 0$	&	50	&	-0.39	&	-0.26	&	0.0003	&	0.0145	&	0.000005	&	435.50	&	4.82	&	0.0022	\\
3	&	$\rightarrow 0$	&	$\rightarrow \infty$	&	-0.36	&	-0.27	&	0.0000	&	3.1057	&	0.000005	&	82809.50	&	0.02	&	0.4133	\\
3	&	0.1	&	1	&	-0.94	&	-0.11	&	0.0272	&	0.0002	&	0.000005	&	32.45	&	62.23	&	0.0005	\\
3	&	0.1	&	50	&	-0.39	&	-0.26	&	0.0003	&	0.0145	&	0.000005	&	429.88	&	4.61	&	0.0022	\\
3	&	0.1	&	$\rightarrow \infty$	&	-0.36	&	-0.28	&	0.0000	&	2.8001	&	0.000005	&	74406.43	&	0.03	&	0.3724	\\
3	&	0.5	&	1	&	-0.93	&	-0.11	&	0.0270	&	0.0002	&	0.000005	&	32.00	&	63.13	&	0.0005	\\
3	&	0.5	&	50	&	-0.39	&	-0.25	&	0.0003	&	0.0144	&	0.000005	&	432.93	&	4.46	&	0.0022	\\
3	&	0.5	&	$\rightarrow \infty$	&	-0.37	&	-0.29	&	0.0000	&	1.6401	&	0.000005	&	40849.87	&	0.05	&	0.2185	\\
\bottomrule
\end{tabular}
\end{table*}
\end{center}

\clearpage
\newpage

\begin{center}
\begin{table*}[t]%
\caption{Average counts and proportions of the 11 genera within the phylum \textit{Actinobacteria} by fibre intake (low, i.e., < 25 grams per day vs. high, i.e. $\geq$ 25 grams per day). The data are based on the STRIP study \Citep{Pahkala2020}. \label{phylocount}}
\centering
\begin{tabular}{llllllllllll}
\toprule
&\multicolumn{2}{@{}c@{}}{\textbf{Count}} & \multicolumn{2}{@{}c@{}}{\textbf{Proportion}} \\\cmidrule{2-3}\cmidrule{4-5}
\textbf{Genus} & \textbf{Fibre} < 25 g/d  & \textbf{Fibre} $\geq$ 25 g/d & \textbf{Fibre}< 25 g/d & \textbf{Fibre}$\geq$ 25 g/d\\
\midrule
\textit{Bifidobacterium}& 3780.22 &  1865.53 & 66.27 & 62.27 \\
\textit{Actinomyces}&  3.30 & 1.48 &  0.18 & 0.16  \\
\textit{Rothia}&  2.04 & 1.44 & 0.15 & 0. 15  \\
\textit{Enterorhabdus}& 8.56 & 4.56  & 0. 41 & 0.56\\
\textit{Eggerthella}& 52.95 & 29.96 & 3.00 & 2.60 \\
\textit{Collinsella}& 697.60 & 433.02 & 20.95 & 25.04 \\
\textit{Adlercreutzia}& 95.59 & 83.05 &  5.22 & 6.14 \\
\textit{Slackia}&  22.42 & 21.24 & 1.35 & 1.29 \\
\textit{Senegalimassilia}& 31.26 & 7.61 & 1.29 & 0.88 \\
\textit{Olsenella}&  14.34 & 7.88 & 0.79 & 0.55 \\
\textit{Gordonibacter}&  7.32 & 4.95 & 0.39 & 0.36  \\ 
\midrule
Total & 4715.61  & 2460.73  & 100 & 100 \\
\bottomrule
\end{tabular}
\end{table*}
\end{center}
\clearpage
\newpage

\begin{center}
\begin{table*}[t]%
\caption{Coefficients for the effect of fibre on the ilr coordinates and  coefficients for the effects of the ilr coordinates
on the response, and the coordinate-wise indirect effects (CIE's) $\beta_{1k} \gamma_{1k}$  for each coordinate  in the analyses for mediation through the genera within the Actinobacteria (section \ref{empwithin}). \label{phylotulos}  } 
\centering
\begin{tabular}{llllllllllll}
\toprule
 &$\hat{\beta}_{1k}$   & (90$\%$ CI)      &  $\hat{\gamma}_{1k}$ &  (90$\%$ CI) & $\hat{\beta}_{1k} \hat{\gamma}_{1k}$ (s.e.) & (90$\%$ CI) \\
\midrule
$M_1$   & -0.025  & [-0.341;  0.291] &    0.031  & [-0.006;   0.067] &    -0.001        (0.006) &            [-0.010;           0.009] \\
$M_2$   & -0.255  & [-0.501; -0.009] &   -0.031  & [-0.078;   0.016] &     0.008        (0.009) &            [-0.006;           0.022] \\
$M_3$   & -0.269  & [-0.740;  0.201] &    0.025  & [0.002;   0.048]  &    -0.007        (0.008) &            [-0.020;           0.007] \\
$M_4$   &  0.409  & [0.060;  0.759]  &   -0.041  & [-0.074;  -0.008] &    -0.017        (0.012) &            [-0.036;           0.003] \\
$M_5$   & -0.176  & [-0.712;  0.359] &    0.038  & [0.015;   0.060]  &    -0.007        (0.012) &            [-0.027;           0.014] \\
$M_6$   &  0.264  & [-0.220;  0.749] &    0.011  & [-0.013;   0.035] &     0.003        (0.005) &            [-0.005;           0.011] \\
$M_7$   &  0.071  & [-0.414;  0.557] &    0.001  & [-0.024;   0.027] &     0.000        (0.001) &            [-0.002;           0.002] \\
$M_8$   &  0.012  & [-0.434;  0.458] &    0.014  & [-0.010;   0.038] &     0.000        (0.004) &            [-0.006;           0.006] \\
$M_9$   & -0.165  & [-0.589;  0.259] &    0.017  & [-0.010;   0.044] &    -0.003        (0.005) &            [-0.011;           0.006] \\
$M_{10}$ & -0.151  & [-0.596;  0.295] &    0.012  & [-0.015;   0.039] &    -0.002        (0.004) &            [-0.009;           0.005] \\
\bottomrule
\end{tabular}
\end{table*}
\end{center}

\begin{center}
\begin{sidewaystable*}[t]%
\caption{Comparison of different approaches to mediation analysis with compositional mediator in \Citep{Zhang2019a}, \Citep{Sohn2019}, \Citep{Wang2020} and \Citep{Fu2023}. Abbreviations: SEM = Structural Equation Model; ilr = isometric logratio transformation; alr = additive logratio transformation; LM = linear regression model. By sieving we refer to testing for statistically significant signals in a large set of potential mediators. $^\dagger$: by overdispersion and sparsity, we refer to investigating the effect of these features on the estimation and performance in the simulation study.  
\label{paperivertailut}}
\centering
\begin{tabular}{l |l| l| l| l | l }
\toprule
 & \textbf{This study} & \textbf{Zhang et al.} &  \textbf{Sohn and Li} & \textbf{Wang} & \textbf{Fu}\\
\midrule
Framework & Counterfactual & SEM & Counterfactual & Counterfactual & Counterfactual\\
Mediator & Ilr & Ilr & Alr & Alr & Taxa \\
Overall IE & Yes & No & Yes & Yes & Yes \\
Mediator-wise IE & Yes & Yes & Yes & Yes & Yes \\
Zero counts & Replace with 0.5 & Replace with 0.5 & Replace with 0.5 & Dirichlet regression & Replace with 0.5 \\ 
Overdispersion$^\dagger$ & Yes&No &No &No &No  \\
Sparsity$^\dagger$ & Yes & No & No & No & No \\
Dimension & Low & High & High & High & High \\
Regularization & No & de-biased Lasso & No & L1 penalty & Bayesian priors  \\
Space & Euclidean & Euclidean & Simplex & Not specified  & Not specified \\
Taxonomic hierarchy & Yes & No & No & Possible & No \\
Model for mediator & LM & LM & Compositional algebra & Dirichlet regression &  Log-linear regression \\
Model for response & LM & LM & Linear log-contrast  & Linear log-contrast &  LM \\
Aim  &  Estimation of causal effects & Sieving & Estimation, sieving & Sieving & Estimation, sieving \\
\bottomrule
\end{tabular}
\end{sidewaystable*}
\end{center}
\clearpage
\newpage

\section*{Figures}
\subsection*{List of figure legends}
\begin{figure}[H]
\centerline{\includegraphics[width=12cm]{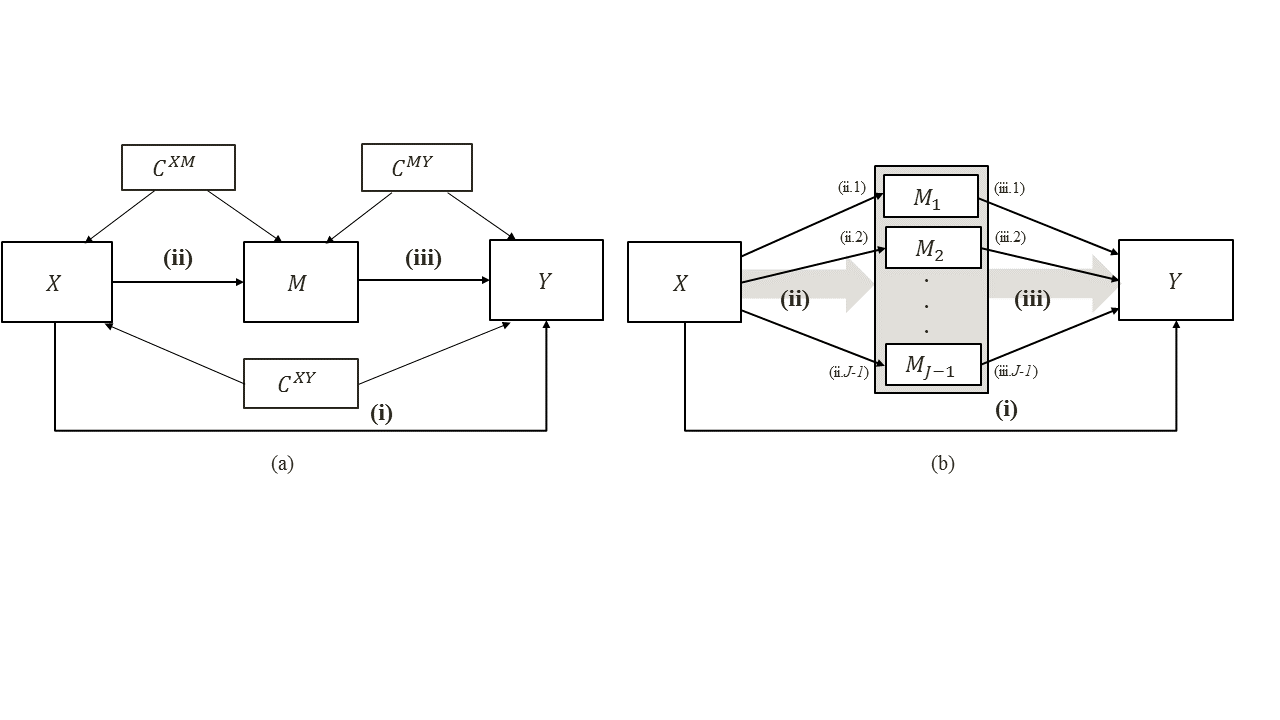}}
\caption{Directed acyclic graphs (DAGs) depicting mediation between exposure $X$ and response $Y$ through (a) a single mediator, and (b) a compositional mediator, such as the gut microbiome. Potential confounders are denoted as $C^{XY}$, $C^{XM}$ and $C^{MY}$ in DAG (a) and omitted for clarity in DAG (b). 
\label{fig1}}
\end{figure}

\begin{figure}[H]
\centerline{\includegraphics[width=12cm]{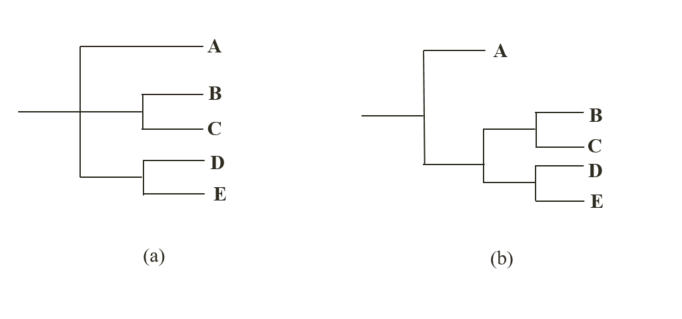}}
\caption{An example of a taxonomic hierarchy among five classes, i.e., five distinct taxa. (a) A taxonomy describing a known relational structure between  five taxa. (b) An example of a hierarchy based on taxonomy (a) and encoded by the sequential binary partition matrix \ref{taxmatrix} of the main text.
\label{fig2}}
\end{figure}

\begin{figure}[H]
\centerline{\includegraphics[width=14cm]{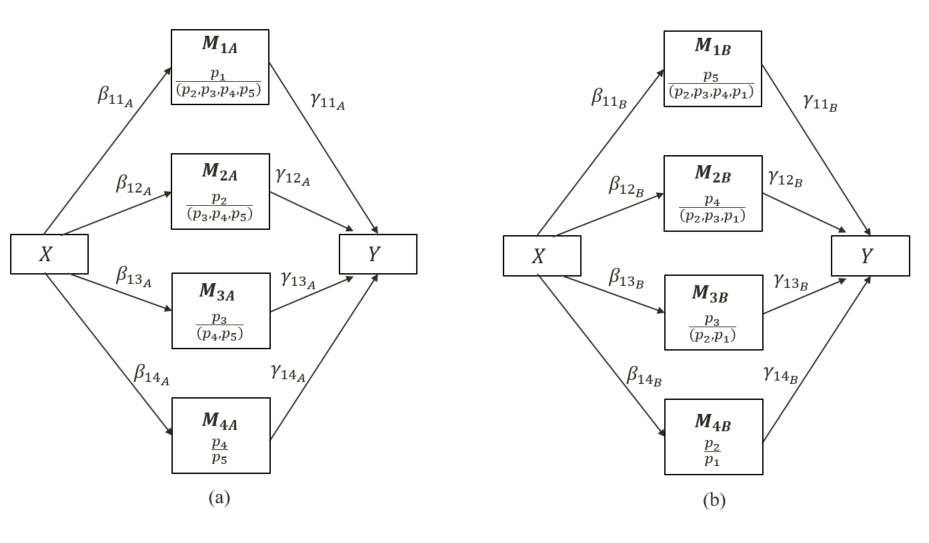}}
\caption{DAGs describing the contrasted parts of the composition in  models 3A and 3B. The correct causal DAG  used in simulating data is presented in (a) and the DAG encoded by the misspecified contrast matrix  is presented in (b).\label{fig3}}
\end{figure}

\begin{figure}[t]
\centerline{\includegraphics[width=14cm]{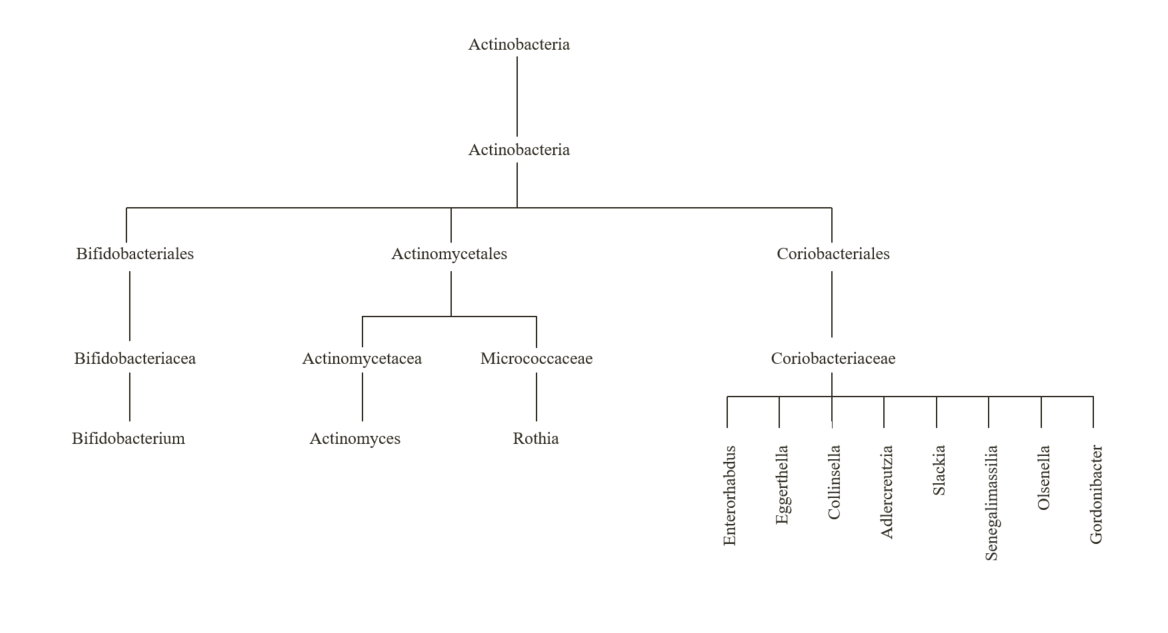}}
\caption{Taxonomic structure within the \textit{Actinobacteria} phylum. Within the phylum, there is one class, which was further divided into three orders consisting of eleven genera within four families. \label{fig4}}
\end{figure}

\begin{enumerate}
    \item[]   Figure  \ref{fig1}   Directed acyclic graphs (DAGs) depicting mediation between exposure $X$ and \break response $Y$ through (a) a single mediator, and (b) a compositional mediator, such as the gut microbiome. 
    \item[]    Figure  \ref{fig2}  An example of a taxonomic hierarchy among five classes, i.e., five distinct taxa. (a) A taxonomy describing a known relational structure between  five taxa. (b) An example of a hierarchy based on taxonomy (a) and encoded by the sequential binary partition matrix \ref{taxmatrix} of the main text. 
    \item[]  Figure  \ref{fig3}  DAGs describing the contrasted parts of the composition in  models 3A and 3B. The correct causal DAG  used in simulating data is presented in (a) and the DAG encoded by the misspecified contrast matrix  is presented in (b).
    \item[]     Figure  \ref{fig4}   Taxonomic structure within the \textit{Actinobacteria} phylum. Within the phylum, there is one class, which was further divided into three orders consisting of eleven genera within four families.  
\end{enumerate}

\newpage

\newpage

\newpage

\noindent {\bf{Acknowledgments}}
\noindent {\it{NK has been financially supported by Emil Aaltonen Foundation and the MATTI programme in The University of Turku Graduate School (UTUGS). The STRIP study has been financially supported by the Academy of Finland (grants 206374, 294834, 251360, 275595, 307996, and 322112), the Juho Vainio Foundation, the Finnish Foundation for Cardiovascular Research, the Finnish Ministry of Education and Culture, the Finnish Cultural Foundation, the Sigrid Jusélius Foundation, Special Governmental grants for Health Sciences Research (Turku University Hospital), the Yrjö Jahnsson Foundation, the Finnish Medical Foundation, and the Turku University Foundation.}}

\noindent {\bf{Data availability statement}} 
\noindent {\it{The details of the simulation study, including the extensive simulation codes, are available upon request from the corresponding author. The empirical dataset comprises health related participant data and their use is therefore
restricted under the regulations on professional secrecy (Act on the Openness of
Government Activities, 612/1999) and on sensitive personal data (Personal Data Act,
523/1999, implementing the EU data protection directive 95/46/EC). Due to these legal
restrictions, the data from this study (can not be stored in public repositories or otherwise
made publicly available. However, data access may be permitted on a case by case basis upon request only. Data sharing requires a data-sharing agreement. Investigators can submit an expression of
interest to the corresponding author.}}

\noindent {\bf{Conflict of Interest}}
\noindent {\it{The authors declare no potential conflict of interests.}}

\noindent {\bf{Financial disclosure}}
\noindent {\it{There are no financial conflicts of interest to disclose. }}

\noindent {\bf{Ethical approval and patient consent}}
\noindent {\it{The empirical study was approved by the associated university and hospital district ethical authorities. Written informed consent was obtained from the participants.  }}

\bibliographystyle{unsrt}  
\bibliography{libraryshort}

\appendix
\beginsupplement
\clearpage

\section*{Supplementary material}

\section{Mediation analysis}
\subsection{Total, direct and indirect effects in compositional mediation }\label{IE_selitys}

In this supplement, we provide the detailed definitions of the direct, indirect and total effects. This includes defining non-parametric estimands for coordinate-wise indirect effects in a situation with multiple correlated mediators, i.e., correlated ilr coordinates. The non-parametric definition of coordinate-wise effects is based on Wang et al. \Citep{Wang2013}. Based on the parametric models as specified in Section 3 of the main text, we then derive parametric expressions for the direct and indirect effects, allowing their estimation from empirical observations under a set of ignorability assumptions. Finally, we briefly discuss other approaches that have been presented to define and empirically estimate effects mediated through individual coordinates of a vector-valued mediator.
 
Denote the levels of a binary exposure $X$ as $x$ and $x^\ast$ and let $Y$ be a continuous response variable. Let $Y(x^\prime,{\bm m})$ denote the response if exposure was set to level $x^\prime$ and a vector-valued mediator ${\bm M}$ to value ${\bf m}$. Then, 
$Y(x^\prime,{\bm M}(x^{\prime\prime}))$ is the response if $X$ was set to level $x^\prime$ and the mediator ${\bm M}$ would obtain the value it would naturally have when exposure is set at  $x^{\prime\prime}$. Here $x^\prime,x^{\prime\prime} \in \lbrace x,x^\ast\rbrace$. 

To define coordinate-wise effects, we also need more detailed manipulations of the vector-valued mediator $\bm M$. Specifically, for any $k=1,\ldots,J$, let $Y(x_0,{\bm M}_{k-}(x_1),M_{k}(x_2),{\bm M}_{k+}(x_3))$ denote the response if exposure is set at $x_0$, and $(M_1,\ldots,M_{k-1})$, $M_k$ and $(M_{k+1},\ldots,M_{J})$ would obtain values they would have when exposure is set at $x_1$, $x_2$, and $x_3$, respectively. Here $x_0,x_1,x_2,x_3\in\lbrace x,x^\ast\rbrace$. Note that with $x_0=x^\prime$ and $x_1=x_2=x_3=x^{\prime\prime}$ (and for any $k$) we obtain as a special case that 
$Y(x^\prime,{\bm M}_{k-}(x^{\prime\prime}),M_k(x^{\prime\prime}),
M_{k+}(x^{\prime\prime}))$ $= Y(x^\prime,{\bm M}(x^{\prime\prime}))$.

Following the standard approach to causal mediation analysis (see e.g. VanderWeele {\sl et al.}, 2014), we  define the natural direct effect (NDE),  the 
natural overall indirect effect (OIE) and the total effect (TE) as follows:
$$
\begin{array}{ll}
\hbox{NDE} &= \hbox{E}[Y(x,{\bm M}(x^\ast))] - \hbox{E}[Y(x^\ast, {\bm M}(x^\ast))],\\[3mm]
\hbox{OIE} &= \hbox{E}[Y(x,{\bm M}(x))] - \hbox{E}[Y(x,{\bm M}(x^\ast))],\\[3mm]
\hbox{TE}
 &=  \hbox{E}[Y(x,{\bm M}(x))] - \hbox{E}[Y(x^\ast, {\bm M}(x^\ast))].
\end{array}
$$
As usual, the above effects are average direct and indirect natural effects where the average is taken with respect to the population of units (individuals) \Citep{Pearl2001}. The total effect equals the sum of the direct and indirect effects, i.e.,
$\hbox{TE} = \hbox{NDE} + \hbox{OIE}$. The overall indirect effect refers to the effect on exposure mediated through the entire vector ${\bm M}$.

Following Wang et al. \Citep{Wang2013}, we next define natural coordinate-wise indirect effects in the following manner:
\begin{equation}
\begin{array}{lll}
\hbox{CIE}_k &=  
&\hbox{E}[Y(x,{\bm M}_{k-}(x),M_k(x),{\bm M}_{k+}(x^\ast))]\\[4mm]
&~~-\hskip -11pt &\hbox{E}[Y(x,{\bm M}_{k-}(x),M_k(x^\ast),{\bm M}_{k+}(x^\ast)],~k=1,...,J.\\[3mm]
\end{array}
\end{equation}
It follows from (S1) that the coordinate-wise effects sum up to the overall indirect effect,
i.e., $\hbox{OIE} = \sum_{k=1}^{J}\hbox{CIE}_k$ as follows \Citep{Wang2013}:

 $$\begin{array}{ll}
 \sum_{k=1}^{J} \hbox{CIE}_{k} 
 &= \sum_{k=1}^{J}
 \hbox{E}[Y(x,{\bm M}_{k-}(x),M_k(x),{\bm M}_{k+}(x^\ast))] 
 -\hbox{E}[Y(x,{\bm M}_{k-}(x),M_k(x^\ast),{\bm M}_{k+}(x^\ast)] \\ [3mm] 
 & = \hbox{E}[Y(x, M_1(x), {\bm M}_{1+}(x^\ast))] - \hbox{E}[Y(x, M_1(x^\ast), {\bm M}_{1+}(x^\ast))] \\  [3mm] 
&\hskip 11pt + \hbox{E}[Y(x, M_{1}(x), M_2(x), {\bm M}_{2+}(x^\ast) )] - \hbox{E}[Y(x, M_1(x), M_2(x^\ast), {\bm M}_{2+}(x^\ast))] \\[3mm]
 &\hskip 11pt +\ldots\\[3mm]
 &\hskip 11pt + \hbox{E}[Y(x, {\bm M}_{J-}(x), M_J(x))] - 
  \hbox{E}[Y(x, {\bm M}_{J-}(x), M_J(x^\ast))] \\[3mm]
  & = - \hbox{E}[Y(x, M_1(x^\ast), {\bm M}_{1+}(x^\ast))]  + \hbox{E}[Y(x, {\bm M}_{J-}(x), M_J(x))] \\ [3mm]
& = \hbox{E}[Y(x, {\bm M}(x))] - \hbox{E}[Y(x, {\bm M}(x^\ast))],\\[3mm]
 \end{array}
 $$
 since the first and second terms on each succesive two rows cancel out. 

An alternative definition consideres each coordinate with all the remainig coordinates
set at their reference level (cf. Kim et al., 2019):
\begin{equation}
\begin{array}{lll}
\overline{\hbox{CIE}}_k&=  
&\hbox{E}[Y(x,{\bm M}_{k-}(x),M_k(x),{\bm M}_{k+}(x))]\\[4mm]
&~~-\hskip -11pt &\hbox{E}[Y(x,{\bm M}_{k-}(x),M_k(x^\ast),{\bm M}_{k+}(x)],~k=1,...,J. \\[3mm]
\end{array}
\end{equation}
Of note, with definition (S2), the coordinate-wise effects do not generally sum up to the overall indirect effect.

{\bf Identifiability of the direct and indirect effects.} To enable the estimation of direct and indirect effects from empirical observations, a set of assumptions is needed \Citep{Pearl2014}. Here, we require the existence of set of variables $C$ such that the following sequentical ignorability assumptions hold  \Citep{Wang2013}: 

$$\begin{array}{ll}
\hbox{(i)}~\lbrace{Y(x,{\bf m}),{\bm M}}\rbrace \indep X\vert C,\\[2mm]
\hbox{(ii)}~Y(x^\prime,{\bm m})\indep M_k(x^\prime)\vert X=x^\prime,C;~~ k=1,\ldots,J; x^\prime\in\lbrace x,x^\ast\rbrace .\\[2mm]
\end{array}
$$

Let $H$ denote the set of strata pertaining to the variables $C = \lbrace C_{XM},C_{XY},C_{MY}\rbrace$ (see the main text) and $P(h)$ be the probability of a unit belonging to stratum $h\in H$. As shown in \Citep{Wang2013}, the following then holds under the above ignorability assumptions:
\begin{equation}
\begin{array}{ll}
&\hskip -55pt \hbox{E}[Y(x_0,{\bf M}_{k-}(x_1),M_k(x_2),{\bf M}_{k+}(x_3)\vert h]\\[4mm]
&\hskip -55pt = \int\hbox{E}[Y\vert x_0,{\bm M}_{k-}(x_1)={\bf m}_{k-},M_k(x_2)= m_k,{\bm M}_{k+}(x_3) 
= {\bm m}_{k+}(x_3),h]dF_{{\bm M}_{k-}(x_1),M_k(x_2),{\bm M}_{k+}(x_3)}({\bm m}_{k-},m_k,{\bm m} _{k+}),\\[4mm]
\end{array}
\end{equation}
where, separately within each stratum, the integration is with respect to the joint distribution of the mediators $(M_1,\ldots,M_J)$. 

In general, the integration here is a ``cross-world" problem, which is not tractable without further assumptions. Wang et al.  \Citep{Wang2013} suggested employing the empirically estimable marginal distributions of $M_k$, $k=1,\ldots,J$, along with a pre-spefied correlation structure, e.g. a uniform correlation between all elements of $\bm M$.
In our application, however, with a linear dependence of  $\hbox{E}[Y\vert x,{\bm m}]$ on the mediator and in absence of interaction between exposure and the mediator, the integral only depends on the marginal distributions of $M_k$, $k=1,\ldots,J$. Under the ignorability assumptions and the linear parametric models (9) and (10) of the main text, the overall indirect effect can thus be estimated as

$$
\begin{array}{ll}
\hbox{OIE} &= \sum_{h \in H}\left[\int (\gamma_{h0} + \sum_{k=1}^{J}\gamma_{1k}m_{k} + \gamma_2 x)dF_{{\bm M}(x)}({\bm m})\right.\\[4mm]
&~~~~~~~~~~~~\left. 
- 
\int(\gamma_{h0} + \sum_{k=1}^{J}\gamma_{1k}
m_k + \gamma_2 x)dF_{{\bm M}(x^\ast)}({\bm m})
\right]\hbox{P}(h)\\[4mm]
& = \sum_{h\in H}\sum_{k=1}^{J}\int\gamma_{1k}m_k 
[dF_{{\bm M}(x)}({\bm m})  - dF_{{\bm M}(x^\ast)}({\bm m})]\hbox{P}(h)\\[4mm]
&= \sum_{h\in H}\sum_{k=1}^{J}\gamma_{1k}\beta_{h1k}\hbox{P}(h)\\[4mm]
&= \sum_{k=1}^J\gamma_{1k}\sum_{h\in H}\beta_{h1k}\hbox{P}(h).
\end{array}
$$
The above expression for OIE for multiple mediators was derived by VanderWeele {\sl et al.} (2014) under slightly different identifiability conditions. Note that in our application the last expression follows from the previous, more general one because the parameters for the dependence of the response on the mediator and exposure are assumed to be independent of the strata in $H$.

Similarly to OIE, under the ignorability assumptions and given the parametric models (10) and (11) of the main text, the direct direct effect can be estimated as

$$
\begin{array}{ll}
\hbox{NDE} &= \sum_{h\in H}\left[\int (\gamma_{h0} + \sum_{k=1}^{J}
\gamma_{1k}m_k + \gamma_2 x)dF_{{\bm M}(x^\ast)}({\bm m}) \right.\\[4mm]
 &~~~~~~~~~~~~~~ \left. 
-\int (\gamma_{h0} + \sum_{k=1}^{J}\gamma_{1k}
m_k + \gamma_2 x^\ast )dF_{{\bm M}(x^\ast)}({\bm m})\right] P(h)\\[4mm]
& = \gamma_2(x-x^\ast).
\end{array}
$$

Finally, with parametric models (10) and (11) the coordinate-wise effects
can be estimated as

$$\begin{array}{ll}
\hbox{CIE}_k 
& = \sum_{h\in H}\left\lbrace\int
[\gamma_{h0}+\sum_{k=1}^{J}\gamma_{1k}m_k + 
\gamma_2x]
dF_{{\bm M}_{k-}(x),M_k(x),{\bm M}_{k+}(x^\ast)}\right\rbrace P(h)\\[6mm]
&~~~-\sum_{h\in H}\left\lbrace \int
[\gamma_{h0}+\sum_{k=1}^{J}\gamma_{1k}m_k + \gamma_2x]
dF_{{\bm M}_{k-}(x),M_k(x^\ast),{\bm M}_{k+}(x^\ast)}
\right\rbrace P(h)\\[6mm]
& = \sum_{h\in H}\sum_{k=1}^J\gamma_{1k}m_k
[
dF_{{\bm M}_{k-}(x),M_k(x),{\bm M}_{k+}(x^\ast)}
- dF_{{\bm M}_{k-}(x),M_k(x^\ast),{\bm M}_{k+}(x^\ast)}]\hbox{P}(h)\\[4mm]
& = \sum_{h\in H}\gamma_{1k}\beta_{h1k}P(h)\\[4mm]
& = \gamma_{1k}\sum_{h\in H}\beta_{h1k}\hbox{P}(h). 
\end{array}
$$

We note that with the linear parametric model for the dependence of the response on the mediators, also definition (S2) leads to the same expressions. This means that unlike in the general nonparametric defintion the coordinate-wise effects do not sum up to the overall indirect effect, such additivity still holds with the chosen parametric model.

\vskip 11pt
{\bf Relation to other approaches.} We based the treatment of coordinate-wise effects on the approach by Wang et al.  \Citep{Wang2013}. In particular, the (non-parametric) definition of coordinate-wise causal contrasts were built sequentially (see (S1) above). While the parametric model in Wang et al. was based on a logistic regression of the response on the mediators and exposure, the linear model in our paper lead to expressions that do not require the specification of the correlation structure among the mediators. 

Kim et al. (2019) presented an alternative approach to define coordinate-wise effects.
Their definition is essentially the same as (S2) above. Kim et al. applied the model to 
more complicated problem with multiple dependent mediators under somewhat different
identifiability conditions. In the liner parametric model, as applied in our paper, the parametric expressions turn out to be the same with definition (S1).

VanderWeele (2014) considered indirect effects mediated through a sequence of vector-valued mediators ${\bm M}^{(k)} = (M_1,\ldots,M_k)$, $k=1,\ldots J$. For each model, they then defined the indirect effect as

$$\begin{array}{ll}
&\hbox{c}_k=\hbox{E}[Y(x,{\bm M}^{(k)}(x)] - \hbox{E}[Y(x,{\bm M}^{(k-1)}(x^\ast)],~k=1,\ldots,J,
\end{array}
$$

Moreover, it was shown that if the parametric model for the dependence of $Y$ on
the mediators is linear and there is no interaction between exposure and $\bm M$, 
the different models are coherent in the sense that they all can be derived from
the largest model with $k=J$. Moreover, the identifiability conditions for model with $k=J$ render also models with $k<H$ estimable from empirical data.
This makes it possible to consider coordinate-wise effects as increments in the
mediated effect by inclusion of one more coordinate to the mediator: $c_k-c_{k-1}$, $k=2,\ldots, J$. Nevertheless, each coordinate-wise effect refers to two different causal models as it measures the increment in the mediated effect by inclusion of one more
individual mediator in a pre-set order. In addition, unless the elements of the mediator $\bm M^{(k)}$ are uncorrelated, these effects are not equal to $\gamma_{1k}\beta_{1k}$.

Similarly to our paper, Fu et al. \Citep{Fu2023} considered log-ratios (i.e. balances or ilr coordinates) as mediators. Coordinate-wise effects were defined only for the pivotal SBP matrix and were basically regression-based, i.e., relied on a parametric formulation, thus lacking a proper definition in terms of causal contrasts. Consequently, the parametric forms of the indirect effects were not similar to those of Wang et al. \Citep{Wang2013} or those considered in our paper.

\subsection{Standard error of the indirect effect}\label{s_e_t}

The standard errors of the indirect effects were derived using the delta method  \Citep{Valeri2013}. Let $\Sigma_\beta$ and $\Sigma_\gamma$ denote the variance-covariance matrices of the estimators of $ \boldsymbol{\beta} = (\beta_{1k},\ldots,\beta_{1,J})^\prime$ and 
$\boldsymbol{\gamma} = (\gamma_{1k},\ldots,\gamma_{1,J})^\prime$, respectively. Let  $\boldsymbol{\psi} = (\boldsymbol{\beta}^\prime,\boldsymbol{\gamma}^\prime)^\prime$. The variance of the estimator of
OIE, ($\boldsymbol{\psi}$) = $\sum_{k=1}^{J}\beta_{1k}\gamma_{1k}$, is then

\begin{align}\label{deltamethod}
&{[\nabla \hbox{OIE}(\boldsymbol{\psi})]}^\prime 
\left(
\begin{matrix}
\Sigma_\beta & 0 \\
0 & \Sigma_\gamma \\
\end{matrix}
\right)  \nabla  \hbox{OIE}(\boldsymbol{\psi}) \\ \nonumber
& = \sum_{k=1}^{J} \beta^2_{1k} \hbox{var}(\hat{\gamma}_{1k}) +
 \sum_{k=1}^{J} \gamma^2_{1k} \hbox{var}(\hat{\beta}_{1k}) \\
& + \sum_{k \neq l} \beta_{1k} \beta_{1l} \hbox{cov}(\hat{\gamma}_{1k} + \hat{\gamma}_{1l}) + 
 \sum_{k \neq l} \gamma_{1k} \gamma_{1l} \hbox{cov}(\hat{\beta}_{1k} + \hat{\beta}_{1l}). \nonumber
\end{align}

Correspondingly, the variances of the CIE's IE($\boldsymbol{\psi}$) = $\beta_{1k}\gamma_{1k}$ are 
\begin{equation}\label{cie_se}
\beta_{1k}^2\hbox{var}(\hat\gamma_{1k}) + \gamma_{1k}^2\hbox{var}(\hat\beta_{1k})
= \sigma_{\beta k}^2 \sigma_{\gamma k}^2\left\lbrace
{\left( \frac{\beta_{1k}}{\sigma_{\beta k}}\right)}^2
 + {\left(\frac{\gamma_{1k}}{\sigma_{\gamma k}}\right)}^2
\right\rbrace,~k=1,\ldots,J,
\end{equation}
where $\sigma_{\beta k}^2 = \hbox{var}(\hat\beta_{1k})$ and $\sigma_{\gamma k}^2 = \hbox{var}(\hat\gamma_{1k})$. Standard errors were obtained as square roots of expressions (\ref{deltamethod}) and (\ref{cie_se}) after replacing the parameters with their estimated values.

\clearpage
\newpage

\section{Simulation study}
\subsection{Supplementary tables of the simulation study}
Here, we present some supplementary results for the simulation studies. The tables of this Appendix are as follows:

\begin{tabular}{ll}
\ref{trueparameters_all} & The $\beta_{1k}$ and $\gamma_{1k}$ parameters under different parameter combinations  \\
\ref{tulos3a_10000} & Simulation results for scenario 3, model 3A\\
\ref{tulos3b_10000} & Simulation results for scenario 3, model 3B \\
\ref{scen12_ilr} & The expected values and variances of the ilr coordinates in simulation scenarios 1 and 2. \\
\ref{scen3_ilr} & The expected values and variances of the ilr coordinates in simulation scenarios 3A and 3B. \\
\ref{scen12_betagamma} & Average values of the estimates of 
$\beta_{1k}$ and $\gamma_{1k}$ parameters and their variances  in scenarios 1 and 2\\
\ref{scen3_betagamma} & Average values of the estimates of 
$\beta_{1k}$ and $\gamma_{1k}$ parameters and the their variances  in scenarios 3A and 3B \\ 
\end{tabular}

\clearpage
\newpage

\begin{center}
\begin{table*}[t]%
\caption{The $\beta_{1k}$ parameters and corresponding $\gamma_{1k}$ parameters  in  scenarios 1, 2 and 3 under different values for parameters $\alpha_S$ and $\theta$. 
The product $\beta_{1k} \gamma_{1k}$ equals to the chosen value of CIE as given in  Table \ref{parametritaulu}. The $\beta_{1k}$ parameters are averages over the stratum-specific values  $\beta_{h1k}$. The stratum-specific values for each combination of stratum and $X$ were simulated using  Monte Carlo simulation with $\mu=10000$ per parameter combination. 
\label{trueparameters_all}}
\centering
\begin{tabular}{l l l l l ll}
\toprule
 $\alpha_S$ & $\theta$  & $\boldsymbol{\beta}_1$ & $\boldsymbol{\gamma}_1$  \\ 
\midrule
 	1	&	$\rightarrow 0$	&	(-0.87, -1.742, 1.747, -2.031)	&	(-0.046, -0.006, 0.017, -0.01)	\\
 	50	&	$\rightarrow 0$		&	(-0.363, -0.905, 0.758, -0.953)	&	(-0.11, -0.011, 0.04, -0.021)	\\
 	$\rightarrow \infty$	&	$\rightarrow 0$		&	(-0.341, -0.775, 0.672, -0.832)	&	(-0.117, -0.013, 0.045, -0.024)	\\
 	1	& 0.1 &	(-0.869, -1.731, 1.734, -2.013)	&	(-0.046, -0.006, 0.017, -0.01)	\\
 	50	&	0.1&	(-0.363, -0.904, 0.758, -0.953)	&	(-0.11, -0.011, 0.04, -0.021)	\\
 	$\rightarrow \infty$	&	0.1	&	(-0.341, -0.776, 0.672, -0.832)	&	(-0.117, -0.013, 0.045, -0.024)	\\
 	1	&	0.5	&	(-0.855, -1.674, 1.682, -1.952)	&	(-0.047, -0.006, 0.018, -0.01)	\\
 	50	&	0.5	&	(-0.364, -0.906, 0.758, -0.954)	&	(-0.11, -0.011, 0.04, -0.021)	\\
 	$\rightarrow \infty$	&	0.5	&(-0.341, -0.776, 0.673, -0.832)	&	(-0.117, -0.013, 0.045, -0.024)	\\
\midrule
 	1	&	$\rightarrow 0$	&		(-0.19, 0.024, -0.836, 0.666)	&	(-0.211, 0.408, -0.036, 0.03)	\\
 	50	&	$\rightarrow 0$	&		(-0.073, 0.026, -0.305, 0.331)	&	(-0.548, 0.379, -0.098, 0.06)	\\
 	$\rightarrow \infty$	&	$\rightarrow 0$	&	(-0.07, 0.021, -0.281, 0.287)	&	(-0.57, 0.484, -0.107, 0.07)	\\
 	1	&	0.1	&	(-0.186, 0.023, -0.83, 0.66)	&	(-0.215, 0.426, -0.036, 0.03)	\\
 	50	&	0.1	&	(-0.073, 0.027, -0.305, 0.331)	&	(-0.546, 0.364, -0.098, 0.06)	\\
 	$\rightarrow \infty$	&	0.1	&		(-0.07, 0.021, -0.281, 0.287)	&	(-0.57, 0.486, -0.107, 0.07)	\\
 	1	&	0.5	&	(-0.185, 0.019, -0.81, 0.639)	&	(-0.216, 0.528, -0.037, 0.031)	\\
 	50	&	0.5	&	(-0.074, 0.027, -0.305, 0.332)	&	(-0.543, 0.369, -0.098, 0.06)	\\
 	$\rightarrow \infty$	&	0.5		&	(-0.07, 0.021, -0.281, 0.287)	&	(-0.57, 0.483, -0.107, 0.07)	\\
\midrule
	 	1	&	$\rightarrow 0$	& (-0.95, -0.01, -0.01, 0)	&	(-0.11, 0, 0, 0)	\\
	 	50	&	$\rightarrow 0$		&	(-0.39, 0, 0, 0)	&	(-0.26, 0, 0, 0)	\\
	 	$\rightarrow \infty$	&	$\rightarrow 0$	&	(-0.36, 0, 0, 0)	&	(-0.27, 0, 0, 0)	\\
	 	1	&	0.1	&	(-0.94, -0.01, -0.01, 0)	&	(-0.11, 0, 0, 0)	\\
	 	50	&	0.1	&	(-0.39, 0, 0, 0)	&	(-0.26, 0, 0, 0)	\\
	 	$\rightarrow \infty$	&	0.1	& (-0.36, 0, 0, 0)	&	(-0.27, 0, 0, 0)	\\
	 	1	&	0.5	&	(-0.93, 0, 0, 0)	&	(-0.11, 0, 0, 0)	\\
	 	50	&	0.5	&	(-0.39, 0, 0, 0)	&	(-0.26, 0, 0, 0)	\\
	 	$\rightarrow \infty$	&	0.5	&	(-0.36, 0, 0, 0)	&	(-0.27, 0, 0, 0)	\\
\bottomrule
\end{tabular}
\end{table*}
\end{center}

\begin{center}
\begin{table*}[t]%
\caption{Simulation results for scenario 3, model 3A. The true and estimated indirect effects, bias, average standard errors ($\widehat{\hbox{SE}}$), standard deviation of the estimates ($\hbox{SE}_{\hbox{est}}$), and power and coverage probability of the 90 \% CI 
of the coordinate-wise (CIE) and overall (OIE) indirect effects with varying sparsity $\alpha_S$ and overdispersion $\theta$,  when using the correct pivotal contrast matrix. \label{tulos3a_10000} }
\centering
\begin{tabular}{llllllllllll}
\toprule
$\alpha_S$ & $\theta$  & Eff.& True & Est. & Bias & $\widehat{\hbox{SE}}$ & $\hbox{SE}_{\hbox{est}}$ & Power & Coverage  \\ 
\midrule
 	1	&	$\rightarrow 0$	&	CIE$_1$	&	0.10	&	0.10	&	0.00	&	0.02	&	0.02	&	1.00	&	0.89	\\
 	1	&	$\rightarrow 0$	&	CIE$_2$	&	0.00	&	0.00	&	0.00	&	0.01	&	0.00	&	0.01	&	0.99	\\
 	1	&	$\rightarrow 0$	&	CIE$_3$	&	0.00	&	0.00	&	0.00	&	0.01	&	0.00	&	0.01	&	0.99	\\
 	1	&	$\rightarrow 0$	&	CIE$_4$	&	0.00	&	0.00	&	0.00	&	0.01	&	0.00	&	0.00	&	1.00	\\
 	1	&	$\rightarrow 0$	&	OIE	&	0.10	&	0.10	&	0.00	&	0.03	&	0.02	&	1.00	&	0.91	\\
 	50	&	$\rightarrow 0$	&	CIE$_1$	&	0.10	&	0.10	&	0.00	&	0.05	&	0.05	&	0.71	&	0.90	\\
 	50	&	$\rightarrow 0$	&	CIE$_2$	&	0.00	&	0.00	&	0.00	&	0.01	&	0.00	&	0.01	&	1.00	\\
 	50	&	$\rightarrow 0$	&	CIE$_3$	&	0.00	&	0.00	&	0.00	&	0.01	&	0.00	&	0.01	&	0.99	\\
 	50	&	$\rightarrow 0$	&	CIE$_4$	&	0.00	&	0.00	&	0.00	&	0.01	&	0.00	&	0.00	&	1.00	\\
 	50	&	$\rightarrow 0$	&	OIE	&	0.10	&	0.10	&	0.00	&	0.05	&	0.05	&	0.68	&	0.90	\\
 	$\rightarrow \infty$	&	$\rightarrow 0$	&	CIE$_1$	&	0.10	&	0.10	&	0.00	&	0.64	&	0.66	&	0.11	&	0.89	\\
 	$\rightarrow \infty$	&	$\rightarrow 0$	&	CIE$_2$	&	0.00	&	0.00	&	0.00	&	0.02	&	0.03	&	0.10	&	0.90	\\
 	$\rightarrow \infty$	&	$\rightarrow 0$	&	CIE$_3$	&	0.00	&	0.00	&	0.00	&	0.02	&	0.02	&	0.09	&	0.91	\\
 	$\rightarrow \infty$	&	$\rightarrow 0$	&	CIE$_4$	&	0.00	&	0.00	&	0.00	&	0.01	&	0.00	&	0.01	&	0.99	\\
 	$\rightarrow \infty$	&	$\rightarrow 0$	&	OIE	&	0.10	&	0.10	&	0.00	&	0.64	&	0.66	&	0.11	&	0.89	\\
 	1	&	0.1	&	CIE$_1$	&	0.10	&	0.10	&	0.00	&	0.02	&	0.02	&	1.00	&	0.88	\\
 	1	&	0.1	&	CIE$_2$	&	0.00	&	0.00	&	0.00	&	0.01	&	0.00	&	0.01	&	0.99	\\
 	1	&	0.1	&	CIE$_3$	&	0.00	&	0.00	&	0.00	&	0.01	&	0.00	&	0.01	&	0.99	\\
 	1	&	0.1	&	CIE$_4$	&	0.00	&	0.00	&	0.00	&	0.01	&	0.00	&	0.01	&	0.99	\\
 	1	&	0.1	&	OIE	&	0.10	&	0.10	&	0.00	&	0.03	&	0.02	&	1.00	&	0.90	\\
 	50	&	0.1	&	CIE$_1$	&	0.10	&	0.10	&	0.00	&	0.05	&	0.05	&	0.69	&	0.91	\\
 	50	&	0.1	&	CIE$_2$	&	0.00	&	0.00	&	0.00	&	0.01	&	0.00	&	0.01	&	0.99	\\
 	50	&	0.1	&	CIE$_3$	&	0.00	&	0.00	&	0.00	&	0.01	&	0.00	&	0.00	&	1.00	\\
 	50	&	0.1	&	CIE$_4$	&	0.00	&	0.00	&	0.00	&	0.01	&	0.00	&	0.00	&	1.00	\\
 	50	&	0.1	&	OIE	&	0.10	&	0.10	&	0.00	&	0.05	&	0.05	&	0.67	&	0.91	\\
 	$\rightarrow \infty$	&	0.1	&	CIE$_1$	&	0.10	&	0.10	&	0.00	&	0.61	&	0.63	&	0.12	&	0.89	\\
 	$\rightarrow \infty$	&	0.1	&	CIE$_2$	&	0.00	&	0.00	&	0.00	&	0.02	&	0.02	&	0.09	&	0.91	\\
 	$\rightarrow \infty$	&	0.1	&	CIE$_3$	&	0.00	&	0.00	&	0.00	&	0.02	&	0.02	&	0.11	&	0.89	\\
 	$\rightarrow \infty$	&	0.1	&	CIE$_4$	&	0.00	&	0.00	&	0.00	&	0.01	&	0.00	&	0.00	&	1.00	\\
 	$\rightarrow \infty$	&	0.1	&	OIE	&	0.10	&	0.10	&	0.00	&	0.61	&	0.63	&	0.13	&	0.89	\\
 	1	&	0.5	&	CIE$_1$	&	0.10	&	0.10	&	0.00	&	0.02	&	0.02	&	1.00	&	0.90	\\
 	1	&	0.5	&	CIE$_2$	&	0.00	&	0.00	&	0.00	&	0.01	&	0.00	&	0.00	&	1.00	\\
 	1	&	0.5	&	CIE$_3$	&	0.00	&	0.00	&	0.00	&	0.01	&	0.00	&	0.00	&	1.00	\\
 	1	&	0.5	&	CIE$_4$	&	0.00	&	0.00	&	0.00	&	0.01	&	0.00	&	0.01	&	0.99	\\
 	1	&	0.5	&	OIE	&	0.10	&	0.10	&	0.00	&	0.03	&	0.02	&	1.00	&	0.91	\\
 	50	&	0.5	&	CIE$_1$	&	0.10	&	0.10	&	0.00	&	0.05	&	0.05	&	0.67	&	0.91	\\
 	50	&	0.5	&	CIE$_2$	&	0.00	&	0.00	&	0.00	&	0.01	&	0.00	&	0.01	&	0.99	\\
 	50	&	0.5	&	CIE$_3$	&	0.00	&	0.00	&	0.00	&	0.01	&	0.00	&	0.01	&	0.99	\\
 	50	&	0.5	&	CIE$_4$	&	0.00	&	0.00	&	0.00	&	0.01	&	0.00	&	0.01	&	0.99	\\
 	50	&	0.5	&	OIE	&	0.10	&	0.10	&	0.00	&	0.05	&	0.05	&	0.65	&	0.92	\\
 	$\rightarrow \infty$	&	0.5	&	CIE$_1$	&	0.10	&	0.11	&	-0.01	&	0.47	&	0.48	&	0.11	&	0.90	\\
 	$\rightarrow \infty$	&	0.5	&	CIE$_2$	&	0.00	&	0.00	&	0.00	&	0.02	&	0.02	&	0.10	&	0.90	\\
 	$\rightarrow \infty$	&	0.5	&	CIE$_3$	&	0.00	&	0.00	&	0.00	&	0.02	&	0.02	&	0.08	&	0.92	\\
 	$\rightarrow \infty$	&	0.5	&	CIE$_4$	&	0.00	&	0.00	&	0.00	&	0.01	&	0.00	&	0.01	&	0.99	\\
 	$\rightarrow \infty$	&	0.5	&	OIE	&	0.10	&	0.11	&	-0.01	&	0.47	&	0.48	&	0.12	&	0.90	\\
\bottomrule
\end{tabular}
\end{table*}
\end{center}

\begin{center}
\begin{table*}[t]%
\caption{Simulation results for scenario 3, model 3B. The estimated indirect effects, average standard errors ($\widehat{\hbox{SE}}$) and standard deviation of the estimates ($\hbox{SE}_{\hbox{est}}$)
of the coordinate-wise (CIE) and overall (OIE) indirect effects with varying sparsity $\alpha_S$ and overdispersion $\theta$,  when using the incorrect pivotal contrast matrix. \label{tulos3b_10000} }
\centering
\begin{tabular}{llllllllllll}
\toprule
 $\alpha_S$ & $\theta$  & Eff.&  Est. & $\widehat{\hbox{SE}}$ & $\hbox{SE}_{\hbox{est}}$   \\ 
\midrule
 	1	&	$\rightarrow 0$	&	CIE$_1$	&	0.01	&		0.01	&	0.01	\\
 	1	&	$\rightarrow 0$	&	CIE$_2$	&	0.01	&		0.01	&	0.01	\\
 	1	&	$\rightarrow 0$	&	CIE$_3$	&	0.02	&		0.01	&	0.01	\\
 	1	&	$\rightarrow 0$	&	CIE$_4$	&	0.06	&		0.02	&	0.02	\\
 	1	&	$\rightarrow 0$	&	OIE	&	0.10	&		     0.03	&	0.02	\\
 	50	&	$\rightarrow 0$	&	CIE$_1$	&	0.01	&		0.01	&	0.01	\\
 	50	&	$\rightarrow 0$	&	CIE$_2$	&	0.01	&		0.01	&	0.01	\\
 	50	&	$\rightarrow 0$	&	CIE$_3$	&	0.02	&		0.02	&	0.02	\\
 	50	&	$\rightarrow 0$	&	CIE$_4$	&	0.06	&		0.03	&	0.03	\\
 	50	&	$\rightarrow 0$	&	OIE	&	0.10	&		0.05	&	0.05	\\
 	$\rightarrow \infty$	&	$\rightarrow 0$	&	CIE$_1$	&	0.01	&		0.11	&	0.11	\\
 	$\rightarrow \infty$	&	$\rightarrow 0$	&	CIE$_2$	&	0.01	&		0.14	&	0.14	\\
 	$\rightarrow \infty$	&	$\rightarrow 0$	&	CIE$_3$	&	0.02	&		0.20	&	0.20	\\
 	$\rightarrow \infty$	&	$\rightarrow 0$	&	CIE$_4$	&	0.06	&		0.44	&	0.45	\\
 	$\rightarrow \infty$	&	$\rightarrow 0$	&	OIE	&	0.10	      &		0.64	&	0.66	\\
 	1	&	0.1	&	CIE$_1$	&	0.01	&		0.01	&	0.01	\\
 	1	&	0.1	&	CIE$_2$	&	0.01	&		0.01	&	0.01	\\
 	1	&	0.1	&	CIE$_3$	&	0.02	&		0.01	&	0.01	\\
 	1	&	0.1	&	CIE$_4$	&	0.06	&		0.02	&	0.02	\\
 	1	&	0.1	&	OIE	&	0.10	&		0.03	&	0.02	\\
 	50	&	0.1	&	CIE$_1$	&	0.01		&	0.01	&	0.01	\\
 	50	&	0.1	&	CIE$_2$	&	0.01		&	0.01	&	0.01	\\
 	50	&	0.1	&	CIE$_3$	&	0.02		&	0.02	&	0.01	\\
 	50	&	0.1	&	CIE$_4$	&	0.06		&	0.03	&	0.03	\\
 	50	&	0.1	&	OIE	&	0.10	&		0.05	&	0.05	\\
 	$\rightarrow \infty$	&	0.1	&	CIE$_1$	&	0.01	&		0.10	&	0.10	\\
 	$\rightarrow \infty$	&	0.1	&	CIE$_2$	&	0.01	&		0.14	&	0.14	\\
 	$\rightarrow \infty$	&	0.1	&	CIE$_3$	&	0.02	&		0.19	&	0.19	\\
 	$\rightarrow \infty$	&	0.1	&	CIE$_4$	&	0.07	&		0.42	&	0.43	\\
 	$\rightarrow \infty$	&	0.1	&	OIE	&	0.10	&		0.61	&	0.63	\\
 	1	&	0.5	&	CIE$_1$	&	0.01	&		0.01	&	0.01	\\
 	1	&	0.5	&	CIE$_2$	&	0.01	&		0.01	&	0.01	\\
 	1	&	0.5	&	CIE$_3$	&	0.02	&		0.01	&	0.01	\\
 	1	&	0.5	&	CIE$_4$	&	0.06	&		0.02	&	0.02	\\
 	1	&	0.5	&	OIE	&	0.10	&		0.03	&	0.02	\\
 	50	&	0.5	&	CIE$_1$	&	0.01		&	0.01	&	0.01	\\
 	50	&	0.5	&	CIE$_2$	&	0.01		&	0.01	&	0.01	\\
 	50	&	0.5	&	CIE$_3$	&	0.02		&	0.02	&	0.02	\\
 	50	&	0.5	&	CIE$_4$	&	0.06		&	0.03	&	0.03	\\
 	50	&	0.5	&	OIE	&	0.10	&		0.05	&	0.05	\\
 	$\rightarrow \infty$	&	0.5	&	CIE$_1$	&	0.01		&	0.08	&	0.08	\\
 	$\rightarrow \infty$	&	0.5	&	CIE$_2$	&	0.02		&	0.10	&	0.11	\\
 	$\rightarrow \infty$	&	0.5	&	CIE$_3$	&	0.02		&	0.14	&	0.14	\\
 	$\rightarrow \infty$	&	0.5	&	CIE$_4$	&	0.07		&	0.32	&	0.32	\\
 	$\rightarrow \infty$	&	0.5	&	OIE	&	0.11	&		0.47	&	0.48	\\
\bottomrule
\end{tabular}
\end{table*}
\end{center}

\begin{center}
\begin{sidewaystable*}[t]%
\caption{The expected values and variances of the ilr coordinates in simulation scenarios 1 and 2  when $\mu=10000$. \label{scen12_ilr}}
\centering
\begin{tabular}{lllllllllllllllllll}
\toprule
 & &  \multicolumn{4}{c}{Avg. ilr, $X=0$} & \multicolumn{4}{c}{Avg. ilr, $X=1$} & \multicolumn{4}{c}{Var of ilr, $X=0$} & \multicolumn{4}{c}{Var of ilr, $X=1$}\\
 $\alpha_S$ & $\theta$ & $M_1$ & $M_2$ &$M_3$ & $M_4$ & $M_1$ & $M_2$ &$M_3$ & $M_4$ &  $M_1$ & $M_2$ &$M_3$ & $M_4$ &  $M_1$ & $M_2$ &$M_3$ & $M_4$ \\ 
\midrule
1	&	$\rightarrow 0$	&	4.82	&	1.61	&	0.76	&	1.15	&	3.96	&	-0.12	&	2.50	&	-0.87	&	5.14	&	12.53	&	13.62	&	11.55	&	6.12	&	12.53	&	11.53	&	13.45	\\
50	&	$\rightarrow 0$	&	1.80	&	0.74	&	0.28	&	0.63	&	1.44	&	-0.15	&	1.03	&	-0.33	&	0.09	&	0.30	&	0.19	&	0.41	&	0.09	&	0.25	&	0.33	&	0.19	\\
$\rightarrow \infty$	&	$\rightarrow 0$	&	1.70	&	0.65	&	0.26	&	0.53	&	1.36	&	-0.12	&	0.92	&	-0.30	&	0.00	&	0.01	&	0.01	&	0.00	&	0.01	&	0.01	&	0.03	&	0.00	\\
1	&	0.1	&	4.80	&	1.60	&	0.76	&	1.14	&	3.95	&	-0.12	&	2.47	&	-0.87	&	5.14	&	12.41	&	13.49	&	11.40	&	6.06	&	12.43	&	11.37	&	13.37	\\
50	&	0.1	&	1.80	&	0.74	&	0.28	&	0.62	&	1.44	&	-0.15	&	1.03	&	-0.33	&	0.09	&	0.30	&	0.19	&	0.41	&	0.09	&	0.25	&	0.33	&	0.19	\\
$\rightarrow \infty$	&	0.1	&	1.70	&	0.65	&	0.26	&	0.53	&	1.36	&	-0.12	&	0.92	&	-0.30	&	0.00	&	0.01	&	0.01	&	0.00	&	0.01	&	0.01	&	0.03	&	0.00	\\
1	&	0.5	&	4.72	&	1.55	&	0.73	&	1.11	&	3.88	&	-0.10	&	2.41	&	-0.85	&	5.07	&	11.90	&	12.98	&	10.93	&	5.98	&	11.99	&	11.02	&	12.83	\\
50	&	0.5	&	1.80	&	0.74	&	0.28	&	0.62	&	1.44	&	-0.15	&	1.03	&	-0.33	&	0.09	&	0.30	&	0.19	&	0.41	&	0.09	&	0.25	&	0.33	&	0.20	\\
$\rightarrow \infty$	&	0.5	&	1.70	&	0.65	&	0.26	&	0.53	&	1.36	&	-0.12	&	0.92	&	-0.30	&	0.00	&	0.01	&	0.01	&	0.01	&	0.01	&	0.01	&	0.03	&	0.00	\\
\hline
1	&	$\rightarrow 0$	&	4.80	&	1.56	&	0.76	&	0.24	&	4.62	&	1.59	&	-0.08	&	0.90	&	5.21	&	12.78	&	13.65	&	12.15	&	5.38	&	12.89	&	13.78	&	12.11	\\
50	&	$\rightarrow 0$	&	1.76	&	0.65	&	0.28	&	0.12	&	1.69	&	0.69	&	-0.03	&	0.45	&	0.08	&	0.27	&	0.19	&	0.34	&	0.08	&	0.26	&	0.18	&	0.35	\\
$\rightarrow \infty$	&	$\rightarrow 0$	&	1.67	&	0.58	&	0.26	&	0.10	&	1.60	&	0.61	&	-0.03	&	0.39	&	0.00	&	0.01	&	0.01	&	0.00	&	0.00	&	0.01	&	0.01	&	0.00	\\
1	&	0.1	&	4.78	&	1.55	&	0.76	&	0.23	&	4.60	&	1.61	&	-0.09	&	0.89	&	5.19	&	12.74	&	13.47	&	12.05	&	5.37	&	12.78	&	13.71	&	11.96	\\
50	&	0.1	&	1.76	&	0.65	&	0.28	&	0.12	&	1.69	&	0.69	&	-0.03	&	0.45	&	0.08	&	0.26	&	0.19	&	0.34	&	0.08	&	0.26	&	0.18	&	0.35	\\
$\rightarrow \infty$	&	0.1	&	1.67	&	0.58	&	0.26	&	0.10	&	1.60	&	0.61	&	-0.03	&	0.39	&	0.00	&	0.01	&	0.01	&	0.00	&	0.00	&	0.01	&	0.01	&	0.00	\\
1	&	0.5	&	4.69	&	1.50	&	0.74	&	0.22	&	4.51	&	1.54	&	-0.09	&	0.86	&	5.13	&	12.19	&	12.99	&	11.50	&	5.33	&	12.29	&	13.16	&	11.50	\\
50	&	0.5	&	1.76	&	0.65	&	0.28	&	0.12	&	1.69	&	0.69	&	-0.03	&	0.45	&	0.08	&	0.27	&	0.19	&	0.34	&	0.08	&	0.26	&	0.18	&	0.35	\\
$\rightarrow \infty$	&	0.5	&	1.67	&	0.58	&	0.26	&	0.10	&	1.60	&	0.61	&	-0.03	&	0.39	&	0.00	&	0.01	&	0.01	&	0.00	&	0.00	&	0.01	&	0.01	&	0.00	\\
\bottomrule
\end{tabular}
\end{sidewaystable*}
\end{center}

\begin{center}
\begin{sidewaystable*}[t]%
\caption{The expected values and variances of the ilr coordinates in simulation scenarios 3A and 3B  when $\mu=10000$. \label{scen3_ilr}}
\centering
\begin{tabular}{lllllllllllllllllll}
\toprule
 & &  \multicolumn{4}{c}{Avg. ilr, $X=0$} & \multicolumn{4}{c}{Avg. ilr, $X=1$} & \multicolumn{4}{c}{Var of ilr, $X=0$} & \multicolumn{4}{c}{Var of ilr, $X=1$}\\
 $\alpha_S$ & $\theta$ & $M_1$ & $M_2$ &$M_3$ & $M_4$ & $M_1$ & $M_2$ &$M_3$ & $M_4$ &  $M_1$ & $M_2$ &$M_3$ & $M_4$ &  $M_1$ & $M_2$ &$M_3$ & $M_4$ \\ 
\midrule
 	1	&	$\rightarrow 0$	&	4.73	&	0.05	&	0.21	&	0.01	&	3.78	&	0.03	&	0.21	&	0.01	&	5.34	&	13.61	&	13.57	&	13.38	&	6.41	&	13.88	&	13.75	&	13.79	\\
 	50	&	$\rightarrow 0$	&	1.70	&	0.02	&	0.08	&	0.00	&	1.32	&	0.01	&	0.07	&	0.00	&	0.07	&	0.24	&	0.23	&	0.24	&	0.07	&	0.19	&	0.18	&	0.18	\\
 	$\rightarrow \infty$	&	$\rightarrow 0$	&	1.62	&	0.02	&	0.08	&	0.00	&	1.26	&	0.01	&	0.07	&	0.00	&	0.00	&	0.01	&	0.01	&	0.00	&	0.00	&	0.01	&	0.00	&	0.00	\\
 	1	&	0.1	&	4.71	&	0.04	&	0.21	&	0.00	&	3.78	&	0.03	&	0.22	&	-0.02	&	5.32	&	13.48	&	13.43	&	13.29	&	6.35	&	13.72	&	13.65	&	13.66	\\
 	50	&	0.1	&	1.70	&	0.02	&	0.08	&	0.00	&	1.32	&	0.01	&	0.07	&	0.00	&	0.07	&	0.24	&	0.23	&	0.24	&	0.07	&	0.19	&	0.18	&	0.18	\\
 	$\rightarrow \infty$	&	0.1	&	1.62	&	0.02	&	0.08	&	0.00	&	1.26	&	0.01	&	0.07	&	0.00	&	0.00	&	0.01	&	0.01	&	0.00	&	0.00	&	0.01	&	0.00	&	0.00	\\
 	1	&	0.5	&	4.63	&	0.04	&	0.20	&	0.00	&	3.70	&	0.03	&	0.20	&	0.01	&	5.27	&	12.94	&	12.89	&	12.71	&	6.28	&	13.22	&	13.13	&	13.12	\\
 	50	&	0.5	&	1.70	&	0.02	&	0.08	&	0.00	&	1.32	&	0.02	&	0.07	&	0.00	&	0.08	&	0.24	&	0.23	&	0.24	&	0.07	&	0.19	&	0.18	&	0.18	\\
 	$\rightarrow \infty$	&	0.5	&	1.62	&	0.02	&	0.08	&	0.00	&	1.26	&	0.01	&	0.07	&	0.00	&	0.00	&	0.01	&	0.01	&	0.00	&	0.00	&	0.01	&	0.01	&	0.00	\\
\midrule 
 	1	&	$\rightarrow 0$	&	-1.30	&	-1.67	&	-2.04	&	-3.71	&	-1.06	&	-1.36	&	-1.60	&	-2.97	&	12.90	&	12.60	&	11.97	&	8.42	&	13.32	&	13.00	&	12.32	&	9.19	\\
 	50	&	$\rightarrow 0$	&	-0.47	&	-0.61	&	-0.73	&	-1.34	&	-0.37	&	-0.48	&	-0.56	&	-1.03	&	0.23	&	0.22	&	0.20	&	0.13	&	0.18	&	0.17	&	0.16	&	0.11	\\
 	$\rightarrow \infty$	&	$\rightarrow 0$	&	-0.45	&	-0.57	&	-0.70	&	-1.27	&	-0.35	&	-0.45	&	-0.54	&	-0.99	&	0.00	&	0.00	&	0.02	&	0.00	&	0.00	&	0.00	&	0.01	&	0.00	\\
 	1	&	0.1	&	-1.29	&	-1.66	&	-2.03	&	-3.70	&	-1.03	&	-1.37	&	-1.60	&	-2.97	&	12.79	&	12.50	&	11.87	&	8.36	&	13.23	&	12.86	&	12.19	&	9.09	\\
 	50	&	0.1	&	-0.47	&	-0.61	&	-0.73	&	-1.34	&	-0.37	&	-0.48	&	-0.56	&	-1.03	&	0.23	&	0.22	&	0.20	&	0.13	&	0.18	&	0.17	&	0.16	&	0.11	\\
 	$\rightarrow \infty$	&	0.1	&	-0.45	&	-0.57	&	-0.70	&	-1.27	&	-0.35	&	-0.45	&	-0.54	&	-0.99	&	0.00	&	0.00	&	0.02	&	0.00	&	0.00	&	0.00	&	0.01	&	0.00	\\
 	1	&	0.5	&	-1.26	&	-1.63	&	-2.00	&	-3.63	&	-1.03	&	-1.32	&	-1.57	&	-2.91	&	12.28	&	11.99	&	11.42	&	8.13	&	12.70	&	12.41	&	11.79	&	8.85	\\
 	50	&	0.5	&	-0.47	&	-0.61	&	-0.73	&	-1.34	&	-0.37	&	-0.47	&	-0.56	&	-1.03	&	0.23	&	0.22	&	0.21	&	0.13	&	0.18	&	0.17	&	0.16	&	0.11	\\
 	$\rightarrow \infty$	&	0.5	&	-0.45	&	-0.57	&	-0.70	&	-1.27	&	-0.35	&	-0.45	&	-0.54	&	-0.99	&	0.00	&	0.00	&	0.02	&	0.00	&	0.00	&	0.00	&	0.01	&	0.00	\\
\bottomrule
\end{tabular}
\end{sidewaystable*}
\end{center}

\begin{center}
\begin{sidewaystable*}[t]%
\caption{Average values of the estimates of 
$\beta_{1k}$ and $\gamma_{1k}$ parameters and the corresponding variances over the 1000 replications in scenarios 1 (top) and 2 (bottom). \label{scen12_betagamma}}
\centering
\begin{tabular}{lllllllllllllllllll}
\toprule
 & & \multicolumn{2}{c}{$\beta_{11}$} &  \multicolumn{2}{c}{$\beta_{12}$} & 
  \multicolumn{2}{c}{$\beta_{13}$} &    \multicolumn{2}{c}{$\beta_{14}$} & 
       \multicolumn{2}{c}{$\gamma_{11}$} &      \multicolumn{2}{c}{$\gamma_{12}$} &       \multicolumn{2}{c}{$\gamma_{13}$} &        \multicolumn{2}{c}{$\gamma_{14}$} \\
 $\alpha_S$ & $\theta$ & Avg. & Var & Avg. & Var & Avg. & Var & Avg. & Var & Avg. & Var & Avg. & Var & Avg. & Var & Avg. & Var \\
\midrule
	1	&	$\rightarrow 0$	&	-0.87	&	0.03	&	-1.76	&	0.06	&	1.75	&	0.06	&	-2.02	&	0.06	&	-0.05	&	0.00	&	-0.01	&	0.00	&	0.02	&	0.00	&	-0.01	&	0.00	\\
 	50	&	$\rightarrow 0$	&	-0.36	&	0.00	&	-0.91	&	0.00	&	0.76	&	0.00	&	-0.95	&	0.00	&	-0.11	&	0.01	&	-0.01	&	0.00	&	0.04	&	0.01	&	-0.02	&	0.00	\\
 	$\rightarrow \infty$	&	$\rightarrow 0$	&	-0.34	&	0.00	&	-0.78	&	0.00	&	0.67	&	0.00	&	-0.83	&	0.00	&	-0.15	&	3.00	&	0.01	&	0.99	&	0.05	&	1.22	&	-0.04	&	0.77	\\
 	1	&	0.1	&	-0.86	&	0.03	&	-1.75	&	0.06	&	1.73	&	0.06	&	-2.02	&	0.06	&	-0.05	&	0.00	&	-0.01	&	0.00	&	0.02	&	0.00	&	-0.01	&	0.00	\\
 	50	&	0.1	&	-0.36	&	0.00	&	-0.90	&	0.00	&	0.76	&	0.00	&	-0.95	&	0.00	&	-0.10	&	0.01	&	-0.01	&	0.00	&	0.04	&	0.01	&	-0.02	&	0.00	\\
 	$\rightarrow \infty$	&	0.1	&	-0.34	&	0.00	&	-0.78	&	0.00	&	0.67	&	0.00	&	-0.83	&	0.00	&	-0.15	&	2.71	&	0.01	&	0.90	&	0.06	&	1.10	&	0.00	&	0.70	\\
 	1	&	0.5	&	-0.85	&	0.03	&	-1.68	&	0.06	&	1.69	&	0.06	&	-1.96	&	0.06	&	-0.05	&	0.00	&	-0.01	&	0.00	&	0.02	&	0.00	&	-0.01	&	0.00	\\
 	50	&	0.5	&	-0.36	&	0.00	&	-0.91	&	0.00	&	0.76	&	0.00	&	-0.95	&	0.00	&	-0.10	&	0.01	&	-0.01	&	0.00	&	0.04	&	0.01	&	-0.02	&	0.00	\\
 	$\rightarrow \infty$	&	0.5	&	-0.34	&	0.00	&	-0.78	&	0.00	&	0.67	&	0.00	&	-0.83	&	0.00	&	-0.14	&	1.59	&	-0.01	&	0.52	&	0.04	&	0.65	&	-0.04	&	0.40	\\
\midrule
	1	&	$\rightarrow 0$	&	-0.19	&	0.03	&	0.01	&	0.06	&	-0.84	&	0.07	&	0.67	&	0.06	&	-0.21	&	0.00	&	0.41	&	0.00	&	-0.04	&	0.00	&	0.03	&	0.00	\\
 	50	&	$\rightarrow 0$	&	-0.07	&	0.00	&	0.03	&	0.00	&	-0.30	&	0.00	&	0.33	&	0.00	&	-0.55	&	0.01	&	0.38	&	0.00	&	-0.10	&	0.01	&	0.06	&	0.00	\\
 	$\rightarrow \infty$	&	$\rightarrow 0$	&	-0.07	&	0.00	&	0.02	&	0.00	&	-0.28	&	0.00	&	0.29	&	0.00	&	-0.64	&	3.07	&	0.53	&	0.99	&	-0.10	&	1.30	&	0.07	&	0.73	\\
 	1	&	0.1	&	-0.19	&	0.03	&	0.04	&	0.06	&	-0.84	&	0.07	&	0.66	&	0.06	&	-0.22	&	0.00	&	0.43	&	0.00	&	-0.04	&	0.00	&	0.03	&	0.00	\\
 	50	&	0.1	&	-0.07	&	0.00	&	0.02	&	0.00	&	-0.31	&	0.00	&	0.33	&	0.00	&	-0.54	&	0.01	&	0.36	&	0.00	&	-0.10	&	0.01	&	0.06	&	0.00	\\
 	$\rightarrow \infty$	&	0.1	&	-0.07	&	0.00	&	0.02	&	0.00	&	-0.28	&	0.00	&	0.29	&	0.00	&	-0.53	&	2.77	&	0.48	&	0.89	&	-0.09	&	1.18	&	0.08	&	0.65	\\
 	1	&	0.5	&	-0.19	&	0.03	&	0.02	&	0.06	&	-0.82	&	0.06	&	0.64	&	0.06	&	-0.22	&	0.00	&	0.53	&	0.00	&	-0.04	&	0.00	&	0.03	&	0.00	\\
 	50	&	0.5	&	-0.07	&	0.00	&	0.03	&	0.00	&	-0.31	&	0.00	&	0.33	&	0.00	&	-0.54	&	0.01	&	0.37	&	0.00	&	-0.10	&	0.01	&	0.06	&	0.00	\\
 	$\rightarrow \infty$	&	0.5	&	-0.07	&	0.00	&	0.02	&	0.00	&	-0.28	&	0.00	&	0.29	&	0.00	&	-0.51	&	1.62	&	0.48	&	0.52	&	-0.15	&	0.69	&	0.06	&	0.38	\\
\bottomrule
\end{tabular}
\end{sidewaystable*}
\end{center}

\begin{center}
\begin{sidewaystable*}[t]%
\caption{Average values of the estimates of 
$\beta_{1k}$ and $\gamma_{1k}$ parameters and the corresponding variances over the 1000 replications in scenario 3 with correct contrast matrix (top) and with a misspecified contrast matrix (bottom). \label{scen3_betagamma}}
\centering
\begin{tabular}{lllllllllllllllllll}
\toprule
& & \multicolumn{2}{c}{$\beta_{11}$} &  \multicolumn{2}{c}{$\beta_{12}$} & 
  \multicolumn{2}{c}{$\beta_{13}$} &    \multicolumn{2}{c}{$\beta_{14}$} & 
       \multicolumn{2}{c}{$\gamma_{11}$} &      \multicolumn{2}{c}{$\gamma_{12}$} &       \multicolumn{2}{c}{$\gamma_{13}$} &        \multicolumn{2}{c}{$\gamma_{14}$} \\
 $\alpha_S$ & $\theta$ & Avg. & Var & Avg. & Var & Avg. & Var & Avg. & Var & Avg. & Var & Avg. & Var & Avg. & Var & Avg. & Var \\
\midrule
 	1	&	$\rightarrow 0$	&	-0.95	&	0.03	&	-0.02	&	0.07	&	-0.02	&	0.07	&	0.00	&	0.07	&	-0.11	&	0.00	&	0.00	&	0.00	&	0.00	&	0.00	&	0.00	&	0.00	\\
 	50	&	$\rightarrow 0$	&	-0.39	&	0.00	&	0.00	&	0.00	&	-0.02	&	0.00	&	0.00	&	0.00	&	-0.26	&	0.01	&	0.00	&	0.00	&	0.00	&	0.01	&	0.00	&	0.00	\\
 	$\rightarrow \infty$	&	$\rightarrow 0$	&	-0.36	&	0.00	&	0.00	&	0.00	&	-0.02	&	0.00	&	0.00	&	0.00	&	-0.27	&	3.11	&	-0.01	&	1.08	&	0.03	&	1.10	&	0.00	&	1.04	\\
 	1	&	0.1	&	-0.94	&	0.03	&	-0.01	&	0.07	&	-0.02	&	0.07	&	-0.03	&	0.06	&	-0.11	&	0.00	&	0.00	&	0.00	&	0.00	&	0.00	&	0.00	&	0.00	\\
 	50	&	0.1	&	-0.39	&	0.00	&	0.00	&	0.00	&	-0.02	&	0.00	&	0.00	&	0.00	&	-0.26	&	0.01	&	0.00	&	0.00	&	0.00	&	0.01	&	0.00	&	0.00	\\
 	$\rightarrow \infty$	&	0.1	&	-0.36	&	0.00	&	0.00	&	0.00	&	-0.02	&	0.00	&	0.00	&	0.00	&	-0.28	&	2.80	&	0.02	&	0.98	&	-0.02	&	1.00	&	0.01	&	0.94	\\
 	1	&	0.5	&	-0.93	&	0.03	&	-0.02	&	0.06	&	-0.03	&	0.06	&	0.01	&	0.06	&	-0.11	&	0.00	&	0.00	&	0.00	&	0.00	&	0.00	&	0.00	&	0.00	\\
 	50	&	0.5	&	-0.39	&	0.00	&	0.00	&	0.00	&	-0.02	&	0.00	&	0.00	&	0.00	&	-0.25	&	0.01	&	0.00	&	0.00	&	0.00	&	0.00	&	0.00	&	0.00	\\
 	$\rightarrow \infty$	&	0.5	&	-0.37	&	0.00	&	0.00	&	0.00	&	-0.02	&	0.00	&	0.00	&	0.00	&	-0.29	&	1.64	&	0.02	&	0.57	&	-0.04	&	0.58	&	0.02	&	0.55	\\
\hline
	1	&	$\rightarrow 0$	&	0.26	&	0.06	&	0.33	&	0.06	&	0.43	&	0.06	&	0.74	&	0.04	&	0.03	&	0.00	&	0.03	&	0.00	&	0.05	&	0.00	&	0.08	&	0.00	\\
 	50	&	$\rightarrow 0$	&	0.11	&	0.00	&	0.14	&	0.00	&	0.17	&	0.00	&	0.30	&	0.00	&	0.06	&	0.01	&	0.09	&	0.01	&	0.12	&	0.01	&	0.21	&	0.01	\\
 	$\rightarrow \infty$	&	$\rightarrow 0$	&	0.10	&	0.00	&	0.13	&	0.00	&	0.16	&	0.00	&	0.29	&	0.00	&	0.06	&	1.17	&	0.07	&	1.25	&	0.15	&	1.56	&	0.20	&	2.36	\\
 	1	&	0.1	&	0.27	&	0.06	&	0.30	&	0.06	&	0.42	&	0.06	&	0.74	&	0.04	&	0.03	&	0.00	&	0.03	&	0.00	&	0.05	&	0.00	&	0.08	&	0.00	\\
 	50	&	0.1	&	0.11	&	0.00	&	0.14	&	0.00	&	0.17	&	0.00	&	0.30	&	0.00	&	0.06	&	0.01	&	0.08	&	0.01	&	0.12	&	0.01	&	0.20	&	0.01	\\
 	$\rightarrow \infty$	&	0.1	&	0.10	&	0.00	&	0.13	&	0.00	&	0.16	&	0.00	&	0.29	&	0.00	&	0.07	&	1.05	&	0.10	&	1.13	&	0.11	&	1.41	&	0.24	&	2.13	\\
 	1	&	0.5	&	0.24	&	0.06	&	0.33	&	0.06	&	0.42	&	0.06	&	0.72	&	0.04	&	0.03	&	0.00	&	0.04	&	0.00	&	0.05	&	0.00	&	0.09	&	0.00	\\
 	50	&	0.5	&	0.11	&	0.00	&	0.14	&	0.00	&	0.17	&	0.00	&	0.31	&	0.00	&	0.06	&	0.01	&	0.08	&	0.01	&	0.11	&	0.01	&	0.20	&	0.01	\\
 	$\rightarrow \infty$	&	0.5	&	0.10	&	0.00	&	0.13	&	0.00	&	0.16	&	0.00	&	0.29	&	0.00	&	0.07	&	0.62	&	0.12	&	0.66	&	0.09	&	0.83	&	0.24	&	1.24	\\
\bottomrule
\end{tabular}
\end{sidewaystable*}
\end{center}

\clearpage
\newpage

\section{Empirical data}

\subsection{The Special Turku Coronary Risk  factor Intervention Project}\label{STRIP}
The STRIP study (The Special Turku Coronary Risk  factor Intervention Project) is an  infancy-onset, longitudinal randomised controlled intervention study aiming to prevent atherosclerosis beginning in infancy\Citep{Pahkala2020}. 
Families of 5-month-old infants, born between July 1989 and December 1991, were recruited by nurses at well-baby clinics in Turku, Finland. At age 7 months, 1062 infants (56.5 \% of the eligible age-cohort) were randomly allocated to a dietary intervention (n=540) or a control (n=522) group.  A group of 45 children, born from March to July 1989, was also similarly recruited and randomised (intervention n=22, control n=23) to first test the study protocols and thus serve as a ‘pilot’ group.  In addition, the cohort included 2 children with Down syndrome (both control), 2 with familial hypercholesterolemia (intervention and control), and 5 children who had been randomised to the intervention group, but who missed the first study visits prior to age 13 months and were later treated as controls.

The study started in 1990, and altogether 1116 babies were recruited in well-baby clinics into the intervention (n=564) and control (n=552) groups. For 20 years, the intervention group received biannual dietary counselling focusing especially on the quality of dietary fat, whereas the control group visited the clinic annually for measurements but only received the basic health education given at Finnish well-baby clinics and school. 
When the participants were 26 years of age, i.e., 6 years after the end of the intervention, a follow-up study on the effects of the 20-year intervention was conducted with 551 participants from the original STRIP study population. 

During both the intervention phase and the follow-up, data have been collected on anthropometric measures, serum lipid profile and insulin sensitivity. Nutritional data were collected by completing a food diary on four consecutive days, including 1-2 weekend days. Portion sizes were estimated using household measures or a food picture booklet. A dietary technician  revised the diaries for completeness and accuracy. Food on nutritient intakes were quantified using the Micro Nutrica programme 
developed at the Research and Development Centre of the Social Insurance Institution (Turku, Finland).
The programme calculates values of 66 nutritiens in more than 4000 foods and dishes.
Serum insulin was measured with a chemiluminescent micro particle immunoassay (Architect insulin assay, Abbott, Chicago, IL, USA) on an Architect ci8200 analyzer (Abbott).  At the 26-year follow-up visit, the participants were asked for the first time to provide a faecal sample for assessing the gut microbiota.
 The raw sequence data were processed into an amplicon sequence variant (ASV) table using the \textit{DADA2} pipeline \Citep{Callahan2016}.

The study has shown that during the 20-year intervention period, intervention has resulted in phenotypic changes related to reduced risk of metabolic syndrome, cardiovascular disease and type 2 diabetes \Citep{Nupponen, Oranta2013, Niinikoski2012, Pahkala2013, Lehtovirta2018}. Recently, the study group reported certain effects of the intervention that were maintained 6 years post-intervention \Citep{Pahkala2020}. For example, those who received intervention had more favourable intake of saturated and unsaturated fat and  more often met the ideal total cholesterol and LDL cholesterol concentration. Males in the intervention group had higher fibre intake than males in the control group, and in total the intervention group  more often met the strict dietary target for fibre (3g/MJ). In the 26-year follow-up study the intervention group had a slightly but non-significantly lower insulin level than the control group, however the life-course insulin level between ages 7 and 26 was lower in the intervention group. Tehe prevalence of overweight or obesity did not differ between the groups. 

The STRIP study has been financially supported by the Academy of Finland (grants 206374, 294834, 251360, 275595, 307996, and 322112), the Juho Vainio Foundation, the Finnish Foundation for Cardiovascular Research, the Finnish Ministry of Education and Culture, the Finnish Cultural Foundation, the Sigrid Jusélius Foundation, Special Governmental grants for Health Sciences Research (Turku University Hospital), the Yrjö Jahnsson Foundation, the Finnish Medical Foundation, and the Turku University Foundation. 

\clearpage
\newpage

\subsection{Data exclusions}\label{strip_restrict}
\begin{center} 
\begin{figure}
  \includegraphics[width=18cm]  {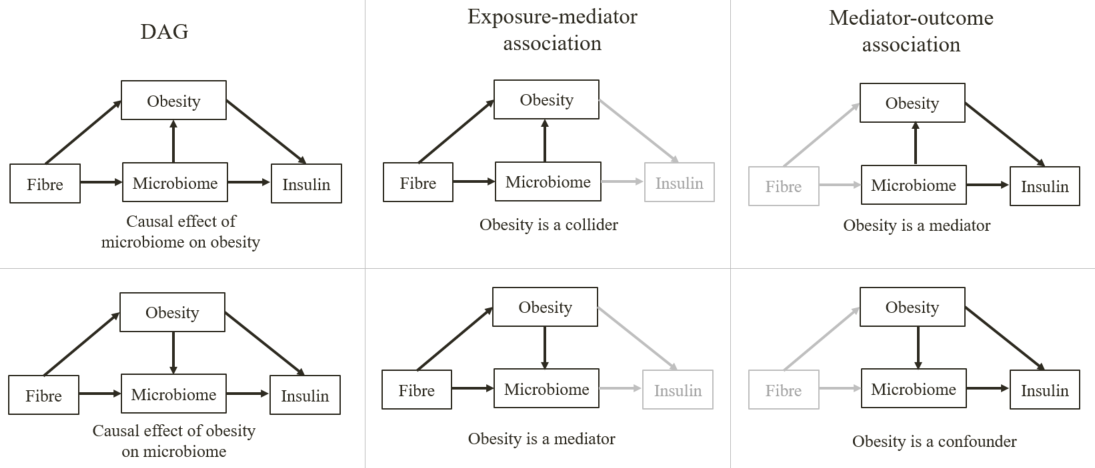}
\caption{DAGs describing the role of obesity in the mediation analysis question. If the microbiome had a causal role on obesity (top row), obesity would simultaneously function as a collider in the association between fibre intake and microbiome and a mediator in the association between microbiome and insulin level. If obesity had a causal role on the microbiome (bottom row), obesity would function as a mediator in the association between fibre intake and the microbiome and a confounder in the association between microbiome and insulin level. }
\label{problem}
\end{figure} 
\end{center}

In the empirical analyses, we excluded obese participants  (BMI $\geq 30 \hbox{kg/m}^2$) to avoid potential biases. As higher fibre intake is associated with lower overall food consumption, fibre intake may decrease body weight and risk of adiposity   \Citep{Barber2020}. Furthermore, obesity is linked to insulin resistance via the metabolism modulated by adipose tissue \Citep{Kahn206}. Thus, adiposity may function as a mediator between fibre intake and insulin level. 
In addition, gut microbiome may have an affect obesity causally \Citep{Gerard2016, Ridaura2014}. If such  link exists, obesity would function as a collider for the fibre-microbiome association and, simultaneously, as a mediator for the microbiome-insulin association (top row of  Figure \ref{problem}).  If there was reverse causality, i.e. if obesity influenced the microbiome, obesity would function simultaneously as both mediator and confounder  (bottom row of  Figure \ref{problem}). Controlling for obesity would thus either cause collider bias or prevent interpreting the estimates  as direct and indirect effects \Citep{VanderWeele2016}. 
 
\clearpage
\newpage
 
\subsection{Stratum-specific results}
We stratified the STRIP sample into four distinct strata based on sex and intervention group status.  The results presented in the main text are pooled over the strata using the stratum-specific sample sizes as weights. Of the 264 eligible participants, 68 were intervention girls, 59 intervention boys, 82 control girls and 55 control boys. Of these groups 17 (25\%), 17 (29\%), 
18 (22\%) and 11 (20\%) respectively, had sufficient fibre intake. 

We here present the results for both taxonomic and pivotal questions for each stratum separately. 

\subsubsection{Mediation through the genera within the \textit{Actinobacteria} phylum }
The stratum-specific results on mediation via the genera within the \textit{Actinobacteria} phylum are presented in Table \ref{phylocountit_strata} (average counts and proportions of the genera), 
Table \ref{ilr_avg_strata} (average values of the ilr coordinates),
Table \ref{medresult_strata} (the total effects, direct effects and OIEs of the mediation analysis) and 
Table \ref{phylotulos_strata} (the CIE's).  

The results are not as clear-cut in the individual strata as they are in the entire population. Interestingly, for the intervention girls the total effect is mostly due to the indirect effects (OIE -0.055, NDE -0.21). Furthermore, the fibre intake increases the fourth mediator (i.e. ilr coordinate) $M_4$ ($\hat{\beta}_{14} = 1.32 \hbox{ }(90\hbox{ } \% \hbox{ CI }  [0.73; 1.90]$). The mediating effect of $M_4$ for the intervention girls is -0.067 (90 \% CI [-0.17; 0.04]. In addition, the mediator $M_2$ has an effect on the log(insulin) levels ($\hat{\gamma}_{12} = -0.087$, 90 \% CI [-0.16; -0.01]).

For the intervention boys, the point estimate of the OIE via the genera within the \textit{Actinobacteria} phyla is 0.038, 90 \% CI [-0.11; 0.19]. In this group, fibre intake has an effect on the mediator $M_2$ and mediator $M_9$ (relative proportion of \textit{Senegalimassilia} contrasted to \textit{Olsenella} and \textit{Gordonibacter}). Furthermore, the level of the mediator $M_3$ increases 
whereas the level of the mediator $M_4$ decreases the log(insulin) levels. The single largest mediated effect is through mediator $M_2$, $\hat{\beta}_{12} \hat{\gamma}_{12} = 0.04$, 90 \% CI [-0.025; 0.11]. 

For the control girls, the total effect of fibre intake on log(insulin) level is -0.14, 90 \% CI [-0.31; 0.04]. However, the majority of the total effect is via direct effects. Fibre intake does not have a notable effect on any of the mediators. However, the fifth mediator has an effect on the log(insulin) level: $\hat{\gamma}_{15} = 0.05$, 90 \% CI (0.01; 0.09). 

For the control boys, the point estimate for the OIE is also positive (OIE = 0.037, 90 \% CI [-0.06; 0.14]). The fibre intake does not have an effect on the mediators, and furthermore, the mediators do not affect log(insulin) level. 

\begin{center}
\begin{table*}[t]%
\caption{\label{phylocountit_strata}Average counts and proportions of the genera within the \textit{Actinobacteria} by fibre intake (low/high). \label{phylocountit_strata}}
\centering
\begin{tabular}{l l l l l l }
\toprule
& & \multicolumn{2}{c}{Count} & \multicolumn{2}{c}{Proportion} \\
& Genus & Fibre < 25 & Fibre $\geq$ 25 &  Fibre < 25 & Fibre $\geq$ 25 \\
\midrule
IG& \textit{Bifidobacterium}&  3442.75 & 1393.56 & 68.70  & 57.37 \\
IG& \textit{Actinomyces}&         3.02   & 1.97 &  0.17 &  0.20 \\
IG& \textit{Rothia}&              2.52  &  2.44 &  0.16 &  0.21 \\
IG& \textit{Enterorhabdus}&       0.70   & 9.91  & 0.08  & 0.92 \\
IG& \textit{Eggerthella}&        40.88 &  41.65 &  3.05&   4.25 \\
IG& \textit{Collinsella}&       612.51 & 495.12 & 18.47 & 26.30 \\
IG& \textit{Adlercreutzia}&      89.31 & 125.88 &  5.72  & 7.33 \\
IG& \textit{Slackia}&            18.30 &  59.44 &  0.67   &1.22 \\
IG& \textit{Senegalimassilia}&   50.35  & 10.76 &  1.33  & 1.40 \\
IG& \textit{Olsenella}&          14.94  &  2.44 &  1.01  & 0.31 \\
IG& \textit{Gordonibacter}&       8.30  &  5.00 &  0.62  & 0.49 \\ 
\midrule
IG&  Total &           4283.59& 2148.18 & 100 & 100 \\
\midrule
IB& \textit{Bifidobacterium}&  3104.05&  3274.29 &  61.10&  74.95 \\
IB& \textit{Actinomyces}&         3.08&    1.09&   0.25&   0.15 \\
IB& \textit{Rothia}&              0.77&    0.50&   0.21&   0.13 \\
IB& \textit{Enterorhabdus}&      13.42&    0.85&   0.68&   0.30 \\
IB& \textit{Eggerthella}&        21.17&   22.88&   2.07&   1.84 \\
IB& \textit{Collinsella}&       734.55&  329.06&  26.30&  11.32 \\
IB& \textit{Adlercreutzia}&      57.36&  101.38&   4.80&   8.06 \\
IB& \textit{Slackia}&            26.83&    7.91&   1.59&   1.74 \\
IB& \textit{Senegalimassilia}&   36.36&    4.74&   2.10&   0.75 \\
IB& \textit{Olsenella}&          16.25&   10.59&   0.73&   0.43 \\
IB& \textit{Gordonibacter}&       3.39&    4.41&   0.19&   0.33 \\
\midrule
IB& Total   &         4017.23& 3757.71  & 100 & 100 \\
\midrule
CG& \textit{Bifidobacterium}&  4200.06&  788.22&  67.66 & 53.77 \\
CG& \textit{Actinomyces}&         3.56&    0.97&   0.14 &  0.12 \\
CG& \textit{Rothia}&              1.55&    1.50&   0.08 &  0.15 \\
CG& \textit{Enterorhabdus}&      14.81&    5.56&   0.55&   0.80 \\
CG& \textit{Eggerthella}&        69.95&   21.28&   4.77&   1.92 \\
CG& \textit{Collinsella}&      626.64&  428.47&  18.81 & 35.75 \\
CG& \textit{Adlercreutzia}&      62.89&   42.67&   4.36&   4.60 \\
CG& \textit{Slackia}&            16.09 &   7.36&   1.38&   1.43 \\
CG& \textit{Senegalimassilia}&   17.14 &   1.39&   1.11&   0.80 \\
CG& \textit{Olsenella}&          13.33&    1.42&   0.70&   0.29 \\
CG& \textit{Gordonibacter}&      10.15&    7.17&   0.44&   0.38 \\
\midrule
CG&  Total   &         5036.18 &1306.00  & 100 & 100 \\
\midrule
CB& \textit{Bifidobacterium}&  4206.14 & 2180.64 &  66.38 & 64.15 \\
CB& \textit{Actinomyces}&         3.45 &   2.18 &  0.17 &  0.18 \\
CB& \textit{Rothia}&              3.40 &   1.23 &  0.17 &  0.06 \\
CB& \textit{Enterorhabdus}&       3.97 &   0.50 &  0.34 &  0.05 \\
CB& \textit{Eggerthella}&        72.56 &  37.05 &  1.26 &  2.34 \\
CB& \textit{Collinsella}&       864.17 & 505.14 & 21.81 & 26.81 \\
CB& \textit{Adlercreutzia}&     186.91 &  54.59 &  6.29 &  3.86 \\
CB& \textit{Slackia}&            32.19 &   5.50 &  1.88 &  0.48 \\
CB& \textit{Senegalimassilia}&   24.82 &  17.36 &  0.73 &  0.38 \\
CB& \textit{Olsenella}&          13.28 &  22.68 &  0.75 &  1.54 \\
CB& \textit{Gordonibacter}&       5.80 &   2.09 &  0.21 &  0.14 \\
\midrule
CB& Total       &     5416.68 &2828.95  & 100 & 100 \\
\bottomrule
\end{tabular}
\end{table*}
\end{center}

\begin{center}
\begin{table*}[t]%
\caption{Average values of the ilr coordinates for the fibre groups in each strata. \label{ilr_avg_strata}}
\centering
\begin{tabular}{llll}
\toprule
& & \multicolumn{2}{c}{Average ilr} \\
Ilr & Group & Fibre <25 & Fibre $\geq$ 25 \\
\midrule
$M_1$&       IG& -2.41& -2.62 \\
$M_2$&       IG&  0.02& -0.16 \\
$M_3$&       IG&  5.18&  4.50 \\
$M_4$&       IG& -2.15& -0.83 \\
$M_5$&       IG&  0.44&  0.40 \\
$M_6$&       IG&  3.35&  3.72 \\
$M_7$&       IG&  2.26&  2.42 \\
$M_8$&       IG& -0.47& -0.14 \\
$M_9$&       IG& -0.28&  0.37 \\
$M_{10}$&      IG&  0.31& -0.42 \\
\midrule
$M_1$&     IB& -2.75 &-2.89 \\
$M_2$&    IB & 0.66 & 0.20 \\
$M_3$&     IB&  4.91&  5.23 \\
$M_4$&     IB& -1.44& -1.66 \\
$M_5$&     IB& -0.26&  0.03 \\
$M_6$&     IB&  4.07&  3.51 \\
$M_7$&     IB&  1.92&  2.67 \\
$M_8$&     IB&  0.21& -0.24 \\
$M_9$&     IB&  0.69& -0.42 \\
$M_{10}$&    IB&  0.35&  0.16 \\
\midrule
$M_1$&     CG& -2.77& -2.62 \\
$M_2$&     CG&  0.25& -0.11 \\
$M_3$&    CG & 5.03 & 4.42 \\
$M_4$&     CG& -1.55& -1.08 \\
$M_5$&     CG&  0.86&  0.14 \\
$M_6$&     CG&  3.39&  4.15 \\
$M_7$&     CG&  2.20&  1.99 \\
$M_8$&     CG&  0.02&  0.33 \\
$M_9$&     CG& -0.17& -0.34 \\
$M_{10}$&    CG& -0.13& -0.47 \\
\midrule
$M_1$&    CB &-2.52& -2.47 \\
$M_2$&     CB&  0.32&  0.36 \\
$M_3$&     CB&  5.23&  5.36 \\
$M_4$&     CB& -1.96& -2.08 \\
$M_5$&     CB& -0.30& -0.33 \\
$M_6$&     CB&  3.86&  4.13 \\
$M_7$&     CB&  3.08&  2.73 \\
$M_8$&     CB&  0.04& -0.30 \\
$M_9$&     CB& -0.43& -0.56 \\
$M_{10}$&    CB&  0.39&  1.28 \\
\bottomrule
\end{tabular}
\end{table*}
\end{center}

\begin{center}
\begin{table*}[t]%
\caption{The total effect, direct effect and overall indirect effect via the genera within the \textit{Actinobacteria} phylum for each stratum. \label{medresult_strata}}
\centering
\begin{tabular}{lllllll}
\toprule
Group & TE & (90 \% CI) & DE & (90 \% CI) & OIE &(90 \% CI) \\
\midrule
IG & -0.076  &(-0.26;  0.11) &-0.021 &(-0.24; 0.20) & -0.055 &(-0.20; 0.093) \\ 
IB & -0.10  &(-0.29;  0.09) &-0.14&(-0.34; 0.058) & 0.038 &(-0.11; 0.19) \\          
CG & -0.14 &(-0.31;  0.04) &-0.093 &(-0.29; 0.10) & -0.042 &(-0.14; 0.05) \\          
CB & -0.11 &(-0.33;  0.11) &-0.14 &(-0.39; 0.10) & 0.037 &(-0.06; 0.14) \\  
\bottomrule
\end{tabular}
\end{table*}
\end{center}

\begin{center}
\begin{table*}[t]%
\caption{\label{phylotulos_strata} Stratum-specific coefficients for the effect of fibre on the ilr coordinates and effects of  ilr coordinates on the response, and the coordinate-wise indirect effects $\beta_{1k} \gamma_{1k}$  for each coordinate. }
\centering
\begin{tabular}{l l l l l l l l} 
\toprule
\midrule
&  & $\hat{\beta}_{1k}$   & (90$\%$ CI)      &  $\hat{\gamma}_{1k}$ &  (90$\%$ CI) & $\hat{\beta}_{1k} \hat{\gamma}_{1k}$ (s.e.) & (90$\%$ CI) \\
\hline
$M_1$      &IG&  -0.21  &(-0.84;  0.43)  &    0.04   &(-0.03;   0.11)  &    -0.008        (0.017)  &         (-0.036;          0.020) \\
$M_2$      &IG&  -0.19  &(-0.73;  0.36)  &   -0.09   &(-0.16;  -0.01)  &     0.016        (0.030)   &        (-0.033;           0.065) \\
$M_3$      &IG&  -0.69  &(-1.62;  0.24)  &    0.02   &(-0.02;   0.06)  &    -0.016        (0.022)   &        (-0.052;          0.020) \\
$M_4$      &IG&   1.32  &( 0.73;  1.90)  &   -0.05   &(-0.13;   0.02)  &    -0.067        (0.063)   &        (-0.170;          0.036) \\
$M_5$      &IG&  -0.05  &(-1.06;  0.97)  &    0.02   &(-0.03;   0.07)  &    -0.001        (0.012)   &        (-0.021;           0.019) \\
$M_6$      &IG&   0.37  &(-0.70;  1.44)  &    0.002   &(-0.04;   0.04)  &     0.001        (0.009)   &        (-0.013;           0.015) \\
$M_7$      &IG&   0.16  &(-0.83;  1.15)  &   -0.003   &(-0.05;   0.04)  &    -0.001        (0.005)   &        (-0.009;           0.007) \\
$M_8$      &IG&   0.32  &(-0.47;  1.12)  &   -0.01   &(-0.06;   0.04)  &   -0.004        (0.011)   &        (-0.022;           0.015) \\
$M_9$      &IG&   0.65  &(-0.18;  1.48)  &    0.03   &(-0.03;   0.08)  &     0.017        (0.026)   &        (-0.025;           0.059) \\
$M_{10}$     &IG&  -0.73  &(-1.65;  0.20)  &   -0.01   &(-0.06;   0.04)  &     0.007        (0.021)   &        (-0.028;           0.042) \\
\midrule
$M_1$     &IB&  -0.14  &(-0.81;  0.54)  &    0.001   &(-0.08;   0.08)  &     0.000        (0.006)   &        (-0.011;           0.010) \\
$M_2$     &IB&  -0.46  &(-0.84; -0.08)  &   -0.09   &(-0.21;   0.04)  &     0.040        (0.040)   &        (-0.025;           0.106) \\
$M_3$     &IB&   0.32  &(-0.48;  1.13)  &    0.1    &(0.04;   0.15)  &     0.032        (0.049)   &        (-0.049;           0.113) \\
$M_4$     &IB&  -0.22  &(-0.96;  0.51)  &   -0.08   &(-0.14;  -0.017)  &     0.017        (0.035)   &        (-0.041;           0.075) \\
$M_5$     &IB&   0.29  &(-0.73;  1.30)  &    0.04   &(-0.005;   0.09)  &     0.012        (0.026)   &        (-0.031;           0.054) \\
$M_6$     &IB&  -0.56  &(-1.31;  0.20)  &    0.02   &(-0.04;   0.08)  &    -0.011        (0.023)   &        (-0.048;           0.026) \\
$M_7$     &IB&   0.75  &(-0.24;  1.74)  &   -0.02   &(-0.06;   0.03)  &    -0.011        (0.024)   &        (-0.050;           0.028) \\
$M_8$     &IB&  -0.44  &(-1.29;  0.41)  &    0.04   &(-0.01;   0.09)  &    -0.017        (0.024)   &        (-0.057;           0.022) \\
$M_9$     &IB&  -1.12  &(-2.00; -0.24)  &    0.02   &(-0.03;   0.07)  &    -0.023        (0.036)   &        (-0.082;           0.036) \\
$M_{10}$    &IB&  -0.19  &(-1.04;  0.65)  &   -0.004   &(-0.06;   0.05)  &     0.001        (0.007)   &        (-0.010;           0.012) \\
\midrule
$M_1$     &CG&   0.16  &(-0.37;  0.68)  &    0.02   &(-0.05;   0.09)  &     0.003        (0.009)   &        (-0.012;           0.018) \\
$M_2$     &CG&  -0.36  &(-0.78;  0.07)  &   -0.003   &(-0.09;   0.08)  &     0.001        (0.018)   &        (-0.029;           0.031) \\
$M_3$     &CG&  -0.61  &(-1.54;  0.31)  &   -0.002   &(-0.04;   0.04)  &     0.001        (0.015)   &        (-0.023;           0.025) \\
$M_4$     &CG&   0.47  &(-0.24;  1.18)  &   -0.008   &(-0.06;   0.04)  &    -0.004        (0.014)   &        (-0.027;           0.020) \\
$M_5$     &CG&  -0.72  &(-1.72;  0.28)  &    0.05   &( 0.01;   0.09)  &    -0.036        (0.035)   &        (-0.093;           0.021) \\
$M_6$     &CG&   0.76  &(-0.15;  1.67)  &    0.007   &(-0.04;   0.05)  &     0.005        (0.020)   &        (-0.028;           0.038) \\
$M_7$     &CG&  -0.21  &(-1.11;  0.70)  &    0.03   &(-0.02;   0.07)  &    -0.006        (0.016)   &        (-0.033;           0.021) \\
$M_8$     &CG&   0.31  &(-0.50;  1.12)  &    0.02   &(-0.02;   0.07)  &     0.007        (0.013)   &        (-0.015;           0.029) \\
$M_9$     &CG&  -0.17  &(-0.93;  0.58)  &    0.01   &(-0.04;   0.07)  &    -0.002        (0.009)   &       (-0.017;           0.012) \\
$M_{10}$    &CG&  -0.34  &(-1.15;  0.47)  &    0.03   &(-0.01;   0.08)  &    -0.011        (0.019)   &        (-0.043;           0.020) \\
\midrule
$M_1$     &CB&   0.05  &(-0.69;  0.79)  &    0.07   &(-0.009;   0.15)  &     0.003        (0.031)   &        (-0.048;           0.054) \\
$M_2$     &CB&   0.03  &(-0.58;  0.65)  &    0.06   &(-0.05;   0.16)  &     0.002        (0.021)   &        (-0.033;           0.037) \\
$M_3$     &CB&   0.13  &(-0.94;  1.19)  &   -0.01   &(-0.07;   0.05)  &    -0.001        (0.008)   &        (-0.015;           0.012) \\
$M_4$     &CB&  -0.13  &(-0.87;  0.62)  &   -0.04   &(-0.12;   0.04)  &     0.005        (0.019)   &        (-0.026;           0.036) \\
$M_5$     &CB&  -0.02  &(-1.30;  1.25)  &    0.04   &(-0.02;   0.09)  &    -0.001        (0.029)   &        (-0.049;           0.047) \\
$M_6$     &CB&   0.28  &(-0.80;  1.35)  &    0.02   &(-0.04;   0.07)  &     0.005        (0.016)   &        (-0.020;           0.031) \\
$M_7$     &CB&  -0.36  &(-1.36;  0.65)  &   -0.02   &(-0.09;   0.05)  &     0.006        (0.018)   &        (-0.024;           0.035) \\
$M_8$     &CB&  -0.33  &(-1.49;  0.82)  &    0.008   &(-0.04;   0.06)  &    -0.003        (0.011)   &        (-0.021;           0.016) \\
$M_9$     &CB&  -0.14  &(-1.11;  0.84)  &    0.004   &(-0.05;   0.06)  &    -0.001        (0.006)   &        (-0.010;           0.008) \\
$M_{10}$    &CB&   0.88  &(-0.12;  1.88)  &    0.02   &(-0.05;   0.09)  &     0.021        (0.040)   &        (-0.045;           0.088) \\
\bottomrule
\end{tabular}
\end{table*}
\end{center}

 \clearpage
\newpage

\subsubsection{Mediation through the \textit{Actinobacteria} phylym}
Table \ref{strata_hyp1_count} presents the stratum-specific total counts on the phyla level, counts of the \textit{Actinobacteria} phylum and the corresponding average proportions. Table \ref{Medhyp1perstratum} presents the mediation analysis coefficients for the pivotal ilr coordinate with relative proportion of the \textit{Actinobacteria} as a pivot element. 
 In all other groups except control boys, the proportion of \textit{Actinobacteria} is smaller in those with sufficient fibre intake. For the girls especially, sufficient fibre intake seems to have a lowering effect on the pivotal ilr coordinate, whereas for the boys the effect is not as clear. However, only for the intervention boys does the pivotal ilr coordinate affects the insulin levels.

The total effect of fibre intake on log(insulin) level was similar for each stratum:  -0.076 (90\% CI [-0.26; 0.11]) for intervention girls, 
 -0.10 (90\% CI [-0.29; 0.09]) for intervention boys,
-0.14 (90\% CI [-0.31; 0.04]) for control girls and 
-0.11 (90\% CI [-0.33; 0.11]) for control boys. 

The corresponding direct effects were 
-0.02 (90\% CI [-0.22; 0.18])  for intervention girls,
-0.10 (90\% CI [-0.30; 0.11]) for intervention boys,
 -0.13 (90\% CI [-0.33; 0.07])  for control girls and 
 -0.15 (90\% CI [-0.49; 0.10]) for control boys.

\begin{center}
\begin{table*}[t]%
\caption{The average total count, count of \textit{Actinobacteria}, and proportion of \textit{Actinobacteria} for the strata in the STRIP sample. 
Abbreviations: IG=intervention girls; IB=intervention boys; CG=control girls; CB=control boys. \label{strata_hyp1_count}}
\centering
\begin{tabular}{l l l l l l l l l}
\toprule
 & \multicolumn{2}{l}{Mean Total count}  & \multicolumn{2}{l}{Mean \textit{Actinobacteria}} & \multicolumn{2}{l}{Prop \textit{Actinobacteria} (\%)}  & &\\
  & Fibre < 25 & Fibre $\geq$ 25 & Fibre < 25 & Fibre $\geq$ 25  &  Fibre < 25 & Fibre $\geq$ 25  \\
\midrule 
IG & 168140 & 216667 & 4333 & 2180 & 2.7 & 1.4 \\
IB & 166344 & 147828 &   4068  & 3772 &   2.6 & 2.7 \\
CG & 190337 & 168182 &  5119 & 1319 & 2.5 & 0.9 \\
CB & 201420 & 171253 &  5468 &  2854 & 3.0 & 1.7 \\ 
\bottomrule
\end{tabular}
\end{table*}
\end{center}

\begin{center}
\begin{table*}[t]%
\caption{Stratum-specific results for mediation analysis via mediator with \textit{Actinobacteria} as the pivot element. Abbreviations: IG=intervention girls; IB=intervention boys; CG=control girls; CB=control boys. \label{Medhyp1perstratum}}
\centering
\begin{tabular}{l l l l l l}
\toprule
&  $\hat{\beta}_{11}$ (90\% CI) & $\hat{\gamma}_{11}$ (90\% CI)  & $\hat{\beta}_{11}\hat{\gamma}_{11}$ (90\% CI) \\
\midrule  
IG&   -0.58  (-1.13; -0.02) &  0.052   (-0.03;   0.14) &   -0.030          (-0.09;           0.03) \\
IB& -0.13  (-0.77;  0.50) &  0.19    (0.08;   0.29) &    -0.025           (-0.14;           0.09) \\
CG& -0.66  (-1.23; -0.09) & -0.011   (-0.09;   0.07) &     0.007           (-0.05;           0.06) \\
CB&  -0.32  (-1.08;  0.44) & -0.047   (-0.15;   0.06) &     0.015           (-0.03;           0.06) \\
\bottomrule
\end{tabular}
\end{table*}
\end{center}

\end{document}